\documentclass{tran-l}

\usepackage{array}
\usepackage{arydshln}

\usepackage{bm}
\usepackage{mathrsfs}
\usepackage{amssymb}
\usepackage{enumitem}
\usepackage{times}
\usepackage{pdfpages}
\usepackage{float}
\usepackage{subfig,graphicx}
\usepackage{csquotes}
\usepackage{hyperref}
\usepackage{amsmath}
\usepackage{ulem}
\usepackage{lipsum} 
\usepackage{nccmath}
\usepackage{bigints}
\usepackage{natbib}
\usepackage{multirow}
\usepackage{adjustbox}
\usepackage{threeparttable}
\usepackage{bbm}

\usepackage{pdfpages}
\usepackage{wrapfig}
\usepackage{color}

\usepackage{float}
\usepackage{tabularx}
\usepackage{caption}
\usepackage{booktabs}
\usepackage{cite}

\usepackage{graphicx}

\newtheorem{theorem}{Theorem}
\newtheorem{lemma}{Lemma}

\newtheorem{remark}{Remark}

\numberwithin{property}{section}
\numberwithin{proposition}{section}
\numberwithin{equation}{section}
\numberwithin{theorem}{section}
\numberwithin{corollary}{section}
\numberwithin{lemma}{section}
\numberwithin{def1}{section}
\numberwithin{corollary}{section}
\numberwithin{example}{section}
\numberwithin{remark}{section}

\begin{document}

\title[Bootstrapping Lasso in Generalized Linear Models]{Bootstrapping Lasso in Generalized Linear Models}

\author{Mayukh Choudhury}
\address{Department of Mathematics, Indian Institute of Technology Bombay, Mumbai 400076, India}
\email{214090002@iitb.ac.in}
\thanks{The authors thank Prof. Shuva Gupta for many helpful discussions.}

\author{Debraj Das}
\address{Department of Mathematics, Indian Institute of Technology Bombay, Mumbai 400076, India}
\email{debrajdas@math.iitb.ac.in}




\keywords{gamma regression, GLM, Lasso, logistic regression, PB, PRB.}

\begin{abstract}
Generalized linear model or GLM constitutes a large class of models and essentially extends the ordinary linear regression by connecting the mean of the response variable with the covariate through appropriate link functions. On the other hand, Lasso is a popular and easy-to-implement penalization method in regression when not all covariates are relevant. However, the asymptotic distributional properties the Lasso estimator in GLM is still unknown. In this paper, we show that the Lasso estimator in GLM does not have a tractable form and subsequently, we develop two Bootstrap methods, namely the Perturbation Bootstrap and Pearson's Residual Bootstrap methods, for approximating the distribution of the Lasso estimator in GLM. As a result, our Bootstrap methods can be used to draw valid statistical inferences for any sub-model of GLM. We support our theoretical findings by showing good finite-sample properties of the proposed Bootstrap methods through a moderately large simulation study. We also implement one of our Bootstrap methods on a real data set.
\end{abstract}

\maketitle

\section{Introduction}\label{sec:intro}
Generalized Linear Model (or GLM) is a uniform modelling technique, formulated by \citet{nelder1972generalized}. GLM encompasses several submodels such as linear regression, logistic regression, probit regression, Poisson regression, gamma regression, etc. GLM maps the response variables with the covariates through a link function, different choices of which gives rise to different submodels. As a result, many real-life scenarios can be modeled using GLM. Now we define the general structure of GLM. Let $\{y_{1},..,y_{n}\}$ be independent responses and $\{\bm{x}_1,\dots, \bm{x}_n\}$ be non-random covariates. Assume that $y_i$ has density $f_{\theta_i}(y_{i}) = \mbox{exp}\big\{y_{i}\theta_{i} - b(\theta_{i})\big\}c(y_{i})$, for all $i=1,\dots, n$, with respect to a common measure. $(\theta_1,\dots, \theta_n)^\top$ is the canonical parameter vector which lies in an open set. The dependency of the response $y_i$ on the covariate $\bm{x}_i$ is characterized by a link function $g(\cdot)$, defined by $g(\mu_{i}) = \bm{x}^\top_{i}\bm{\beta}$, where $\bm{\beta}$ is the regression parameter. Note that for all  $i=1,\dots, n$, $\mu_{i} = \mathbbm{E}(y_{i}) = b^\prime(\theta_{i})$ and $\mathbbm{v}\text{ar}(y_i)=b^{\prime\prime}(\theta_i)$, and  
hence $\theta_{i} = h(\bm{x}^\top_{i}\bm{\beta})$ where $h = (g \circ b^{\prime})^{-1}$, assuming its existence. Therefore, the log-likelihood as a function of $\bm{\beta}$ is given by $$\sum_{i=1}^{n}\ell_{ni}(\bm{\beta}) = \sum_{i=1}^{n}\big[y_ih(\bm{x}_i^\top \bm{\beta}) - b(h(\bm{x}_i^\top \bm{\beta}))\big].$$

The popularity of GLM lies in the fact that many real-life scenarios can be modeled with one of the sub-models of GLM. Linear regression evaluates the relationship between two variables: a continuous dependent variable and one (usually continuous) independent variable, with the dependent variable expressed as a linear function of the independent variable. Here, the link function is the identity function. One of the most useful methods in the field of medical sciences, clinical trials, surveys etc. is the logistic regression when the response variable is dichotomous or binary. \citet{berkson1944application} introduced the `logit' link function as a pivotal instrument and later,  in his seminal paper, \citet{cox1958regression} familiarized it in the field of regression when the response variable is binary. In risk modelling or insurance policy pricing, Poisson regression is ideal provided response variable is the number of claim events per year. On the other hand, duration of interruption as a response variable lead to gamma regression in predictive maintenance. In both Poisson and gamma regression, generally the `log' link function is utilized. Usual forms of the different components present in the GLM for important submodels are presented in Table \ref{tab1}.
\begin{table*}[ht]
\centering
\caption{Some Common Types of GLM and their Components}
\label{tab1}
\begin{tabular}{@{}lcrcrrr@{}}
\hline
 &\multicolumn{4}{c}{Components of GLM} \\
\cline{2-5}
Regression Type &
\multicolumn{1}{c}{$\mu=b^\prime (\theta)$} &
\multicolumn{1}{c}{Link Function $(g(\cdot))$}&
\multicolumn{1}{c}{$h(u)$}&
\multicolumn{1}{c@{}}{$b(h(u))$}\\
\hline
Linear & $\theta$ & identity & $u$ & $u^2/2$\\
Logistic & $e^{\theta}(1+e^{\theta})^{-1}$  & logit & $u$ & $\log(1+e^u)$ \\
Probit & $e^{\theta}(1+e^{\theta})^{-1}$ & probit & $\log\big\{\frac{\Phi(u)}{1-\Phi(u)}\big\}$ &$-\log\{1-\Phi(u)\}$ \\
Poisson    & $e^{\theta}$ & $\log$ & $u$ & $e^u$ \\
Gamma & $-\alpha\theta^{-1}$ & $\log$ & $-\alpha e^{-u}$ & $\alpha u$    \\
\hline
\end{tabular}
\begin{tablenotes}%
\item $\alpha:$ known shape parameter in gamma distribution.
\item[$^{1}$] $\Phi:$ cumulative distribution function of the standard normal distribution.
\end{tablenotes}
\end{table*} 
When we try to draw an inference about the parameter in a regression set-up, the first thing we generally check is whether all the covariates are relevant, i.e. whether the parameter $\bm{\beta}$ actually sits in a lower dimensional space. The most popular way to identify this and simultaneously draw inferences about the underlying unknown parameters is to employ Lasso, introduced by \citet{tibshirani1996regression}. In this paper, our aim is to perform statistical inference on the regression parameter $\bm{\beta}$ based on Lasso which remains valid for any submodel of GLM. Following \citet{friedman2010regularization} and \citet{buhlmann2011statistics}, the Lasso estimator in GLM is defined as
\begin{align}\label{eqn:def}
\hat{\bm{\beta}}_{n} \equiv \hat{\bm{\beta}}_{n}(\lambda_n) \in \mbox{Argmin}_{\bm{\beta}} \Big\{-\sum_{i=1}^{n}\ell_{ni}(\bm{\beta}) +\lambda_{n}\sum_{j=1}^{p}|\beta_{j}|\Big\},
\end{align}
which is nothing but the $l_1-$penalized negative log-likelihood. Here $\lambda_n \;(>0)$ is the penalty parameter which is essentially helping to point out the relevant covariates, i.e., inducing sparsity in the model. Under the convexity of $-h(\cdot)$ and $b(h(\cdot))$, which we assume in Section \ref{sec:assump}, $\hat{\bm{\beta}}_n$ is unique in (\ref{eqn:def}). It is natural to carry out statistical inference on $\bm{\beta}$ based on the asymptotic distribution of $\hat{\bm{\beta}}_n$. The first result in this direction was due to \citet{knight2000asymptotics} who found that the asymptotic distribution of the usual Lasso estimator in linear regression does not have a closed form. In the same spirit, we also show that the asymptotic distribution of $\hat{\bm{\beta}}_n$ is intractable; see Theorem \ref{thm:lassolimit} for details. Therefore, it is essential to explore an alternative approximation technique for the distribution of $\hat{\bm{\beta}}_n$, a natural choice being the Bootstrap.

In this paper, we explore two Bootstrap methods, namely the Perturbation Bootstrap and the Pearson's Residual Bootstrap, for approximating the distribution of $\hat{\bm{\beta}}_n$. The idea of Perturbation Bootstrap is based on perturbing the underlying objective function using random quantities. It was first defined by \citet{jin2001simple} in a U-process structure and later popularized by the works \citet{minnier2011perturbation} and \citet{das2019perturbation}. On the other hand, Pearson's Residual Bootstrap is an analogue of classical Residual Bootstrap of \citet{freedman1981bootstrapping} and is based on resampling the standardized Pearson residuals. The classical Residual Bootstrap generally fails in GLM due to heteroscedastic nature of the responses. Hence standardization of the residuals is essential before resampling. Pearson's Residual Bootstrap in GLM was first devised by \citet{moulton1986bootstrapping} and its asymptotic properties were explored by \citet{simonoff1988jackknifing}, \citet{moulton1989,moulton1991}, \citet{lee1990bootstrapping}; among others. One of the nice features of Pearson's Residual Bootstrap method in GLM is that the Bootstrap estimator is essentially a least square estimator and hence can easily be extended to Lasso. However, Pearson's Residual Bootstrap may not remain valid when the covaraites are random whereas the Perturbation Bootstrap method works irrespective of the nature of the covariates. This important feature of the Perturbation Bootstrap method is pointed out in Remark \ref{rem:randomassump}.  For details on the construction of the Perturbation Bootstrap estimator $\hat{\bm{\beta}}_n^{*(PB)}$ and the Pearson's Residual Bootstrap estimator $\hat{\bm{\beta}}_n^{*(PRB)}$, see Section \ref{sec:desboot}.

We are interested in the Bootstrap approximation of the distribution of $\bm{T}_n =n^{1/2}(\hat{\bm{\beta}}_n -\bm{\beta})$, the centered and scaled version of the Lasso estimator. In either of the Bootstrap approximations, most natural is to define the Bootstrap estimators based on Lasso residuals and then to define the Bootstrap versions of $\bm{T}_n$ as $\hat{\bm{T}}_n^{*(PB)} =n^{1/2}(\hat{\bm{\beta}}_n^{*(PB)} - \hat{\bm{\beta}}_n)$ and $\hat{\bm{T}}_n^{*(PRB)} =n^{1/2}(\hat{\bm{\beta}}_n^{*(PRB)} -\hat{\bm{\beta}}_n)$, respectively for Perturbation and Pearson's Residual Bootstrap methods. However, we show in  Theorem \ref{prop:ntheo} that the Bootstrap distribution of $\hat{\bm{T}}_n^{*(PB)}$ converges to some non-degenerate random measure instead of the distributional limit of $\bm{T}_n$ and hence $\hat{\bm{T}}_n^{*(PB)}$ can not be used a Bootstrap pivotal quantity. This failure is due to the inability of $\hat{\bm{\beta}}_n$ to capture the true signs of the zero components of $\bm{\beta}$, as pointed out in Section \ref{sec:naivepb}. Due to the same reason, $\hat{\bm{T}}_n^{*(PRB)}$ can also be shown to fail in approximating the distribution of $\bm{T}_n$. The failure of such `naive' Bootstrap methods was already observed in the literature by \citet{chatterjee2010asymptotic}, \citet{camponovo2015validity} and \citet{das2019distributional} in linear regression. As a remedy, following the prescription of \citet{chatterjee2011bootstrapping}, we define a thresholded Lasso estimator $\tilde{\bm{\beta}}_n$ (see Section \ref{sec:modpb}) and consider residuals based on $\tilde{\bm{\beta}}_n$. Subsequently, we consider the Bootstrap estimators $\hat{\bm{\beta}}_n^{*(PB)}$ and $\hat{\bm{\beta}}_n^{*(PRB)}$ based on those modified residuals and define the modified Bootstrap pivotal quantities as $\tilde{\bm{T}}_n ^{*(PB)} = n^{1/2}(\hat{\bm{\beta}}_n^{*(PB)}-\tilde{\bm{\beta}}_n)$ and $\tilde{\bm{T}}_n ^{*(PRB)} = n^{1/2}(\hat{\bm{\beta}}_n^{*(PRB)}-\tilde{\bm{\beta}}_n)$. We establish in Theorem \ref{thm:ptheo} and \ref{thm:prbpos} that these modified Bootstrap pivotal quantities can correctly approximate the distribution of $\bm{T}_n$. 

The main difficulty in handling the Lasso estimator in GLM over the same in linear regression is that the objective function does not have a closed polynomial form and a suitable quadratic approximation of it through Taylor's theorem is necessary in order to perform asymptotic analysis. The approximation error also needs to be handled carefully so that the $Argmin$'s of the original and the approximate objective functions are close in almost sure sense. See lemma \ref{lem:asargmin} and the lemma \ref{lem:asconcentration} for details. Furthermore, we have established a version of the convergence of the distribution of $Argmin$ of convex stochastic processes in Lemma \ref{lem:argmin} and used it to establish the main results. The distribution convergence of convex stochastic processes has been studied by many authors including \citet{pollard1990empirical}, \citet{davis1992m}, \citet{hjort1993asymptotics}, \citet{geyer1996asymptotics}, \citet{kato2009asymptotics} and \citet{ferger2021continuous}; among others. We have established the distribution convergence of $Argmin$ of a convex stochastic process under finite-dimensional convergence and stochastic equicontinuity on compact sets, provided the limiting process has a unique minimizer. This result is in contrast with the epiconvergence tools, generally used under convexity. Lemma \ref{lem:asargmin} may be of independent interest in other related problems, since stochastic equicontinuity on compact sets generally holds when the underlying convex process has a nice form, as in case of Lasso.

We conclude this section with a brief literature review on the Bootstrap methods in Lasso. \citet{knight2000asymptotics} was the first to explore the Residual Bootstrap in Lasso in linear regression. \citet{chatterjee2010asymptotic} considered the linear regression model with non-random covariates and homoscedastic errors and showed that Residual Bootstrap fails. Subsequently, under the same settings, \citet{chatterjee2011bootstrapping} developed a Residual Bootstrap method which can approximate the distribution of the Lasso estimator. Later \citet{camponovo2015validity} developed a Paired Bootstrap method to handle the random design scenario in Lasso in linear regression.  \citet{das2019distributional} and \citet{ng2022random} explored the Perturbation Bootstrap method for Lasso in linear regression and showed that it works irrespective of whether the design is random or nonrandom and also when the errors are heteroscedastic. Recently, \citet{das2025higher} explored the second and higher order correctness of Residual and Perturbation Bootstrap methods for the Lasso estimator in linear regression. 

The rest of the paper is organized as follows. The regularity conditions necessary for our results are stated and explained in section \ref{sec:assump}. The asymptotic distributional result of properly centered and scaled Lasso is provided in section \ref{sec:mainlasso}. In section \ref{sec:des}, we describe the Perturbation Bootstrap method and in section \ref{sec:prbdef}, the construction of Pearson's Residual Bootstrap is provided. The results on the Perturbation Bootstrap approximation of the distribution of GLM are presented in section \ref{sec:main}. The PRB-based approximation results have been provided to section \ref{sec:prbmain}. Section \ref{sec:sim} contains a moderately large simulation study, whereas a real data example is presented in section \ref{sec:clinical}. Proofs of the main results and requisite lemmas are provided in the Appendix \ref{sec:proofmain} and \ref{sec:prooflemma} respectively. Additional simulation results are also relegated to the Appendix \ref{sec:appA}.

\section{Regularity Conditions}\label{sec:assump}
In this section, we present the set of regularity conditions required to establish our main results. 
Recall that GLM consists of responses $\{y_1,\dots, y_n\}$, and covariates $\{\bm{x}_1\dots, \bm{x}_n\}$, which are connected through the regression parameter $\bm{\beta} = (\beta_1,\dots, \beta_p)^\top$. For simplicity, we assume throughout the paper that all the design vectors are non-random.  
Note that $\mu_i=\mathbf{E}(y_i)= b^\prime(\theta_i)=g^{-1}(\bm{x}_i^\top \bm{\beta})$, $i\in \{1,\dots,n\}$, where $g(\cdot)$ is the link function. Hence for all $i\in \{1,\dots,n\}$, $\theta_i = h(\bm{x}_i^\top\bm{\beta})$ where $h = (g \circ b^{\prime})^{-1}$.  Let 
 $\bm{W}_n= n^{-1/2}\sum_{i=1}^{n}(y_i-\mu_i)\bm{x}_ih^\prime(\bm{x}_i^\top\bm{\beta})$ be the leading random term that appears in the Lasso objective function. Define the matrices $\bm{L}_n=n^{-1}\sum_{i=1}^{n}\bm{x}_i\bm{x}_i^\top\Big[\big\{(g^{-1})^\prime(\bm{x}_i^\top\bm{\beta})\big\}h^\prime(\bm{x}_i^\top\bm{\beta})-(y_i-\mu_i)h^{\prime\prime}(\bm{x}_i^\top\bm{\beta})\Big]$ and $\bm{S}_n=n^{-1}\sum_{i=1}^{n}\bm{x}_i\bm{x}^\top_i\big\{h^\prime(\bm{x}_i^\top\bm{\beta})\big\}^2\mathbbm{E}(y_i-\mu_i)^2$. Clearly, $\bm{S}_n = \mathbbm{E}(\bm{L}_n)$ and when $h(\cdot)$ is the identity map, $\bm{S}_n=\bm{L}_n$. Note also that $\bm{S}_n$ is the variance of $\bm{W}_n$. Let $\|\cdot\|$ denote the Euclidean norm.
 Now we list the regularity conditions. 
   \begin{enumerate}[label=(C.\arabic*)]
    \item $y_i\in \mathbb{R}$ for all $i$, $h$ is the identity function and $b(h(u))=u^2/2$ (the linear regression case) or,  $y_i\geq 0$ for all $i$ and $-h$ \& $h_1$ are convex where $h_1(u)=b(h(u))$ (covers other sub-models of GLM).
    \item $\mathbbm{E}(\bm{L}_n)$ ($=\bm{S}_n$) converges to positive definite (p.d) matrix $\bm{L}$.
  \item $g^{-1}$ is twice continuously differentiable and $h$ is thrice continuously differentiable.

\item $\max(\|\bm{x}_i\|:i\in \{1,\dots,n\})=O(1)$, as $n\rightarrow \infty$.

\item $n^{-1}\sum_{i=1}^{n}\mathbbm{E}(|y_i|^6) =O(1)$, as $n\rightarrow \infty$.

\item $n^{-1/2}\lambda_n\rightarrow \lambda_0 \in [0,\infty)$, as $n\rightarrow \infty$.
\end{enumerate}

We briefly explain the regularity conditions. Our asymptotic analysis requires $-y_ih(u)$ and $b(h(u))$ to be convex as a function of $u$ for all $i$. For GLM other than linear regression (e.g Logistic, Poisson, Gamma etc.) $y_i$'s are usually non-negative and hence convexity of $-h$ and $h_1$ are enough. Assumption (C.1) states such requirements. Condition (C.2) ensures that the limiting Lasso objective function has unique minimum and that the variances of the random parts of the original and the Bootstrapped objective functions are close. Assumption (C.2) is also quite standard in the literature; see for example \citet{freedman1981bootstrapping} and \citet{ma2005robust}. Conditions (C.1) and (C.2) are crucial to apply $Argmin$ theorem (cf. Lemma \ref{lem:argmin}). To derive a suitable Taylor's approximation of the log-likelihood and then to handle the log-likelihood over any compact set (required to derive the asymptotic distribution of Lasso), assumption (C.3) is required. Assumption (C.4) is generally needed to establish asymptotic normality of $\bm{W_n}$ and its Bootstrapped versions. Assumption (C.5) is just a moment condition on the responses, which is essential to have a quadratic approximation of the Lasso objective function. When $h$ is identity, then the condition (C.3) can be dropped and (C.4) - (C.5) can be relaxed, as pointed out in Remark \ref{rem:assump} below. The regularity condition (C.6) is a standard one in the literature (see \citet{knight2000asymptotics}, \citet{camponovo2015validity},  \citet{das2019distributional} and the references therein) and is needed for $n^{1/2}$-consistency of the Lasso estimator. Now we highlight some sub-models of GLM and point out why the above regularity conditions are natural to assume. \\\\
{\textbf{\underline{Example 1 (Linear regression)}:}} 
Here the response variables $y_i \in \mathbb{R}$, and the log-likelihood function is given by $\sum_{i=1}^{n}\ell_{ni}(\bm{\beta}) = \sum_{i=1}^{n}\Big\{y_{i}(\bm{x}_{i}^\top\bm{\beta}) -(\bm{x}_{i}^\top\bm{\beta})^2 /2 \Big\}.$ Here, $h(u)=u$, $h_1(u)=b\{h(u)\}=b(u)=u^2 /2$ and $g^{-1}(u)=u$. Also note that in the notations defined earlier, $\mu_i=\mathbbm{E}(y_i)=g^{-1}(\bm{x}_i^\top \bm{\beta})= \bm{x}_i^\top \bm{\beta} $, $i\in \{1,\dots,n\}$, $\bm{W}_n= n^{-1/2}\sum_{i=1}^{n}(y_i-\mu_i)\bm{x}_i$ and $\bm{L}_n=n^{-1}\sum_{i=1}^{n}\bm{x}_i\bm{x}_i^\top.$ The variance of $\bm{W}_n$ is $\bm{S}_n=n^{-1}\sum_{i=1}^{n}\bm{x}_i\bm{x}^\top_i$ which is same as $\bm{L}_n$.
Note that, (C.1) is clearly satisfied here. And the assumptions (C.2),  (C.4) and (C.5) are very natural to assume and also present in the literature (cf. \citet{knight2000asymptotics}).\\\\
{\textbf{\underline{Example 2 (Logistic regression)}:}} 
Here the response variables are binary and assumption (C.1) is satisfied. The log-likelihood is given by, $\sum_{i=1}^{n}\ell_{ni}(\bm{\beta})  =\sum_{i=1}^{n}\Big\{y_i(\bm{x}_i^\top \bm{\beta}) - \log (1+e^{\bm{x}_i^\top \bm{\beta}})\Big\}$. Here note that $h(u)=u$, $h_1(u)=b\{h(u)\}=b(u)=\log (1+e^u)$ and $g^{-1}(u)= e^u (1+e^u)^{-1}$. Clearly, $\mu_i=g^{-1}(\bm{x}_i^\top \bm{\beta})= e^{\bm{x}_i^\top\bm{\beta}}(1+e^{\bm{x}_i^\top\bm{\beta}})^{-1} $, $\bm{W}_n= n^{-1/2}\sum_{i=1}^{n}(y_i-\mu_i)\bm{x}_i $ and 
$\bm{L}_n=\bm{S}_n=n^{-1}\sum_{i=1}^{n}\Big\{e^{\bm{x}_i^\top\bm{\beta}}(1+e^{\bm{x}_i^\top\bm{\beta}})^{-2}\Big\}\bm{x}_i\bm{x}_i^\top$. Here the assumptions (C.2), (C.4) and (C.5) are true since the all the derivatives of $g^{-1}(\cdot)$ are bounded and responses are binary. See also \citet{bunea2008honest}. \\\\
{\textbf{\underline{Example 3 (Gamma regression)}:}} Here $y_i \sim Gamma(\alpha, \theta_i)$ independently where $\alpha>0$ is the known shape parameter and $\theta_i$'s are the unknown positive scale parameters. Clearly, $\mu_i =E(y_i)=\alpha\theta_i$. The standard link function generally used here is the log link function, i.e $g(x)= \log (x)$, which in turn implies $\theta_i=\alpha^{-1}e^{\bm{x_i}^\top\bm{\beta}}\ \text{for all}\ i\in \{1,..,n\}$. Here the log-likelihood function is given by
$\sum_{i=1}^{n}\ell_{ni}(\bm{\beta})= \sum_{i=1}^{n}\Big\{-\alpha y_{i}e^{-\bm{x_i}^\top\bm{\beta}} -\alpha(\bm{x}_{i}^\top\bm{\beta}) \Big\}.$
Clearly here $h(u)= -\alpha e^{-u}$, $h_1(u)= \alpha u$ and $g^{-1}(u)= e^u$. Therefore, (C.1) and (C.2) both are satisfied here. Here $\bm{W}_n= n^{-1/2}\sum_{i=1}^{n}(y_i-e^{\bm{x_i}^\top\bm{\beta}})(\alpha e^{-\bm{x_i}^\top\bm{\beta}})\bm{x}_i $, $\bm{L}_n=n^{-1}\sum_{i=1}^{n}y_i(\alpha e^{-\bm{x}_i^\top\bm{\beta}})\bm{x}_i\bm{x}_i^\top$ and $\bm{S}_n=n^{-1}\sum_{i=1}^{n}\alpha \bm{x}_i\bm{x}_i^\top$. Note that $\bm{L}_n$ and $\bm{S}_n$ are not the same. The assumptions (C.3), (C.4) and (C.5) are natural to consider here as well.

\begin{remark}\label{rem:assump}
When the function $h(\cdot)$ is the identity function, the regularity condition (C.3) can be dropped and conditions (C.4) and (C.5) can be replaced by the following relaxed conditions : as $n \to \infty$, \\
(C.4-5)(i) $n^{-1}\sum_{i=1}^{n}\big\{\sup_{|z_i-\bm{x_i^\top \beta}|< \delta} |(g^{-1})^{\prime\prime}(z_i)|^2 \big\} = O(1)$, for some $\delta>0$ 
\\
(C.4-5)(ii) $n^{-1}\sum_{i=1}^{n}\big\{ |(g^{-1})^{\prime}(\bm{x_i^\top \beta})|^2 \big\} = O(1)$ 
\\
(C.4-5)(iii) $n^{-1}\sum_{i=1}^{n}\|\bm{x}_i\|^6 = O(1)$ 
\\
(C.4-5)(iv) $n^{-1}\sum_{i=1}^{n}\mathbbm{E}(|y_i|^{7}) =O(1)$.\\
Clearly, the regularity conditions (C.4-5)(i)-(ii) are local whereas (C.3) is a global condition. Moreover, (C.4-5)(iii) is a condition on mean, whereas (C.4) is on max and hence more restrictive.  
\end{remark}

\section{Asymptotic distribution of the Lasso estimator}\label{sec:mainlasso}
In this section, we address the asymptotic distribution of the Lasso estimator $\hat{\bm{\beta}}_n$.  Let us first define some notations. Let $\mathcal{B}(\mathbbm{R}^p)$ denote the Borel sigma-field defined on $\mathbbm{R}^p$. Define  the Prokhorov metric $\rho(\cdot,\cdot)$ on the collection of all probability measures on $\big(\mathbbm{R}^p,\mathcal{B}(\mathbbm{R}^p)\big)$  as
\begin{align*}
\rho(\mu,\nu)=\inf\big{\{}\epsilon:\mu(B)\leq \nu(B^{\epsilon})+\epsilon\; \text{and}\; \nu(B)\leq \mu(B^{\epsilon})+\epsilon\; \text{for all}\; B \in \mathcal{B}(\mathbbm{R}^p) \big{\}},
\end{align*}
where $B^{\epsilon}$ is the $\epsilon$-neighborhood of the set $B$. Without loss of generality assume that the set of relevant covariates is $\mathcal{A}=\{j: \beta_{j}\neq 0\}=\{1,\ldots,p_0\}$. Further, denote the distribution of $\bm{T}_n=n^{1/2}(\hat{\bm{\beta}}_n-\bm{\beta})$ by $F_n$. Suppose that $\bm{Z}_1$ ($\sim N(\bm{0}, \bm{L})$) is defined on the probability space where $y_1,\dots, y_n$ are defined. Recall that $\bm{L}$ is the limit of the matrix $\mathbbm{E}(\bm{L}_n)$, defined in Section \ref{sec:assump}.
Then for any $\bm{u}=(u_1,\dots,u_p)^\top\in \mathbbm{R}^p$, define
\begin{align}\label{eqn:oassym}
V(\bm{u})=\frac12\bm{u}^\top\bm{L}\bm{u} -\bm{u}^\top \bm{Z}_1 + \lambda_0\Big\{\sum_{j=1}^{p_0}u_jsgn({{\beta}_{j}})+\sum_{j=p_0+1}^{p}|u_j|\Big\}.
\end{align}
where, $sgn(x)$ is $1,0,-1$ respectively when $x>0,x=0$ and $x<0$. Let $F_{\infty}(\cdot)$ denote the distribution of $\mbox{Argmin}_{\bm{u}} V(\bm{u})$. Now we are ready to state the weak convergence of the Lasso estimator.
\begin{theorem}\label{thm:lassolimit}
Under assumptions (C.1)-(C.6), we have
\[
\rho\Big(F_n(\cdot),F_{\infty}(\cdot)\Big)\to 0\quad\text{as}\; n\to\infty.
\]
\end{theorem}
Theorem \ref{thm:lassolimit} provides the asymptotic distribution of $\bm{T}_n$. Similar to the usual Lasso estimator in linear regression,
the limiting distribution of Lasso in GLM also contains both 
continuous and discrete parts, unless $p=p_0$ or $\lambda_0 = 0$. Clearly, the asymptotic distribution is not tractable for the purpose of inference. Therefore, it is essential to devise an alternative approximation of the distribution of the Lasso estimator. To that end, we define two Bootstrap methods in the next section. 

\section{Description of the Bootstrap methods}\label{sec:desboot}
In this section, we define the Perturbation Bootstrap (PB) and Pearson's Residual Bootstrap (PRB) schemes for approximating the distribution of the Lasso estimator.
\subsection{Perturbation Bootstrap Method}\label{sec:des}
Perturbation Bootstrap (hereafter referred to as PB) estimators are obtained by attaching random weights to the original objective function.  The random weights are generally a collection of independent copies $G_1^*,\ldots, G_n^*$ of a non-negative and non-degenerate random variable $G^*$. $G^*$ is independent of the data generation process and has the properties that the mean of $G^*$ is $\mu_{G^*}$, $Var(G^*)=\mu_{G^*}^2$  and $\mathbbm{E}(G_1^{*3})< \infty$. Some immediate choices of the distribution of $G^*$ are Exp ($\kappa$) for any $\kappa>0$ , Poisson (1), Beta($\alpha,\beta$) with $\alpha=(\beta-\alpha)(\beta + \alpha)^{-1}$ etc. In GLM, the main objective function is the negative log-likelihood and hence we attach random weights to the negative log-likelihood. Therefore, the PB version of the Lasso estimator in GLM at penalty $\lambda_n^*$ ($> 0$) can be defined as
\begin{align}\label{eqn:defpb}
\hat{\bm{\beta}}_{n}^{*(PB)} \equiv \hat{\bm{\beta}}_n^{*(PB)}(\lambda_n^*) \in \operatorname*{arg\,min}_{\bm{\beta}}\Bigg[-\sum_{i=1}^{n}\ell_{ni}(\bm{\beta})G^*_i\mu_{G^*}^{-1} + n^{1/2}\bm{\beta}^\top \big\{\mathbbm{E}_*(\bm{W}_n^*)\big\} +\lambda_n^*\sum_{j=1}^{p}|\beta_{j}|\Bigg],
\end{align} 
where $\ell_{ni}(\bm{\beta})=\Big[y_{i}h(\bm{x}_{i}^\top\bm{\beta}) -b\{h(\bm{x}_{i}^\top\bm{\beta})\}\Big]$, as before, and $\mathbbm{E}_*(\cdot)$ denote the conditional expectation given the data. Here, $\bm{W}_n^*=n^{-1/2}\sum_{i=1}^{n}\big\{y_i-g^{-1}\big(\bm{x}_i^\top\check{\bm{\beta}}_n\big)\big\}h^\prime(\bm{x}_i^\top\check{\bm{\beta}}_n)\bm{x}_iG_i^*\mu_{G^*}^{-1}$ where $\check{\bm{\beta}}_n$ is the estimator around which we want to center $\hat{\bm{\beta}}_{n}^*$.  
The convexity of $-h(\cdot)$ and $b(h(\cdot))$ (see Section \ref{sec:assump}) essentially implies that $\hat{\bm{\beta}}_{n}^{*(PB)}$ is unique. The linear term $n^{1/2}\bm{\beta}^\top \big\{\mathbbm{E}_*(\bm{W}_n^*)\big\}$ is essential in (\ref{eqn:defpb}), as discussed in Remark \ref{rem:PB centering} below.

\subsection{Pearson's Residual Bootstrap Method}\label{sec:prbdef}
In contrast to the PB method, the Pearson's Residual Bootstrap (hereafter refereed to as PRB) for Lasso in GLM, can be formalized based on resampling from Pearson's standardized residuals. The basic idea is to obtain a quadratic approximation of the log-likelihood based on with replacement sampling of the standardized Pearson's residuals and then to define the PRB estimator as minimizer of $l_1-$penalized  least square objective function. Now we introduce some quantities before stating the precise definition of the PRB estimator in Lasso GLM. Let
\[
\check{V}_n = \mathrm{diag}\Big(b^{\prime\prime}[h(\bm{x}_1^\top\check{\bm{\beta}}_n)], \dots, b^{\prime\prime}[h(\bm{x}_n^\top\check{\bm{\beta}}_n)]\Big), \check{\Delta}_n = \mathrm{diag}\Big(h'(\bm{x}_1^\top\check{\bm{\beta}}_n), \dots, h'(\bm{x}_n^\top\check{\bm{\beta}}_n)\Big)\]\[\;\text{and}\; \check{G}_n=\check{V}_n^{1/2} \check{\Delta}_n X.
\]
Subsequently, define the $i$-th standardized Pearson's residual and its mean as
\[
e_i^\dagger := \frac{y_i - g^{-1}(\bm{x}_i^\top\check{\bm{\beta}}_n)}{\sqrt{b^{\prime\prime}[h(\bm{x}_i^\top\check{\bm{\beta}}_n)]}}, \quad i = 1, \dots, n\quad\text{and}\quad \bar{e}^\dagger=\frac1n\sum_{i=1}^ne_i^\dagger.
\]
As before, $\check{\bm{\beta}}_n$ is the estimator around which we want to center the PRB estimator. Now we resample $\{e_1^{*},\dots, e_n^*\}$ with replacement from the set of centered residuals $\{e_1^\dagger-\bar{e}^\dagger,....,e_n^\dagger-\bar{e}^\dagger\}$ and define the PRB version of the Lasso estimator in GLM at penalty $\lambda_n^*$ ($> 0$) as
\begin{align}\label{eqn:prb}
\hat{\bm{\beta}}_n^{*(PRB)}\equiv \hat{\bm{\beta}}_n^{*(PRB)}(\lambda_n^*)\in \operatorname*{arg\,min}_{\bm{\beta}}\Bigg\{\frac12\Big\|\Big(\check{G}_n\check{\bm{\beta}}_n + \bm{e}^*\Big) -\check{G}_n\bm{\beta}\Big\|_2^2+\lambda_n^*\sum_{j=1}^{p}|\beta_j|\Bigg\}.   
\end{align}
Note that under the assumption (C.1), $\hat{\bm{\beta}}_n^{*(PRB)}$ is unique. Clearly, $\hat{\bm{\beta}}_n^{*(PRB)}$ is actually the Lasso solution of a  linear regression problem with response vector $\big[\bm{e}^*+\check{G}_n\check{\bm{\beta}}_n\big]$ and the design matrix $\check{G}_n$. This enables us to implement computationally fast algorithms of \citet{efron2004least} and one-at-a-time coordinate wise descent methods of \citet{friedman2007pathwise} to compute PRB version of the Lasso estimator in any sub model in GLM. 

\begin{remark}\label{rem:PB centering}
 It is worth mentioning that the second term $n^{1/2}\bm{\beta}^\top \big\{\mathbbm{E}_*(\bm{W}_n^*)\big\}$ in the definition (\ref{eqn:defpb}) is necessary. To elaborate it more formally, suppose we define the PB-Lasso estimator $\bar{\bm{\beta}}_n^{*(PB)}$ without this second term as follows:
 \begin{align}\label{eqn:defnpb}
 \bar{\bm{\beta}}_n^{*(PB)} \in \operatorname*{arg\,min}_{\bm{\beta}}\Bigg\{-\sum_{i=1}^{n}\ell_{ni}(\bm{\beta})G^*_i\mu_{G^*}^{-1} + \lambda_n\sum_{j=1}^{p}|\beta_{j}|\Bigg\}.
\end{align}
Suppose that we want to center $\bar{\bm{\beta}}_n^{*(PB)}$ around some estimator $\check{\bm{\beta}}_n$ which is $n^{1/2}-$consistent for the parameter $\bm{\beta}$. Suppose we denote, $\bar{u}_n^*=n^{1/2}(\bar{\bm{\beta}}_n^{*(PB)}-\check{\bm{\beta}}_n)$. Then through appropriate Taylor's expansion, it's easy to verify that,
\begin{align}\label{eqn:defncpb}
 \bar{\bm{u}}_n^{*}= \operatorname*{arg\,min}_{\bm{u}}\Bigg\{\frac12\bm{u}^{\top}\check{\bm{L}}_n^*\bm{u} - \bm{u}^{\top}\bm{W}_n^{*}+\bm{r}_n^*+\lambda_n\sum_{j=1}^{p}\Big(|\check{\beta}_{j,n}+\dfrac{u_{j}}{n^{1/2}}|-|\check{\beta}_{j,n}|\Big)\Bigg\},
\end{align}
where $\bm{W}_n^*$ is given as above, $\bm{r}_n^*$ is the remainder term \ and $$\check{\bm{L}}_n^*=n^{-1}\sum_{i=1}^{n}\bm{x}_i\bm{x}_i^\top\Big[\big\{(g^{-1})^\prime(\bm{x}_i^\top\check{\bm{\beta}}_n)\big\}h^\prime(\bm{x}_i^\top\check{\bm{\beta}}_n)-\big\{y_i-g^{-1}\big(\bm{x}_i^\top\check{\bm{\beta}}_n\big)\big\}h^{\prime\prime}(\bm{x}_i^\top\check{\bm{\beta}}_n)\Big]G_i^*\mu_{G^*}^{-1}.$$
Clearly $\bm{W}_n^*$ is a sequence of non-centered random vectors and hence its asymptotic mean  is not necessarily $\mathbf{0}$. This will imply that the asymptotic distribution of $\bar{\bm{u}}_n^*$ has a random mean resulting failure of the PB method. To circumvent this issue, we need the second term in the definition of $\hat{\bm{\beta}}_n^{*(PB)}$ in (\ref{eqn:defpb}).
\end{remark}

\section{Main Results on PB approximation} \label{sec:main}
In this section, we explore the validity of the PB approximation of $F_n(\cdot)$, the distribution of $\bm{T}_n = n^{1/2}(\hat{\bm{\beta}}_n - \bm{\beta})$. Recall that $\big(\Omega,\mathcal{F},\mathbb{P}\big)$ is the main the probability space where $\{y_1,\dots, y_n\}$ and $\{G_1^*,\dots,G_n^*\}$ all are defined. Let $\mathscr{E}\subseteq \mathcal{F}$ be the sigma-field generated by $\{y_i:i\geq 1\}$. Then define the PB version of $\bm{T}_n$ as $$\check{\bm{T}}_n^{*(PB)}=n^{1/2}(\hat{\bm{\beta}}_n^{*(PB)}-\check{\bm{\beta}}_n),$$ and denote the conditional distribution of $\check{\bm{T}}_n^{*(PB)}$ given $\mathscr{E}$ by $\check{F}_n^{(PB)}$. When $\check{\bm{\beta}}_n = \hat{\bm{\beta}}_n$, we denote $\check{F}_n^{(PB)}$ by $\hat{F}_n^{(PB)}$ and $\check{\bm{T}}_n^{*(PB)}$ by $\hat{\bm{T}}_n^{*(PB)}$. This section is divided into two sub-sections where first we study the validity of $\hat{F}_n^{(PB)}$. We show that it fails and subsequently we define a modified version.

\subsection{Failure of naive PB, i.e. when $\check{\bm{\beta}}_n=\hat{\bm{\beta}}_n$}\label{sec:naivepb}
In this sub-section, we study the asymptotic behavior of $\hat{\bm{T}}_n^{*(PB)}=n^{1/2}\big(\hat{\bm{\beta}}_n^{*(PB)}-\hat{\bm{\beta}}_n\big)$, where $\hat{\bm{\beta}}_n^{*(PB)}$ is defined in (\ref{eqn:defpb}) with $\check{\bm{\beta}}_n = \hat{\bm{\beta}}_n$. We show that the conditional distribution of $\hat{\bm{T}}_n^{*(PB)}$ can not be used to approximate the distribution of $\bm{T}_n$. Let $\bm{Z}_2\sim N(\bm{0}, \bm{L})$, defined on $\big(\Omega,\mathcal{F},\mathbbm{P}\big)$, and define
\begin{align}\label{eqn:wbassym}
&V_{\infty}(\bm{t}; \bm{u}) = (1/2)\bm{u}^\top \bm{L}\bm{u} - \bm{u}^\top \bm{Z}_2 + \lambda_0\sum_{j=1}^{p_0}u_j sgn(\beta_j) \nonumber\\
& + \lambda_0\sum_{j=p_0+1}^{p} \Bigg[sgn(t_j) \Big[u_j -2\{u_j+t_j\}\mathbbm{1}\Big\{sgn(t_j)(u_j+t_j)<0\Big\}\Big]+|u_j|\mathbbm{1}(t_j=0)\Bigg],
\end{align}
for $\bm{u}=(u_1,\dots,u_p)^\top$, $\bm{t}=(t_1,\dots,t_p)^\top$ $\in \mathbbm{R}^p$. For a fixed $\bm{t}\in \mathbbm{R}^p$, we define the probability distribution of $\bm{T}_{\infty}(\bm{t})=\mbox{Argmin}_{\bm{u}}V_{\infty}(\bm{t}; \bm{u})$ as $G_{\infty}(\bm{t},\cdot)$. Recall that $\lambda_0$ is the limit of $n^{-1/2}\lambda_n$. 
Then we have the following result on naive PB approximation.
\begin{theorem}\label{prop:ntheo}
Suppose that $n^{-1/2}\lambda_n^*\rightarrow \lambda_0$, as $n\rightarrow \infty$. Then under the assumptions (C.1)-(C.6), we have
$$\mathbbm{P}\Big[\lim_{n\rightarrow \infty}\rho\big(\hat{F}_n^{(PB)}(\cdot),G_{\infty}(\hat{\bm{T}}_{\infty},\cdot)\big)= 0\Big]=1,$$ where $\hat{\bm{T}}_{\infty}$ is defined on $\big(\Omega,\mathcal{F},\mathbbm{P}\big)$ and has the distribution $F_{\infty}(\cdot)$. 
\end{theorem}
Theorem \ref{prop:ntheo} shows that $\hat{F}_n^{(PB)}(\cdot)$, the Bootstrap distribution of $n^{1/2}\big(\hat{\bm{\beta}}_n^{*(PB)}-\hat{\bm{\beta}}_n\big)$, converges to $G_{\infty}(\hat{\bm{T}}_{\infty},\cdot)$ instead of $F_{\infty}(\cdot)$. Whereas $F_{\infty}(\cdot)$ is a fixed probability measure, $G_{\infty}(\hat{\bm{T}}_{\infty},\cdot)$ is a random probability measure with the randomness being driven by $\hat{\bm{T}}_{\infty}$. The random quantity $\hat{\bm{T}}_{\infty}$, having the distribution $F_{\infty}(\cdot)$, appears in the picture through Skorokhod's representation theorem applied on the weak convergence of the sequence $\{\bm{T}_n\}_{n\geq 1}$. Again, $F_{\infty}(\cdot)$ is the distribution of $\mbox{Argmin}_{\bm{u}}V(u)$ and $G_{\infty}(\hat{\bm{T}}_{\infty},\cdot)$ is the distribution of $\mbox{Argmin}_{\bm{u}}V_{\infty}(\hat{\bm{T}}_{\infty}; \bm{u})$. Therefore, Theorem \ref{prop:ntheo} implies that the distribution of $\bm{T}_n$ and the Bootstrap distribution of $\hat{\bm{T}}_n^{*(PB)}$ are close, for large $n$, only when $V(u)$ and $V_{\infty}(\hat{\bm{T}}_{\infty}; \bm{u})$ are equal. Clearly, the difference between $V(u)$ and $V_{\infty}(\hat{\bm{T}}_{\infty}; \bm{u})$ is due to the anomaly in the expressions corresponding to last $(p-p_0)$ components of the respective $Argmin$s. More precisely, the difference disappears if last $(p-p_0)$ components of $\hat{\bm{T}}_{{\infty}}$ are $0$ with probability $1$. This happens when the last $(p-p_0)$ components of $\hat{\bm{\beta}}_{n}$ converges to $0$ almost surely, i.e., if the Lasso estimator $\hat{\bm{\beta}}_{n}$ is variable selection consistent. However, \citet{lahiri2021necessary} in his Theorem 4.1 showed that $n^{-1/2}\lambda_n$ must diverge to $\infty$ for $\hat{\bm{\beta}}_{n}$ to perform variable selection in linear regression. Similar result can be established in the GLM setting as well. Therefore, under the condition (C.6) with $\lambda_0 \in (0, \infty)$, the last $(p-p_0)$ components of $\hat{\bm{T}}_{{\infty}}$ may be nonzero with positive probability implying that $V(u)$ and $V_{\infty}(\hat{\bm{T}}_{\infty}; u)$ cannot match when $p \neq p_0$. Therefore, we cannot employ naive PB, i.e. with $\check{\bm{\beta}}_{n} = \hat{\bm{\beta}}_{n}$, to draw valid inferences in Lasso GLM.

\subsection{Modified PB approximation}\label{sec:modpb}
In the previous sub-section, we show that the conditional distribution of $n^{1/2}\big(\hat{\bm{\beta}}_n^{*(PB)}-\hat{\bm{\beta}}_n\big)$ fails to approximate the distribution of  $n^{1/2}\big(\hat{\bm{\beta}}_n-\bm{\beta}_n\big)$. As mentioned above, the primary reason behind the failure of naive PB is that the Lasso estimator $\hat{\bm{\beta}}_n$ is not variable selection consistent under the regularity condition (C.6). Therefore, we need to define $\check{\bm{\beta}}_n$ which remains variable selection consistent even under (C.6). A natural way to make $\hat{\bm{\beta}}_n$ variable selection consistent under (C.6) is by thresholding, following the prescription of \citet{chatterjee2011bootstrapping}. To that end, define the thresholded version of $\hat{\bm{\beta}}_n$ by $\tilde{\bm{\beta}}_n=(\tilde{\beta}_{n,1},\dots,\tilde{\beta}_{n,p})^\top$ with $\tilde{\beta}_{n,j}=\hat{\beta}_{n,j}\mathbbm{1}(|\hat{\beta}_{n,j}|>a_n)$. Here $\mathbbm{1}(\cdot)$ is the indicator function and the sequence
$\{a_n\}_{n\geq 1}$ of positive constants is such that $a_n+(n^{-1/2}\log n)a_n^{-1} \rightarrow 0$ as $n\rightarrow \infty$. Note that due to Lemma \ref{lem:asconcentration}, $\tilde{\beta}_{n,j}$ is $0$ with probability close to $1$ when $\beta_j =0$, i.e. $\tilde{\bm{\beta}}_n$ is variable selection consistent under (C.6). Subsequently, consider $\hat{\bm{\beta}}_n^{*(PB)}$ with $\check{\bm{\beta}}_n=\tilde{\bm{\beta}}_n$ and define $\tilde{\bm{T}}_n^{*(PB)}=n^{1/2}(\hat{\bm{\beta}}_n^{*(PB)}-\tilde{\bm{\beta}}_n)$. Then denote the conditional distribution of $\tilde{\bm{T}}_n^{*(PB)}$ by $\tilde{F}_n^{(PB)}$. Now we are ready to state the theorem on the validity of the modified PB method. 
\begin{theorem}\label{thm:ptheo}
Suppose that the assumptions (C.1)-(C.6) hold. Also assume that $n^{-1/2}\lambda_n^*\rightarrow \lambda_0$, as $n\rightarrow \infty$. Then  we have $$\mathbbm{P}\Big\{\lim_{n\rightarrow \infty}\rho(\tilde{F}_n^{(PB)},F_n)=0\Big\}=1.$$
\end{theorem}
Theorem \ref{thm:ptheo} shows that the modified PB can be used to approximate the distribution of the Lasso estimator in GLM. Therefore, valid inferences can be drawn using the pivotal quantities $n^{1/2}(\hat{\bm{\beta}}_n-\bm{\beta})$ and $n^{1/2}(\hat{\bm{\beta}}_n^{*(PB)}-\tilde{\bm{\beta}}_n)$ for all the sub-models of GLM. For example, confidence regions for $\bm{\beta}$ can be constructed based on Euclidean norms of the pivotal quantities $\bm{T}_n$ and $\tilde{\bm{T}}_n^{*(PB)}$. For some $\alpha \in (0,1)$, let $\big(\|\tilde{\bm{T}}_n^*\|\big)_{\alpha}$ be the $\alpha$th quantile of the conditional distribution of $\|\tilde{\bm{T}}_n^*\|$. Then the nominal $100(1-\alpha) \%$ confidence region of $\bm{\beta}$ is given by the set $C_{1 - \alpha}\subset \mathcal{R}^p$ where
$$C_{1-\alpha} = \bigg\{\bm{\beta}:\|\bm{T}_n\|\leq \big(\|\tilde{\bm{T}}_n^*\|\big)_{1-\alpha}\bigg\},$$ provided the set $\mathcal{A} = \{j: \beta_j \neq 0\}$ is non-empty. This follows from Theorem \ref{thm:ptheo} and the fact that the limiting distribution of $\|\bm{T}_n\|$ is absolutely continuous when $\mathcal{A}$ is non-empty. 
The relationship between confidence region and hypothesis testing can be utilized to perform tests on $\bm{\beta}$.

\begin{remark}\label{rem:randomassump}
Theorem \ref{thm:ptheo} remains valid even when the design vectors are random, i.e. PB works irrespective of whether the design vectors are random or non-random. However, we need the following regularity conditions in addition to the conditions (C.1),(C.3),(C.5) and (C.6):\\ 
(D.1) $\{(y_i, \bm{x}_i)\}_{i=1}^{n}$ are independent and identically distributed.\\
(D.2) $E\{(y_i-g^{-1}(\bm{x}_i^\top \bm{\beta}))|\bm{x}_i\}=0$ for all $i\in\{1,...,n\}.$\\
(D.3) $\mathbbm{E}\Big[n^{-1}\sum_{i=1}^{n}\bm{x}_i\bm{x}^\top_i\big\{h^\prime(\bm{x}_i^\top\bm{\beta})\big\}^2(y_i-\mu_i)^2\Big]$ 
converges to some positive definite matrix $\bm{L}$.\\
(D.4) $\mathbbm{P}(\max_{i\in \{1,..,n\}}||\bm{x}_i||\leq M)=1$ for some $M>0.$\\
Condition (D.4) requires the random design to be bounded. This condition can be relaxed to some moment type conditions on $\|\bm{x}_i\|$, provided that (C.3) is improved to h\"older continuity of $(g^{-1})^{\prime \prime}$ and $h^{\prime \prime \prime}$. 
\end{remark}

\section{Main Results on PRB approximation}\label{sec:prbmain}
We now turn our attention to the distributional approximation properties of PRB method for the Lasso estimator in GLM. Similar to PB, PRB can also be shown to fail when defined based on the original Lasso estimator $\hat{\bm{\beta}}_n$. More precisely, we can show that the naive PRB pivotal quantity $\hat{\bm{T}}_n^{*(PRB)}=n^{1/2}(\hat{\bm{\beta}}_n^{*(PRB)}-\hat{\bm{\beta}}_n)$ converges in distribution to a random probability measure instead of $F_{\infty}$, the distributional limit of $T_n =n^{1/2}(\hat{\bm{\beta}}_n - \beta)$. The failure of the naive PRB is exactly analogous to Theorem \ref{prop:ntheo} and hence is omitted. Subsequently, we consider the thresholded Lasso estimator $\tilde{\bm{\beta}}_n$, defined in Section \ref{sec:modpb}, and define the modified PRB pivotal quantity by $\tilde{\bm{T}}_n^{*(PRB)}=n^{1/2}(\hat{\bm{\beta}}_n^{*(PRB)}-\tilde{\bm{\beta}}_n)$ with $\check{\bm{\beta}}_n = \tilde{\bm{\beta}}_n$. We denote the conditional distribution (conditional on $\mathscr{E}$, the sigma algebra generated by $\{y_1,\dots, y_n\}$) of $\tilde{\bm{T}}_n^{*(PRB)}$ by $\tilde{F}_n^{(PRB)}$. Now we are ready to state the validity result on modified PRB approximation. 

\begin{theorem}\label{thm:prbpos}
Suppose that the assumptions (C.1)-(C.6) hold. Also assume that $n^{-1/2}\lambda_n^*\rightarrow \lambda_0$, as $n\rightarrow \infty$. Then  we have
\[
\mathbbm{P}\Big\{\lim_{n\rightarrow \infty}\rho(\tilde{F}_n^{(PRB)},F_n)=0\Big\}=1.
\]
\end{theorem}
The proof of Theorem \ref{thm:prbpos} is presented in Section \ref{appAthm6.1}. Theorem \ref{thm:prbpos} shows that we can use modified PRB approximation to draw valid inferences on the regression parameter $\bm{\beta}$ based on Lasso in any sub model of GLM when the design is non-random. 

\begin{remark}\label{rem:3}
 We would like to point out that the classical
Residual Bootstrap of \citet{freedman1981bootstrapping} is inconsistent in heteroscedastic linear regression models; see for example, \citet{liu1988bootstrap, das2019distributional}. 
The underlying reason is that the regression errors have heterogeneous and
unknown variances, so resampling residuals fails to reproduce the correct
covariance structure of the score process. 
For generalized linear models, however, the conditional variance is fully
specified by the model and is given by
\[
\mathrm{var}(y_i)=b''\!\big(h(x_i^\top\beta)\big).
\]
The corresponding Pearson residuals
\(
r_i^\dagger
=
\frac{y_i-\mu_i}{\sqrt{b''(h(x_i^\top\beta))}}
\)
satisfy $\mathrm{var}(r_i^\dagger)=1$ for all $i$. This variance standardization removes heteroscedasticity at the observation
level and yields a score process of the form
\(
\frac{1}{\sqrt{n}}\sum_{i=1}^n x_i r_i^\dagger,
\)
whose covariance structure essentially depends only on the design matrix and not on
individual variances. As a consequence, PRB consistently mimics the the distribution of GLM even under Lasso penalty. However, PRB is likely to fail when the design vectors are random. The primary reason is that the design vectors are not resampled while defining the PRB estimator and hence PRB is unable to capture the randomness of the covariates.  

\end{remark}

\section{Simulation Study}\label{sec:sim}
In this section, through the simulation study, we try to capture the finite sample performances of both our proposed PB and PRB methods in terms of empirical coverages of nominal 90\% one sided and both sided confidence intervals. The confidence intervals are obtained for individual regression coefficients as well as the entire regression vector corresponding to some sub-models of GLM, namely logistic regression, gamma regression and linear regression. First, we will reproduce the simultaneous comparative analysis between PB and PRB methods in case of logistic regression under following set-up:
\begin{itemize}
    \item Fix the thresholding parameter $a_n=n^{-1/3}$, vary over $n \in \{50, 100,150,300,500\}$ for varying choices of $(p,p_0)\in \{(5,2),(7,4),(8,3)\}$.
\end{itemize}

Apart from these, we also provide the following extensive comparative analyses of PB method in Appendix \ref{sec:appA}:

\begin{itemize}
\item[(i)]  Fix $(p,p_0)=(7,4)$, fix the thresholding parameter $a_n=n^{-1/3}$ and we vary over $n \in \{50, 100,150,300,500\}$ in case of gamma and linear regression.
\item[(ii)] Same set-up as in (i), but instead of choosing the penalty parameter $\lambda_n$ through $K-$fold CV, we have manually predefined the choice as $\lambda_n=n^{1/2}\lambda_0$ with $\lambda_0=0.025$ for logistic, gamma and linear regression. 
\item[(iii)] Fix $(p,p_0)=(7,4)$, vary over $n \in \{50, 100,150,300,500\}$ for varying choices of $a_n=n^{-c}$ with $c\in \{0.0015,1/6,1/5,1/4,0.485\}$ in logistic regression.
\end{itemize}

The confidence intervals are constructed to be Bootstrap percentile intervals. 
Now for each of $(n,p,p_0)$, the design matrix is once and initially generated from some structure outside the loop (before resampling iteration starts) and kept fixed throughout the entire simulation. Now any $(n\times p)$ real-valued matrix will work for initialization. Without loss of generality, we initiate with $n$ i.i.d design vectors say, $\bm{x}_i = (x_{i1},\ldots, x_{ip})^\top \ \text{for  all} \ i \in \{1,\ldots, n\}$ from zero mean $p$-variate normal distribution such that it has following covariance structure for all $i\in \{1,...,n\}$ and $1\leq j,k\leq p$:
$cov(x_{ij},x_{ik})= \mathbbm{1}(j=k)+0.3^{|j-k|}  \mathbbm{1}(j \neq k).$

We consider the regression parameter $\bm{\beta}=(\beta_1, \dots, \beta_{p})^\top$ as
$\beta_j=0.5(-1)^j j \mathbbm{1}(1 \leq j \leq p_0) $ . Based on those $\bm{x}_i \ \text{and} \ \bm{\beta}$, with appropriate choices of link functions, we pull out $n$ independent copies of response variables namely, $y_1,\ldots,y_n$ from Bernoulli , gamma with shape parameter $1$ and standard Gaussian distribution respectively. To get hold of the penalty parameter, $\lambda_n$ is chosen through 10-fold (nfolds=10 argument in cv.glmnet and h2o.glm in R) cross-validation method and same optimal $\lambda_n$ is used later for finding PB-Lasso and PRB-Lasso estimator as in (\ref{eqn:defpb}) and (\ref{eqn:prb}). The PB-quantities are generated from $Exp(1)$ distribution whereas the PRB resamples are drawn with replacement from the Pearson's residuals as in section \ref{sec:prbdef}. Now keeping that design matrix same for each stage, the entire data set is generated 500 times to compute empirical coverage probability of one-sided and both sided confidence intervals and average width of the both sided confidence intervals over those above mentioned settings of $(n,p,p_0)$.

We also observe the empirical coverage probabilities of 90\% confidence intervals of $\bm{\beta}$ using the Euclidean norm of the vectors $\bm{T}_n = n^{1/2}(\hat{\bm{\beta}}_n-\bm{\beta})$ with $\tilde{\bm{T}}_n^{*(PB)} = n^{1/2}(\hat{\bm{\beta}}_n^{*(PB)}-\tilde{\bm{\beta}}_n)$ and $\tilde{\bm{T}}_n^{*(PRB)} = n^{1/2}(\hat{\bm{\beta}}_n^{*(PRB)}-\tilde{\bm{\beta}}_n)$ and displayed the results in Table \ref{tab:2}. Overall, PB consistently achieves coverage close to the nominal level across all configurations, with slight over-coverage in smaller samples that diminishes as $n$ increases, reflecting stable finite-sample performance. 
In contrast, PRB exhibits pronounced under-coverage for small and moderate sample sizes. 
As the sample size increases, the coverage of PRB improves steadily and approaches the nominal level, suggesting asymptotic validity but slower convergence relative to PB. 
These results highlight the superior finite-sample reliability of PB for joint inference on high-dimensional parameter vectors in generalized linear models.

\begin{table}[H]
\centering
\caption{Empirical coverage probabilities of 90\% confidence regions for the parameter vector $\beta$ under PB and PRB, for logistic regression with varying $(p,p_0)$ and $n$.}
\label{tab:2}

\vspace{0.2cm}

\setlength{\tabcolsep}{10pt}   
\renewcommand{\arraystretch}{1.25} 

\begin{tabular}{c | ccc | ccc}
\hline\hline
 & \multicolumn{6}{c}{\textbf{Coverage Probability}} \\
$n$ 
& \multicolumn{3}{c}{\textbf{PB}} 
& \multicolumn{3}{c}{\textbf{PRB}} \\
\cline{2-4} \cline{5-7}
 & $(5,2)$ & $(7,4)$ & $(8,3)$
 & $(5,2)$ & $(7,4)$ & $(8,3)$ \\
\hline
50  & 0.976 & 0.988 & 0.970 & 0.798 & 0.810 & 0.824 \\
100 & 0.964 & 0.978 & 0.956 & 0.844 & 0.852 & 0.858 \\
150 & 0.934 & 0.942 & 0.924 & 0.866 & 0.878 & 0.880 \\
300 & 0.914 & 0.910 & 0.912 & 0.886 & 0.888 & 0.892 \\
500 & 0.898 & 0.901 & 0.900 & 0.898 & 0.902 & 0.910 \\
\hline\hline
\end{tabular}

\vspace{0.15cm}
\end{table}

Table~\ref{tab:3} summarizes the empirical coverage probabilities and average interval widths of 90\% confidence intervals under the Perturbation Bootstrap (PB) and Pearson Residual Bootstrap (PRB) for logistic regression with $(p,p_0)=(5,2)$. 
For two-sided intervals, PB consistently attains coverage close to the nominal level across all coefficients and sample sizes, with mild over-coverage for small $n$ that diminishes as $n$ increases, accompanied by steadily shrinking interval widths. 
In contrast, PRB exhibits substantial under-coverage in small and moderate samples. 
For right-sided intervals, the comparative behaviour between PB and PRB remains similar and improves gradually with increasing sample size. 
Overall, these results highlight the superior finite-sample reliability of PB and PRB for both two-sided and one-sided inference in logistic regression by achieving nominal performance. The entire simulation is implemented in {\textbf{R}} (all reproducible codes are available at the repository \url{https://github.com/mayukhc13/Bootstrapping-Lasso-in-GLM.git}).
\begin{table*}[ht]
\centering
\caption{Empirical coverage probabilities and average widths (in parentheses) of 90\% confidence intervals for logistic regression for $(p,p_0)=(5,2)$.}
\label{tab:3}

\renewcommand{\arraystretch}{1.05}
\setlength{\tabcolsep}{3pt}

\begin{minipage}[t]{0.48\textwidth}
\centering
\textbf{(a) PB: Two-sided}

\resizebox{\textwidth}{!}{%
\begin{tabular}{@{}lccccc@{}}
\toprule
$\beta_j$ & $n=50$ & $n=100$ & $n=150$ & $n=300$ & $n=500$ \\
\midrule
{$-0.5$} & 0.952 & 0.940& 0.924 & 0.914 & 0.898 \\ 
& (1.594) & (1.088) & (0.641) & (0.574) & (0.467)  \\ 
{$1.0$} & 0.966   & 0.954 & 0.924 & 0.898 & 0.900  \\ 
& (1.927) & (1.278) & (0.820) & (0.682) & (0.412)   \\ 
{$0$} & 0.976 & 0.954 & 0.926 & 0.910 & 0.902 \\ 
& (2.129) & (1.029) & (0.876) & (0.752) & (0.541)  \\ 
{$0$} & 0.984 & 0.950 & 0.924 & 0.910 & 0.896  \\ 
& (1.273) & (0.943) & (0.801) & (0.643) & (0.417)  \\ 
{$0$} & 0.968 & 0.924 & 0.916 & 0.904 & 0.900 \\ 
& (2.334) & (1.079) & (0.838) & (0.684) & (0.432)  \\
\bottomrule
\end{tabular}}
\end{minipage}
\hfill
\begin{minipage}[t]{0.48\textwidth}
\centering
\textbf{(b) PRB: Two-sided}

\resizebox{\textwidth}{!}{%
\begin{tabular}{@{}lccccc@{}}
\toprule
$\beta_j$ & $n=50$ & $n=100$ & $n=150$ & $n=300$ & $n=500$ \\
\midrule
{$-0.5$} & 0.786 & 0.840& 0.864 & 0.884 & 0.896 \\ 
& (1.269) & (1.002) & (0.843) & (0.672) & (0.458)  \\ 
{$1.0$} & 0.816   & 0.844 & 0.872 & 0.888 & 0.902  \\ 
& (1.547) & (1.108) & (0.831) & (0.673) & (0.405)   \\ 
{$0$} & 0.804 & 0.854 & 0.876 & 0.890 & 0.914 \\ 
& (1.629) & (1.312) & (0.927) & (0.775) & (0.544)  \\ 
{$0$} & 0.784 & 0.826 & 0.868 & 0.882 & 0.894  \\ 
& (1.573) & (0.984) & (0.781) & (0.623) & (0.466)  \\ 
{$0$} & 0.808 & 0.834 & 0.860 & 0.882 & 0.898 \\ 
& (2.004) & (1.779) & (1.208) & (0.884) & (0.632)  \\
\bottomrule
\end{tabular}}
\end{minipage}

\vspace{0.4cm}

\begin{minipage}[t]{0.48\textwidth}
\centering
\textbf{(c) PB: Right-sided}

\resizebox{\textwidth}{!}{%
\begin{tabular}{@{}lccccc@{}}
\toprule
$\beta_j$ & $n=50$ & $n=100$ & $n=150$ & $n=300$ & $n=500$ \\
\midrule
$-0.5$ & 0.946 & 0.934 & 0.914 & 0.910 & 0.902\\ 
$1.0$ & 0.924 & 0.916 & 0.912 & 0.906 & 0.898 \\ 
$0$ & 0.960 & 0.948 & 0.936 & 0.922 & 0.910\\ 
$0$ & 0.966 & 0.930 & 0.914 & 0.908 & 0.894\\ 
$0$ & 0.944 & 0.936 & 0.920 & 0.900 & 0.896\\
\bottomrule
\end{tabular}}
\end{minipage}
\hfill
\begin{minipage}[t]{0.48\textwidth}
\centering
\textbf{(d) PRB: Right-sided}

\resizebox{\textwidth}{!}{%
\begin{tabular}{@{}lccccc@{}}
\toprule
$\beta_j$ & $n=50$ & $n=100$ & $n=150$ & $n=300$ & $n=500$ \\
\midrule
$-0.5$ & 0.794 & 0.844 & 0.868 & 0.880 & 0.894\\ 
$1.0$ & 0.812 & 0.832 & 0.858 & 0.878 & 0.890 \\ 
$0$ & 0.804 & 0.830 & 0.862 & 0.888 & 0.904\\ 
$0$ & 0.796 & 0.836 & 0.874 & 0.886 & 0.900\\ 
$0$ & 0.826 & 0.858 & 0.876 & 0.890 & 0.906\\
\bottomrule
\end{tabular}}
\end{minipage}

\end{table*}

\begin{table}[H]
\centering
\caption{Empirical coverage probabilities and average widths (in parentheses) of 90\% confidence intervals for logistic regression for $(p,p_0)=(7,4)$.}
\label{tab:4}

\renewcommand{\arraystretch}{1.05}
\setlength{\tabcolsep}{3pt}

\begin{minipage}[t]{0.48\textwidth}
\centering
\textbf{(a) PB: Two-sided}

\resizebox{\textwidth}{!}{%
\begin{tabular}{@{}lccccc@{}}
\toprule
$\beta_j$ & $n=50$ & $n=100$ & $n=150$ & $n=300$ & $n=500$ \\
\midrule
{$-0.5$} & 0.972 & 0.950& 0.934 & 0.914 & 0.898 \\   
& (2.594) & (1.288) & (0.941) & (0.575) & (0.427)  \\   
{$1.0$} & 0.962   & 0.960 & 0.938 & 0.898 & 0.912  \\   
& (2.927) & (1.278) & (1.120) & (0.682) & (0.512)   \\   
{$-1.5$} & 0.940 & 0.934 & 0.920 & 0.914 & 0.890 \\   
& (2.118) & (1.621) & (1.195) & (0.762) & (0.608)  \\   
{$2.0$} & 0.954 & 0.942 & 0.926 & 0.912 & 0.904 \\   
& (2.215) & (1.947) & (1.417) & (0.896) & (0.659)  \\   
{$0$} & 0.990 & 0.954 & 0.948 & 0.910 & 0.908 \\   
& (2.329) & (1.129) & (0.876) & (0.652) & (0.441)  \\   
{$0$} & 0.984 & 0.956 & 0.930 & 0.920 & 0.910  \\   
& (2.373) & (1.143) & (0.801) & (0.603) & (0.417)  \\   
{$0$} & 0.988 & 0.954 & 0.936 & 0.914 & 0.906 \\   
& (2.334) & (1.379) & (0.938) & (0.584) & (0.432)  \\
\bottomrule
\end{tabular}}
\end{minipage}
\hfill
\begin{minipage}[t]{0.48\textwidth}
\centering
\textbf{(b) PRB: Two-sided}

\resizebox{\textwidth}{!}{%
\begin{tabular}{@{}lccccc@{}}
\toprule
$\beta_j$ & $n=50$ & $n=100$ & $n=150$ & $n=300$ & $n=500$ \\
\midrule
{$-0.5$} & 0.812 & 0.854& 0.878 & 0.886 & 0.900 \\   
& (1.504) & (1.113) & (0.824) & (0.627) & (0.413)  \\   
{$1.0$} & 0.804   & 0.846 & 0.862 & 0.884 & 0.896  \\   
& (1.327) & (1.071) & (0.852) & (0.651) & (0.517)   \\   
{$-1.5$} & 0.796 & 0.834 & 0.866 & 0.880 & 0.892 \\   
& (1.118) & (0.921) & (0.795) & (0.656) & (0.503)  \\   
{$2.0$} & 0.824 & 0.842 & 0.876 & 0.888 & 0.906 \\   
& (1.218) & (0.927) & (0.701) & (0.626) & (0.519)  \\   
{$0$} & 0.802 & 0.854 & 0.876 & 0.890 & 0.912 \\   
& (1.329) & (1.014) & (0.846) & (0.622) & (0.425)  \\   
{$0$} & 0.832 & 0.868 & 0.880 & 0.894 & 0.914  \\   
& (1.073) & (0.924) & (0.787) & (0.632) & (0.411)  \\   
{$0$} & 0.808 & 0.846 & 0.868 & 0.886 & 0.894 \\   
& (1.304) & (0.879) & (0.635) & (0.544) & (0.417)  \\
\bottomrule
\end{tabular}}
\end{minipage}

\vspace{0.4cm}

\begin{minipage}[t]{0.48\textwidth}
\centering
\textbf{(c) PB: Right-sided}

\resizebox{\textwidth}{!}{%
\begin{tabular}{@{}lccccc@{}}
\toprule
$\beta_j$ & $n=50$ & $n=100$ & $n=150$ & $n=300$ & $n=500$ \\
\midrule
$-0.5$ & 0.976 & 0.966 & 0.941 & 0.936 & 0.900\\  
$1.0$ & 0.934 & 0.926 & 0.910 & 0.894 & 0.898 \\  
$-1.5$ &  0.990 & 0.976 & 0.952 & 0.920 & 0.902\\  
$2.0$ & 0.932 & 0.924 & 0.916 & 0.898 & 0.900\\  
$0$ & 0.970 & 0.940 & 0.926 & 0.912 & 0.898\\  
$0$ & 0.966 & 0.930 & 0.924 & 0.916 & 0.904\\  
$0$ & 0.954 & 0.936 & 0.926 & 0.914 & 0.900\\
\bottomrule
\end{tabular}}
\end{minipage}
\hfill
\begin{minipage}[t]{0.48\textwidth}
\centering
\textbf{(d) PRB: Right-sided}

\resizebox{\textwidth}{!}{%
\begin{tabular}{@{}lccccc@{}}
\toprule
$\beta_j$ & $n=50$ & $n=100$ & $n=150$ & $n=300$ & $n=500$ \\
\midrule
$-0.5$ & 0.806 & 0.836 & 0.862 & 0.878 & 0.890\\  
$1.0$ & 0.814 & 0.846 & 0.872 & 0.890 & 0.908 \\  
$-1.5$ &  0.818 & 0.846 & 0.868 & 0.882 & 0.896\\  
$2.0$ & 0.822 & 0.854 & 0.878 & 0.886 & 0.902\\  
$0$ & 0.798 & 0.838 & 0.866 & 0.880 & 0.892\\  
$0$ & 0.814 & 0.856 & 0.872 & 0.892 & 0.914\\  
$0$ & 0.796 & 0.836 & 0.868 & 0.880 & 0.898\\
\bottomrule
\end{tabular}}
\end{minipage}

\end{table}

Table~\ref{tab:4} reports the empirical coverage probabilities and average widths of 90\% confidence intervals for logistic regression with $(p,p_0)=(7,4)$ and exhibits patterns closely aligned with those observed in Table~\ref{tab:3}. 
For two-sided intervals, the Perturbation Bootstrap (PB) continues to deliver coverage near the nominal level across all coefficients and sample sizes, with mild over-coverage in small samples that attenuates as $n$ increases, alongside steadily contracting interval widths. 
In contrast, the Pearson Residual Bootstrap (PRB) again shows substantial under-coverage for small and moderate $n$, despite producing uniformly shorter intervals. 
For right-sided intervals, PB maintains stable and accurate coverage uniformly across coefficients, whereas PRB exhibits under-coverage for small $n$ but improves gradually with increasing sample size, suggesting slower convergence in more complex models. 
\begin{table}[H]
\centering
\caption{Empirical coverage probabilities and average widths (in parentheses) of 90\% confidence intervals for logistic regression for $(p,p_0)=(8,3)$.}
\label{tab:5}

\renewcommand{\arraystretch}{1.05}
\setlength{\tabcolsep}{3pt}

\begin{minipage}[t]{0.48\textwidth}
\centering
\textbf{(a) PB: Two-sided}

\resizebox{\textwidth}{!}{%
\begin{tabular}{@{}lccccc@{}}
\toprule
$\beta_j$ & $n=50$ & $n=100$ & $n=150$ & $n=300$ & $n=500$ \\
\midrule
{$-0.5$} & 0.962 & 0.945& 0.932 & 0.904 & 0.898 \\
& (2.294) & (1.088) & (0.961) & (0.675) & (0.487)  \\
{$1.0$} & 0.952   & 0.946 & 0.928 & 0.918 & 0.896  \\
& (1.927) & (1.678) & (0.820) & (0.782) & (0.612)   \\
{$-1.5$} & 0.948 & 0.932 & 0.922 & 0.904 & 0.900 \\
& (2.108) & (1.021) & (0.975) & (0.762) & (0.638)  \\
{$0$} & 0.952 & 0.948 & 0.926 & 0.920 & 0.904 \\
& (1.215) & (0.947) & (0.617) & (0.596) & (0.459)  \\
{$0$} & 0.986 & 0.964 & 0.948 & 0.924 & 0.910 \\
& (2.029) & (1.109) & (0.872) & (0.602) & (0.401)  \\
{$0$} & 0.964 & 0.952 & 0.932 & 0.922 & 0.902  \\
& (2.373) & (1.043) & (0.841) & (0.503) & (0.416)  \\
{$0$} & 0.984 & 0.966 & 0.942 & 0.924 & 0.912  \\
& (1.773) & (1.043) & (0.864) & (0.643) & (0.426)  \\
{$0$} & 0.972 & 0.944 & 0.936 & 0.904 & 0.890 \\
& (2.034) & (1.179) & (0.918) & (0.545) & (0.402)  \\
\bottomrule
\end{tabular}}
\end{minipage}
\hfill
\begin{minipage}[t]{0.48\textwidth}
\centering
\textbf{(b) PRB: Two-sided}

\resizebox{\textwidth}{!}{%
\begin{tabular}{@{}lccccc@{}}
\toprule
$\beta_j$ & $n=50$ & $n=100$ & $n=150$ & $n=300$ & $n=500$ \\
\midrule
{$-0.5$} & 0.794 & 0.842& 0.876 & 0.888 & 0.896 \\ 
& (1.191) & (0.856) & (0.769) & (0.637) & (0.528)  \\ 
{$1.0$} & 0.814   & 0.842 & 0.866 & 0.880 & 0.894  \\ 
& (1.117) & (0.975) & (0.813) & (0.702) & (0.512)   \\ 
{$-1.5$} & 0.826 & 0.852 & 0.878 & 0.892 & 0.914 \\ 
& (1.108) & (0.821) & (0.672) & (0.516) & (0.438)  \\ 
{$0$} & 0.802 & 0.848 & 0.864 & 0.882 & 0.898 \\ 
& (1.011) & (0.827) & (0.601) & (0.522) & (0.431)  \\ 
{$0$} & 0.816 & 0.852 & 0.870 & 0.886 & 0.900 \\ 
& (1.029) & (0.809) & (0.772) & (0.613) & (0.502)  \\ 
{$0$} & 0.824 & 0.852 & 0.876 & 0.894 & 0.914  \\ 
& (1.303) & (1.037) & (0.804) & (0.609) & (0.417)  \\ 
{$0$} & 0.794 & 0.848 & 0.872 & 0.890 & 0.906  \\ 
& (1.037) & (0.843) & (0.664) & (0.513) & (0.406)  \\ 
{$0$} & 0.808 & 0.842 & 0.870 & 0.882 & 0.894 \\ 
& (1.031) & (0.879) & (0.710) & (0.585) & (0.413)  \\
\bottomrule
\end{tabular}}
\end{minipage}

\vspace{0.4cm}

\begin{minipage}[t]{0.48\textwidth}
\centering
\textbf{(c) PB: Right-sided}

\resizebox{\textwidth}{!}{%
\begin{tabular}{@{}lccccc@{}}
\toprule
$\beta_j$ & $n=50$ & $n=100$ & $n=150$ & $n=300$ & $n=500$ \\
\midrule
$-0.5$ & 0.970 & 0.956 & 0.940 & 0.926 & 0.906\\ 
$1.0$ & 0.964 & 0.946 & 0.930 & 0.824 & 0.914 \\ 
$-1.5$ &  0.980 & 0.956 & 0.938 & 0.912 & 0.892\\ 
$0$ & 0.954 & 0.942 & 0.934 & 0.918 & 0.900\\ 
$0$ & 0.962 & 0.942 & 0.936 & 0.920 & 0.912\\ 
$0$ & 0.966 & 0.942 & 0.924 & 0.914 & 0.904\\ 
$0$ & 0.936 & 0.920 & 0.910 & 0.901 & 0.898\\ 
$0$ & 0.954 & 0.926 & 0.922 & 0.910 & 0.906\\
\bottomrule
\end{tabular}}
\end{minipage}
\hfill
\begin{minipage}[t]{0.48\textwidth}
\centering
\textbf{(d) PRB: Right-sided}

\resizebox{\textwidth}{!}{%
\begin{tabular}{@{}lccccc@{}}
\toprule
$\beta_j$ & $n=50$ & $n=100$ & $n=150$ & $n=300$ & $n=500$ \\
\midrule
$-0.5$ & 0.814 & 0.856 & 0.872 & 0.888 & 0.904\\ 
$1.0$ & 0.798 & 0.836 & 0.866 & 0.882 & 0.896 \\ 
$-1.5$ &  0.802 & 0.846 & 0.870 & 0.884 & 0.898\\ 
$0$ & 0.824 & 0.842 & 0.874 & 0.886 & 0.902\\ 
$0$ & 0.810 & 0.842 & 0.872 & 0.882 & 0.900\\ 
$0$ & 0.796 & 0.844 & 0.868 & 0.878 & 0.890\\ 
$0$ & 0.816 & 0.846 & 0.872 & 0.888 & 0.902\\ 
$0$ & 0.824 & 0.858 & 0.880 & 0.894 & 0.912\\
\bottomrule
\end{tabular}}
\end{minipage}

\end{table}

 {\textbf{CVXR}} package is used for convex optimization. The package {\textbf{glmnet}} is used for cross-validation to obtain optimal $\lambda_n$ and estimated Lasso coefficients of $\bm{\beta}$ for logistic and linear regression. Same purpose is served through {\textbf{h2o}} package for gamma regression in {\textbf{R}}. Similar to other two set-ups, in Table \ref{tab:5}, PB two-sided intervals exhibit systematic over-coverage for small $n$, with coverage converging toward the nominal level as $n$ increases, while interval widths shrink with increasing $n$. In contrast, PRB two-sided intervals are under-covering for small samples but improve steadily with $n$, yielding shorter intervals across all coefficients. 
For right-sided inference, PB remains conservative whereas PRB achieves coverage closer to the nominal level in moderate to large samples, indicating superior finite-sample calibration. Overall, increasing $(p,p_0)$ amplifies the finite-sample consistency between PB and PRB, reinforcing the robustness of the proposed method for both two-sided and one-sided inference in moderate-dimensional logistic regression.

\section{Application to Clinical Data}\label{sec:clinical}
We have applied our proposed PB method to the real life clinical data set (at \url{https://archive.ics.uci.edu/ml/datasets/Breast+Cancer+Coimbra}) related to presence of breast cancer among women depending upon clinical factors. Breast Cancer occurs when mutations take place in genes that regulate breast cell growth. The mutations let the cells divide and multiply in an uncontrolled way. The uncontrolled cancer cells often invade other healthy breast tissues and can travel to the lymph nodes under the arms. Therefore, screening at early stages needs to be detected for having greater survival probability. The recent biomedical studies investigated how the presence of cancer cells may rely on subjects corresponding to routine blood analysis namely,  Glucose, Insulin, HOMA, Leptin, Adiponectin,
Resistin, MCP-1, Age and Body Mass Index (BMI) etc. (cf. \citet{patricio2018using}). We consider a data set of 116 observed clinical features containing a binary response variable indicating the presence or absence of breast cancer along with the 9 clinical covariates. We have reserved the choice of thresholding parameter as $a_n=n^{-1/3}$ and number of CV-folds to be $K=10$ (cf. Real Data Analysis section at \footnote{ \url{https://github.com/mayukhc13/Bootstrapping-Lasso-in-GLM.git}}).
\begin{table}[H]
\centering
\caption{Estimated Lasso Coefficients \& 90\% Bootstrap Percentile Confidence Intervals}
\label{tab7}
\begin{tabular}{@{}lcrcrrr@{}}
\hline
& &\multicolumn{3}{c}{90\% Confidence Intervals} \\
\cline{3-5}
Covariates & $\hat{\beta}_j$&
\multicolumn{1}{c}{Both Sided} &
\multicolumn{1}{c}{Left Sided}&
\multicolumn{1}{c@{}}{Right Sided}\\
\hline
Age & -0.015 & $[-0.042,0.008]$ & $[-0.037,\infty)$ & $(-\infty,0.004]$ \\
BMI & -0.128 & $[-0.247,-0.038]$ & $[-0.206,\infty)$ & $(-\infty,-0.075]$  \\
Glucose & 0.041 & $[-0.002,0.068]$ & $[0.011,\infty)$ & $(-\infty,0.063]$ \\
Insulin & 0.043 & $[-1.316,0.179]$ & $[-0.312,\infty)$ & $(-\infty,0.155]$ \\
HOMA & 0 & $[-0.554,1.589]$ & $[-0.377,\infty)$ & $(-\infty,0.828]$ \\
Leptin & 0 & $[-0.055,0.021]$ & $[-0.023,\infty)$ & $(-\infty,0.017]$ \\
Adiponectin & -0.010 & $[-0.072,0.047]$ & $[-0.054,\infty)$ & $(-\infty,0.035]$ \\
Resistin & 0.033 & $[-0.005,0.071]$ & $[0.005,\infty)$ & $(-\infty,0.062]$ \\
MCP-1 & 0 & $[-0.001,0.002]$ & $[-0.001,\infty)$ & $(-\infty,0.001]$ \\
\hline
\end{tabular}
\end{table}

We regress the data set regularized through fitting Logistic Lasso here (cf. step 22-67 at Real Data.R of the repository) and get the estimates of those covariates. All the covariates are quantitative. We also, find the 90\% both sided , right and left sided Bootstrap percentile confidence intervals for each of the unknown parameter components (see Table \ref{tab7}). We note down the Lasso estimates of all covariates noting that estimates of HOMA, Leptin and MCP-1 as given by variable selection in \textbf{R} are exactly zero. Despite the fact that, 90\% confidence intervals (both sided) for all the factors (except for BMI) contain zero, however, for Resistin and Glucose, we have 90\% CI (both and left sided) mostly skewed towards positive quadrant, whereas, those of Age and BMI contain the negative quadrant implies that these factors have sincere impact in recognising presence of breast cancer, coinciding with the conclusions of \citet{patricio2018using}.

\appendix

\section{Proofs of Requisite Lemmas}\label{sec:prooflemma}
All the requisite lemmas are provided in this section. We denote few notations.  Recall that $\check{\bm{\beta}}_n$ is the estimator around which we want to center the PB-Lasso estimator. PB versions of $\bm{W}_n$ and $\bm{L}_n$ are respectively given by $\check{\bm{W}}_n^*=n^{-1/2}\sum_{i=1}^{n}\big(y_i-\check{\mu}_i\big)h^\prime(\bm{x}_i^\top\check{\bm{\beta}}_n)\bm{x}_i(G_i^*-\mu_{G^*})\mu_{G^*}^{-1}$ and $\check{\bm{L}}_n^*=\mu_{G^*}^{-1}n^{-1}\sum_{i=1}^{n}\bm{x}_i\bm{x}_i^\top\Big[\big\{(g^{-1})^\prime(\bm{x}_i^\top\check{\bm{\beta}}_n)\big\}h^\prime(\bm{x}_i^\top\check{\bm{\beta}}_n)-(y_i-\check{\mu}_i)h^{\prime\prime}(\bm{x}_i^\top\check{\bm{\beta}}_n)\Big]G_i^*,$ where $\check{\mu}_i=g^{-1}(\bm{x}_i^\top\check{\bm{\beta}}_n)$.\\
The PB variance of $\check{\bm{W}}_n^*$ is defined as the matrix $\check{\bm{S}}_n=n^{-1}\sum_{i=1}^{n}\bm{x}_i\bm{x}^\top_i\big\{h^\prime(\bm{x}_i^\top\check{\bm{\beta}}_n)\big\}^2 (y_i-\check{\mu}_i)^2$. Whenever, the centering term $\check{\bm{\beta}}_n=\hat{\bm{\beta}}_n$, we denote $\check{\mu}_i$, $\check{\bm{W}}_n^*$, $\check{\bm{L}}_n^*$ and $\check{\bm{S}}_n$ respectively by $\hat{\mu}_i$, $\hat{\bm{W}}_n^*$, $\hat{\bm{L}}_n^*$ and $\hat{\bm{S}}_n$. We also denote $\check{\mu}_i$, $\check{\bm{W}}_n^*$, $\check{\bm{L}}_n^*$ and $\check{\bm{S}}_n$ respectively by $\tilde{\mu}_i$, $\tilde{\bm{W}}_n^*$, $\tilde{\bm{L}}_n^*$ and $\tilde{\bm{S}}_n$ when $\check{\bm{\beta}}_n=\tilde{\bm{\beta}}_n$.

\begin{lemma}\label{lem:F-N}
Suppose $Y_1,\dots,Y_n$ are zero mean independent random variables with $\mathbbm{E}(|Y_i|^t)< \infty$ for $i\in \{1,\dots,n\}$ and $S_n = \sum_{i = 1}^{n}Y_i$. Let $\sum_{i = 1}^{n}\mathbbm{E}(|Y_i|^t) = \sigma_t$, $c_t^{(1)}=\big(1+\frac{2}{t}\big)^t$ and $c_t^{(2)}=2(2+t)^{-1}e^{-t}$. Then, for any $t\geq 2$ and $x>0$,
\begin{equation*}
\mathbbm{P}[|S_n|>x]\leq c_t^{(1)}\sigma_t x^{-t} + exp(-c_t^{(2)}x^2/\sigma_2)
\end{equation*}
\end{lemma}

Proof of Lemma \ref{lem:F-N}.
This inequality was proved in \citet{fuk1971probability}. \hfill $\square$

\begin{lemma}\label{lem:pointargmin}
Let $C\subseteq \mathbb{R}^p$ be open convex set and let $f_n:C\rightarrow \mathbb{R}$, $n\geq 1$, be a sequence of convex functions such that $\lim_{n\rightarrow \infty}f_n(x)$ exists for all $x\in C_0$ where $C_0$ is a dense subset of $C$. Then $\{f_n\}_{n\geq 1}$ converges pointwise on $C$ and the limit function $f(x)=\lim_{n\rightarrow \infty}f_n(x)$ is finite and convex on $C$. Moreover, $\{f_n\}_{n\geq 1}$ converges to $f$ uniformly over any compact subset $K$ of $C$, i.e. $$\sup_{x\in K}|f_n(x)-f(x)|\rightarrow 0,\;\;\; \text{as}\;\; n \rightarrow \infty.$$
\end{lemma}
Proof of Lemma \ref{lem:pointargmin}. This lemma is stated as Theorem 10.8 of \citet{rockafellar1997convex}. \hfill $\square$

\begin{lemma}\label{lem:nearness}
Suppose that  $\{f_n\}_{n\geq 1}$ and $\{g_n\}_{n\geq 1}$ are random convex functions on $\mathbb{R}^p$. The sequence of minimizers are $\{\alpha_n\}_{n\geq 1}$ and $\{\beta_n\}_{n\geq 1}$ respectively, where the sequence $\{\beta_n\}_{n\geq 1}$ is unique. For some $\delta>0$, define the quantities $$\Delta_n(\delta)=\sup_{\|s-\beta_n\|\leq \delta}|f_n(s)-g_n(s)|\;\;\text{and}\;\; h_n(\delta)=\inf_{\|s-\beta_n\|=\delta}g_n(s)-g_n(\beta_n).$$ Then we have, $\Big\{\|\alpha_n-\beta_n\|\geq \delta\Big\}\subseteq \Big\{\Delta_n(\delta)\geq \frac{1}{2}h_n(\delta)\Big\}.$
\end{lemma}

Proof of Lemma \ref{lem:nearness}. This lemma follows from Lemma 2 of \citet{hjort1993asymptotics}. \hfill $\square$

\begin{lemma}\label{lem:asargmin}
Consider the sequence of convex functions $\{f_n:\mathbb{R}^p\rightarrow\mathbb{R}\}_{n\geq 1}$ having the form $$f_n(\bm{u})=\bm{u}^\top\bm{\Sigma}_n \bm{u} + R_n(\bm{u}),$$
where $\bm{\Sigma}_n$ converges almost surely to a positive definite matrix $\bm{\Sigma}$ and $\mathbbm{P}\big[\lim_{n\rightarrow \infty}\|R_n(\bm{u})\|=0\big]=1$ for any $u\in \mathbb{R}^p$. Let $\{\alpha_n\}_{n\geq 1}$ be the sequence of minimizers of $\{f_n\}_{n\geq 1}$ over $\mathbb{R}^p$. Then,
\begin{align}\label{eqn:5.5.1}
\mathbbm{P}\big(\lim_{n\rightarrow \infty}\|\alpha_n\|=0\big)=1.
\end{align}
\end{lemma}
Proof of Lemma \ref{lem:asargmin}. Note that the almost sure limit function of $\{f_n\}_{n\geq 1}$ is $f(\bm{u})=\bm{u}^\top \bm{\Sigma} \bm{u}$, for any $\bm{u}\in \mathbb{R}^p$. Since $\bm{\Sigma}$ is p.d, $\operatorname*{arg\,min}_{\bm{u}}f(\bm{u})=0$ and is unique. Hence in the notations of Lemma \ref{lem:nearness}, $$\Delta_n(\delta)=\sup_{\|\bm{u}\|\leq \delta}|f_n(\bm{u})-f(\bm{u})|\;\;\text{and}\;\; h_n(\delta)=\inf_{\|\bm{u}\|=\delta}g_n(\bm{u}).$$ Therefore due to Lemma \ref{lem:nearness},
$\limsup_{n\rightarrow \infty}\Big\{\|\alpha_n\|\geq\delta\Big\}\subseteq \limsup_{n\rightarrow \infty}\Big\{\Delta_n(\delta)\geq \frac{1}{2}h_n(\delta)\Big\},$ for any $\delta >0$. Hence to establish (\ref{eqn:5.5.1}), it's enough to show 
\begin{align}\label{eqn:5.5.2}
\mathbbm{P}\Big[\limsup_{n\rightarrow \infty}\Big\{\Delta_n(\delta)\geq \frac{1}{2}h_n(\delta)\Big\}\Big]  = 0,
\end{align}
for any $\delta>0$. Now fix a $\delta>0$. To show (\ref{eqn:5.5.2}), first we show $\mathbbm{P}\Big[\lim_{n\rightarrow \infty}\Delta_n(\delta)=0\Big]=1$. Since $f$ is the almost sure limit of $\{f_n\}_{n\geq 1}$, for any countable dense set $C\subseteq \mathbb{R}^p$, we have $$\mathbbm{P}\Big[f_n(\bm{u})\rightarrow f(\bm{u})\; \text{for all}\; \bm{u}\in C\Big]=1.$$ Therefore using Lemma \ref{lem:pointargmin}, we can say that $\mathbbm{P}\Big[\lim_{n\rightarrow \infty}\Delta_n(\delta)=0\Big]=1$, since $\big\{\bm{u}\in \mathbb{R}^p:\|\bm{u}\|\leq \delta\big\}$ is a compact set. Therefore we have 
\begin{align}\label{eqn:5.5.3}
\mathbbm{P}\Big[\liminf_{n\rightarrow \infty}\Big\{\Delta_n(\delta) < \epsilon\Big\}\Big]  = 1,
\end{align}
for any $\epsilon>0$. Now let us look into $h_n(\delta)$. Suppose that $\eta_1$ is the smallest eigen value of the non-random matrix $\Sigma$.  Then due to the assumed form of $f_n(\bm{u})$, there exists a natural number $N$ such that for all $n\geq N$,
\begin{align}\label{eqn:5.5.4}
\mathbbm{P}\Big[h_n(\delta)>\frac{\eta_1 \delta^2}{2}\Big]=1.
\end{align}
Taking $\epsilon=\frac{\eta_1\delta^2}{4}$, (\ref{eqn:5.5.2}) follows from (\ref{eqn:5.5.3}) and (\ref{eqn:5.5.4}). \hfill $\square$

\begin{lemma}\label{lem:lil}
Under the conditions (C.3), (C.4) and (C.5), we have
\begin{align*}
\|\bm{W}_n\| = o(\log n)\;\; w.p\; 1.
\end{align*}
\end{lemma}

Proof of Lemma \ref{lem:lil}. This lemma follows exactly through the same line of arguments as in case of Lemma 4.1 of \citet{chatterjee2010asymptotic}, if we consider $(y_i-\mu_i)h^\prime(\bm{x}_i^\top\bm{\beta})$ in place of $\epsilon_i$ for all $i\in \{1,\dots,n\}$. \hfill $\square$

\begin{lemma}\label{lem:asconcentration}
Under the assumptions (C.1)-(C.6), we have
\begin{align}\label{eqn:6.0}
\mathbbm{P}\Big[\|(\hat{\bm{\beta}}_n-\bm{\beta})\|=o\big(n^{-1/2}\log n\big)\Big]=1.
\end{align}
\end{lemma}
Proof of Lemma \ref{lem:asconcentration}. Note that 
\begin{align}\label{eqn:6.1}
(\log n)^{-1}n^{1/2}(\hat{\bm{\beta}}_{n}- \bm{\beta}) = \mbox{Argmin}_{\bm{u}} \Big\{ w_{1n}(\bm{u}) + w_{2n}(\bm{u}) \Big\}
\end{align}
where,
$w_{1n}(\bm{u})= \;  (\log n)^{-2}\bigg[\sum_{i=1}^{n}\Big[-y_{i}\Big\{ h\big\{\bm{x}_{i}^\top(\bm{\beta}+\frac{\bm{u}\log n}{n^{1/2}})\big\}-h\big(\bm{x}_{i}^\top\bm{\beta}\big)\Big\}+\Big\{h_{1}\big\{\bm{x}_{i}^\top(\bm{\beta}+\frac{\bm{u} \log n}{n^{1/2}})\big\}-h_{1}\big(\bm{x}_{i}^\top\bm{\beta}\big)\Big\} \Big]\bigg],$ $h_{1} = b \circ h$ and $w_{2n}(u) = (\log n)^{-2}\lambda_{n}\sum_{j=1}^{p}\left( |\beta_{j}+\frac{u_{j} \log n}{n^{1/2}}|-|\beta_{j}|\right).$
Now, by Taylor's theorem and noting that $h_1^\prime = (g^{-1})h^\prime$ and $h_1^{\prime\prime}=(g^{-1})^\prime h^\prime + (g^{-1})h^{\prime\prime}$, we have
\begin{align*}
w_{1n}(\bm{u})= (1/2)\bm{u}^\top\bm{L}_n\bm{u} - (\log n)^{-1}\bm{W}_n^\top \bm{u} + Q_{1n}(\bm{u}),
\end{align*}
where, $Q_{1n}(\bm{u})=(6n^{3/2})^{-1}(\log n)\sum_{i=1}^{n}\Bigg\{-y_ih^{\prime\prime\prime}(z_i)+h_1^{\prime\prime\prime}(z_i)\Bigg\}(\bm{u}^\top\bm{x}_i)^3,$
 for some $z_i$ such that $|z_i-\bm{x}_i^\top \bm{\beta}|\leq \frac{(\log n)\bm{x}_i^\top\bm{u}}{n^{1/2}}$ for all $i\in \{1,\dots,n\}$. Now using the continuity of $h^{\prime\prime\prime}$ and $\big(g^{-1}\big)^{\prime\prime}$ (cf. assumption (C.3)), boundedness of $\|x\|$ (cf. assumption (C.4)) and assumption (C.5) we have $Q_{1n}(\bm{u})=o(1)\;\; w.p\; 1$ due to Lemma \ref{lem:F-N} with $t=2$. Again Lemma \ref{lem:lil} implies $(\log n)^{-1}\bm{W}_n^\top \bm{u}=o(1)\;\;w.p\; 1$. Since $n^{-1/2}\lambda_n\rightarrow \lambda_0$ as $n\rightarrow \infty$, $w_{2n}(\bm{u})\rightarrow 0$ pointwise as $n\rightarrow \infty$. Therefore  (\ref{eqn:6.1}) reduces to 
\begin{align}\label{eqn:6.2}
(\log n)^{-1}n^{1/2}(\hat{\bm{\beta}}_{n}- \bm{\beta}) = \mbox{Argmin}_{\bm{u}} \Big[ (1/2)\bm{u}^\top \bm{L}_n \bm{u} + Q_{2n} \Big ],  
\end{align}
where $Q_{2n}=o(1)\;\;w.p\; 1$. Again note that $\|\bm{L}_n-\bm{L}\|=o(1)\;\;w.p\;1$ (cf. first part of Lemma \ref{lem:varest}). Therefore, (\ref{eqn:6.2}) is in the setup of Lemma \ref{lem:asargmin} and hence (\ref{eqn:6.0}) follows. \hfill $\square$
\begin{lemma}\label{lem:varest}
Under the assumptions (C.1)-(C.5), we have $$\|\bm{L}_n-\bm{L}\|=o(1)\;\;w.p\;1\;\; \text{and}\;\;\|\tilde{\bm{L}}_n^*-\bm{L}\|=o_{P_*}(1) \;\;\text{w.p}\; 1.$$
\end{lemma}
Proof of Lemma \ref{lem:varest}. First we are going to show $\|\bm{L}_n-\bm{L}\|=o(1)\;\;w.p\;1$. Note that
$$\|\bm{L}_n-\bm{L}\|\leq \|\bm{L}_n-\mathbbm{E}(\bm{L}_n)\|+\|\mathbbm{E}(\bm{L}_n)-\bm{L}\|,$$ where the second term in the RHS is $o(1)$ as $n\rightarrow \infty$, due to assumption (C.2). To show that the first term of RHS is $o(1)\;\; w.p\;1$, we need to show $\big{|}n^{-1}\sum_{i=1}^{n}\{x_{ij}x_{ik} h^{\prime\prime}(\bm{x}_i^\top \bm{\beta})(y_i-\mu_i)\}\big{|} = o(1)\;\;w.p\;1$ for any $j,k \in \{1,\dots,p\}$.
By noting the assumptions (C.3), (C.4) and (C.5), this simply follows due to Lemma \ref{lem:F-N} with $t=3$ and then applying Borel-Cantelli lemma. Therefore, we are done. \hfill $\square$ \\
Now let us look into $\|\tilde{\bm{L}}_n^*-\bm{L}\|$. Now note that
$\|\tilde{\bm{L}}_n^*-\bm{L}\|\leq\; A_{1n} + A_{2n} +A_{3n}\;\;\; \text{(say)},$  
where, $$A_{1n}:=\Big{\|}n^{-1}\sum_{i=1}^{n}\big[\bm{x}_i\bm{x}_i^\top\big\{(g^{-1})^\prime(\bm{x}_i^\top\tilde{\bm{\beta}}_n)\big\}h^\prime(\bm{x}_i^\top\tilde{\bm{\beta}}_n)\frac{G_i^*}{\mu_{G^*}}\big]-\mathbbm{E}(\bm{L}_n)\Big{\|},$$ $A_{2n}:=\Big{\|}n^{-1}\sum_{i=1}^{n}\big\{\bm{x}_i\bm{x}_i^\top(y_i-\tilde{\mu}_i)h^{\prime\prime}(\bm{x}_i^\top\tilde{\bm{\beta}}_n)\frac{G_i^*}{\mu_{G^*}}\big\}\Big{\|} $ and $A_{3n}:=\|\mathbbm{E}(\bm{L}_n)-\bm{L}\|$. Now it's easy to check that,
 
\begin{align}\label{eqn:var.0}
A_{3n}=o(1),\;\;\;\;\text{due to assumption (C.3)}.
\end{align}
Next it's easy to see that,
\begin{align*}
\;A_{2n}
&\leq\; \Big{\|}n^{-1}\sum_{i=1}^{n}\bm{x}_i\bm{x}_i^\top\big\{y_i-g^{-1}(\bm{x}_i^\top \tilde{\bm{\beta}}_n)\big\}h^{\prime\prime}(\bm{x}_i^\top \tilde{\bm{\beta}}_n)\Big(\frac{G_i^*}{\mu_{G^*}}-1\Big)\Big{\|}\\
&\quad\quad\quad\quad+ \Big{\|}n^{-1}\sum_{i=1}^{n}\bm{x}_i\bm{x}_i^\top\big\{y_i-g^{-1}(\bm{x}_i^\top \tilde{\bm{\beta}}_n)\big\}h^{\prime\prime}(\bm{x}_i^\top \tilde{\bm{\beta}}_n)\Big{\|}\\
=\; & A_{21n}+A_{22n}\;\;\; \text{(say)}.
\end{align*}
First we are going to show that $A_{21n}=o_{P_*}(1)\;\;w.p\; 1$. For that we need to show that for  any $j,k \in \{1,\dots,p\}$,
\begin{align}\label{eqn:7.-1}
\Big{|}n^{-1}\sum_{i=1}^{n}x_{ij}x_{ik}\big\{y_i-g^{-1}(\bm{x}_i^\top \tilde{\bm{\beta}}_n)\big\}h^{\prime\prime}(\bm{x}_i^\top \tilde{\bm{\beta}}_n)\Big(\frac{G_i^*}{\mu_{G^*}}-1\Big)\Big{|}=o_{P_*}(1)\;\;w.p\; 1.
\end{align}
Now noting the assumption $\mathbbm{E}(G_1^{*3})< \infty$ and using Markov's inequality, this follows if we have
$n^{-2}\sum_{i=1}^{n}x_{ik}^2x_{ik}^2\big\{y_i-g^{-1}(\bm{x}_i^\top \tilde{\bm{\beta}}_n)\big\}^2\big\{h^{\prime\prime}(\bm{x}_i^\top \tilde{\bm{\beta}}_n)\big\}^2 = o(1)\;\; w.p \;1.$
Now note that due to assumptions (C.3), (C.4) and Lemma \ref{lem:asconcentration}, we have $\max\Big\{\Big(\|\bm{x}_i\|^4+h^{\prime\prime}(\bm{x}_i^\top \tilde{\bm{\beta}}_n) + g^{-1}(\bm{x}_i^\top \tilde{\bm{\beta}}_n)\Big):i\in \{1,\dots,n\}\Big\}=O(1)\;\;w.p\;1$. Therefore to show (\ref{eqn:7.-1}), we need to show that $n^{-2}\sum_{i=1}^{n}\big[\{y_i-g^{-1}(\bm{x}_i^\top \bm{\beta})\}^2-\mathbbm{E}\{y_i-g^{-1}(\bm{x}_i^\top \bm{\beta})\}^2\big]=o(1)\;\;w.p\;1$, due to assumption (C.5). This follows by applying Lemma \ref{lem:F-N} with $t=2$ and then Borel-Cantelli Lemma.
Therefore we have
\begin{align}\label{eqn:var.1}
A_{21n}=o_{P_*}(1)\;\;w.p\; 1.
\end{align}
Again by Taylor's expansion of $h^{\prime\prime}$ and $g^{-1}$, we have
\begin{align*}
A_{22n}\leq \; & \Big{\|}n^{-1}\sum_{i=1}^{n}\bm{x}_i\bm{x}_i^\top(y_i-\mu_i)h^{\prime\prime}(\bm{x}_i^\top \bm{\beta})\Big{\|}+ \Big{\|}n^{-1}\sum_{i=1}^{n}\bm{x}_i\bm{x}_i^\top(y_i-\mu_i)h^{\prime\prime\prime}(z_i^{(2)})\big\{\bm{x}_i^\top(\tilde{\bm{\beta}}_n-\bm{\beta})\big\}\Big{\|}\\
&+\Big{\|}n^{-1}\sum_{i=1}^{n}\big[\bm{x}_i\bm{x}_i^\top\big\{(g^{-1})^\prime(z_i^{(1)})\big\}\big\{\bm{x}_i^\top(\tilde{\bm{\beta}}_n-\bm{\beta})\big\}h^{\prime\prime}(\bm{x}_i^\top \tilde{\bm{\beta}}_n)\big]\Big{\|}\\
&= A_{221n} +A_{222n}+A_{223n}\;\; \text{(say)},
\end{align*}
for some $z_i^{(1)}$ and $z_i^{(2)}$ such that $|z_i^{(1)}-\bm{x}_i^\top\bm{\beta}|\leq |\bm{x}_i^\top(\tilde{\bm{\beta}}_n-\bm{\beta})|$ and $|z_i^{(2)}-\bm{x}_i^\top\bm{\beta}|\leq |\bm{x}_i^\top(\tilde{\bm{\beta}}_n-\bm{\beta})|$, $i\in \{1,\dots,n\}$. Now by applying Lemma \ref{lem:F-N} with $t=3$ , Borel-Cantelli Lemma and noting the assumptions (C.3) \& (C.4) we have $A_{221n}=o(1)\;\;w.p\;1$. Whereas $A_{223n}=o(1)\;\;wp\;1$ follows directly due to the fact that $\max\big\{\big(|(g^{-1})^\prime(z_i^{(1)})|+ |h^{\prime\prime\prime}(z_i^{(2)})|+\|\bm{x}_i\|^3\big):i\in \{1,\dots,n\}\big\} = O(1)\;\;w.p\; 1$ and using Lemma \ref{lem:asconcentration}. Similar arguments and and an application of Markov's inequality together with Borel-Cantelli Lemma imply $A_{222n}=o(1)\;\;w.p\;1$. Therefore, 
\begin{align}\label{eqn:var.2}
A_{22n}=o(1)\;\;w.p\; 1.
\end{align}
Combining (\ref{eqn:var.1}) and (\ref{eqn:var.2}), we have \begin{align}\label{eqn:var.3}
A_{2n}=o_{P_*}(1)\;\;w.p\; 1.
\end{align}
Now let us consider $A_{1n}$. Note that,
\begin{align*}
A_{1n}&\leq\Big{\|}n^{-1}\sum_{i=1}^{n}\bm{x}_i\bm{x}_i^\top\big\{(g^{-1})^\prime(\bm{x}_i^\top\tilde{\bm{\beta}}_n)\big\}h^\prime(\bm{x}_i^\top\tilde{\bm{\beta}}_n)-n^{-1}\sum_{i=1}^{n}\bm{x}_i\bm{x}_i^\top\big\{(g^{-1})^\prime(\bm{x}_i^\top\bm{\beta})\big\}h^\prime(\bm{x}_i^\top\bm{\beta})\Big{\|} \\
&+\Big{\|}n^{-1}\sum_{i=1}^{n}\bm{x}_i\bm{x}_i^\top\big\{(g^{-1})^\prime(\bm{x}_i^\top\tilde{\bm{\beta}}_n)\big\}h^\prime(\bm{x}_i^\top\tilde{\bm{\beta}}_n)\Big(\frac{G_i^*}{\mu_{G^*}}-1\Big)\Big{\|} 
= A_{11n}+A_{12n}\;\; \text{(say)}.
\end{align*}
To prove $A_{12n}=o_{P_*}(1),\;\; w.p\; 1$, we will use Lemma \ref{lem:F-N} with $t=3$ and then Borel-Cantelli Lemma, similar to how we dealt with $A_{21n}$ and hence we are omitting the details. Again note that using Taylor's expansion,
\begin{align*}
A_{11n}\leq\; &  \Big{\|}n^{-1}\sum_{i=1}^{n}\bm{x}_i\bm{x}_i^\top\big\{(g^{-1})^{\prime\prime}(z_i^{(1)})\big\}\big\{\bm{x}_i^\top(\tilde{\bm{\beta}}_n-\bm{\beta})\big\}h^{\prime}(\bm{x}_i^\top \tilde{\bm{\beta}}_n)\Big{\|}\\
&\;\;+ \Big{\|}n^{-1}\sum_{i=1}^{n}\bm{x}_i\bm{x}_i^\top\big\{(g^{-1})^{\prime}(\bm{x}_i^\top\bm{\beta})\big\}\big\{\bm{x}_i^\top(\tilde{\bm{\beta}}_n-\bm{\beta})\big\}h^{\prime\prime}(z_i^{(2)})\Big{\|},
\end{align*}
for some $z_i^{(1)}$ and $z_i^{(2)}$ such that $|z_i^{(1)}-\bm{x}_i^\top\bm{\beta}|\leq |\bm{x}_i^\top(\tilde{\bm{\beta}}_n-\bm{\beta})|$ and $|z_i^{(2)}-\bm{x}_i^\top\bm{\beta}|\leq |\bm{x}_i^\top(\tilde{\bm{\beta}}_n-\bm{\beta})|$, $i\in \{1,\dots,n\}$. Apply Lemma \ref{lem:asconcentration} and the continuity of $(g^{-1})^{\prime\prime}$ and $h^{\prime\prime}$, to conclude $A_{11n}=o(1)\;\; w.p\; 1$, with arguments similar to as in case of $A_{22n}$. Hence 
we have
\begin{align}\label{eqn:var.4}
A_{1n}=o_{P_*}(1)\;\;w.p\; 1.
\end{align}
Now combining (\ref{eqn:var.0}), (\ref{eqn:var.3}) and (\ref{eqn:var.4}), the proof is complete. \hfill $\square$

\begin{lemma}\label{lem:prbvarest}
Under the assumptions (C.1)-(C.5) we have,
\[
\Bigg\|\frac{\tilde G^\top\tilde G}{n}-L\Bigg\|=o(1)\;\quad\text{with probability}\;\; 1,
\]
where $\frac{\tilde G^\top\tilde G}{n}=\frac{1}{n}\sum_{i=1}^{n}[h^\prime(\bm{x}_i^\top\tilde{\bm{\beta}}_n)]^2b^{\prime\prime}[h(\bm{x}_i^\top\tilde{\bm{\beta}}_n)]\bm{x}_i\bm{x}_i^\top=\frac{1}{n}\sum_{i=1}^{n}[h^\prime(\bm{x}_i^\top\tilde{\bm{\beta}}_n)][(g^{-1})^{\prime}(\bm{x}_i^\top\tilde{\bm{\beta}}_n)]\bm{x}_i\bm{x}_i^\top$, since $g^{-1}(u)=b^\prime[h(u)]$.
\end{lemma}

Proof of Lemma \ref{lem:prbvarest} We write that,
\begin{align}\label{eqn:eee}
  &\Bigg\|\frac{\tilde G^\top\tilde G}{n}-L\Bigg\|
 \le \Bigg\|\frac{\tilde G^\top\tilde G}{n}-\mathbbm{E}(L_n)\Bigg\|+\Big\|\mathbbm{E}(L_n)-L\Big\|\nonumber\\
 &\le\Bigg\|\frac{1}{n}\sum_{i=1}^{n}\bm{x}_i\bm{x}_i^\top\Big\{[h^\prime(\bm{x}_i^\top\tilde{\bm{\beta}}_n)][(g^{-1})^{\prime}(\bm{x}_i^\top\tilde{\bm{\beta}}_n)]-[h^\prime(\bm{x}_i^\top\bm{\beta})][(g^{-1})^{\prime}(\bm{x}_i^\top\bm{\beta})]\Big\}\Bigg\|+\Big\|\mathbbm{E}(L_n)-L\Big\|
\end{align}

Due to assumption (C.2), second term in (\ref{eqn:eee}) is $o(1)$ as $n\to\infty$. Now the first term is exactly the same term $A_{11n}$ in the proof of Lemma \ref{lem:varest}. Hence it is $o(1)$ with probability $1$. The proof is complete. \hfill $\square$

\begin{lemma}\label{lem:var2}
Under the assumptions (C.1)-(C.5), we have
$$\|\tilde{\bm{S}}_n-\bm{L}\|=o(1)\;\; w.p\; 1.$$
\end{lemma}
Proof of Lemma \ref{lem:var2}. Since $\bm{S}_n$ converges to $\bm{L}$ as $n\rightarrow \infty$, it's enough to show $\|\tilde{\bm{S}}_n-\bm{S}_n\|=o(1)\;\; wp\; 1.$ Now using Taylor's expansion we have,
\begin{align*}
\|\tilde{\bm{S}}_n-\bm{S}_n\| &\leq\;  A_{3n} +A_{4n}+A_{5n}\;\;\text{(say)}.
\end{align*}
where it's easy to see that, $A_{3n}= \Big{\|}n^{-1}\sum_{i=1}^{n}\bm{x}_i\bm{x}_i^\top\mathbbm{E}(y_i-\mu_i)^2\Big[\big\{h^\prime(\bm{x}_i^\top\tilde{\bm{\beta}}_n)\big\}^2-\big\{h^\prime(\bm{x}_i^\top\bm{\beta})\big\}^2\Big]\Big{\|},$\\ $A_{4n}=\Big{\|}n^{-1}\sum_{i=1}^{n}\bm{x}_i\bm{x}_i^\top\big\{h^\prime(\bm{x}_i^\top\tilde{\bm{\beta}}_n)\big\}^2\Big\{(y_i-\tilde{\mu}_i)^2-(y_i-\mu_i)^2\Big\}\Big{\|}$ and lastly we have that $A_{5n}=\Big{\|}n^{-1}\sum_{i=1}^{n}\bm{x}_i\bm{x}_i^\top\big\{h^\prime(\bm{x}_i^\top\tilde{\bm{\beta}}_n)\big\}^2\Big\{(y_i-\mu_i)^2-\mathbbm{E}(y_i-\mu_i)^2\Big\}\Big{\|}$.
Now by Taylor's expansion, for some $z_i^{(3)}$ with $|z_i^{(3)}-\bm{x}_i^\top\bm{\beta}|\leq |\bm{x}_i^\top(\tilde{\bm{\beta}}_n-\bm{\beta})|$, $i\in \{1,\dots,n\}$, we have
\begin{align*}
A_{3n}&\leq \Bigg[\max_{i=1,\dots,n}\Big\{\|\bm{x}_i\|^3*|h^{\prime\prime}(z_i^{(3)})|*2|h^{\prime}(z_i^{(3)})|\Big\}\Bigg]*\Big\{n^{-1}\sum_{i=1}^{n}\mathbbm{E}(y_i-\mu_i)^2\Big\}*\|\tilde{\bm{\beta}}_n-\bm{\beta}\|\\
&=A_{31n}*A_{32n}*A_{33n}\;\;\text{(say)}.
\end{align*}
Now due to assumptions (C.3), and (C.4) and using Lemma \ref{lem:asconcentration}, $A_{31n}=O(1)$.
Again $A_{33n}=o(1)\;\;w.p\;1$ by Lemma \ref{lem:asconcentration} and $A_{32n}=O(1)$ due to assumption (C.5). Therefore combining all the things we have 
\begin{align}\label{eqn:var2.1}
A_{3n}=o(1)\;\;w.p\;1.
\end{align}

Again by Taylor's expansion, for some $z_i^{(4)}$ with $|z_i^{(4)}-\bm{x}_i^\top\bm{\beta}|\leq |\bm{x}_i^\top(\tilde{\bm{\beta}}_n-\bm{\beta})|$, and for some $z_i^{(5)}$ with $|z_i^{(5)}-\bm{x}_i^\top\bm{\beta}|\leq |\bm{x}_i^\top(\tilde{\bm{\beta}}_n-\bm{\beta})|$, $i\in \{1,\dots,n\}$, we have,
\begin{align*}
A_{4n}\leq\;& A_{41n}+A_{42n}\;\;\text{(say)}.
\end{align*} 
where, $A_{41n}=\Big[2 \max_{i=1,\dots,n}\big\{\|\bm{x}_i\|^3*|h^{\prime}(\bm{x_i^\top\tilde{\beta_n}})|^2*|g^{-1}(z_i^{(4)})|*|(g^{-1})^{\prime}(z_i^{(4)})|\big\}\Big]*\|\tilde{\bm{\beta}}_n-\bm{\beta}\|$ and $A_{42n}=\Big[2 \max_{i=1,\dots,n}\big\{\|\bm{x}_i\|^3*|h^{\prime}(\bm{x_i^\top\tilde{\beta_n}})|^2*|(g^{-1})^{\prime}(z_i^{(5)})|\big\}\Big]*\|\tilde{\bm{\beta}}_n-\bm{\beta}\|*\Big(n^{-1}\sum_{i=1}^{n}|y_i|\Big).$
Note that due to Lemma \ref{lem:asconcentration}, $\|\tilde{\bm{\beta}}_n-\bm{\beta}\|=o(1)\;\;w.p\; 1$ and $n^{-1}\sum_{i=1}^{n}\big(|y_i|\big)=O(1)\;\;$ by Markov Inequality and (A.5). Again due to the assumptions (C.3) and (C.4), the ``max'' terms are bounded $w.p\; 1$. Hence 
\begin{align}\label{eqn:var2.2}
A_{4n}=o(1)\;\;w.p\;1.
\end{align}
Note that $A_{5n}\leq A_{51n}+A_{52n} \;\;\text{(say)}$, where
$A_{51n}:=\Big{\|}n^{-1}\sum_{i=1}^{n}\bm{x}_i\bm{x}_i^\top\big\{h^\prime(\bm{x}_i^\top\bm{\beta})\big\}^2\Big\{(y_i-\mu_i)^2-\mathbbm{E}(y_i-\mu_i)^2\Big\}\Big{\|}$ and $ A_{52n}:=\Big{\|}n^{-1}\sum_{i=1}^{n}\bm{x}_i\bm{x}_i^\top\Big[\big\{h^\prime(\bm{x}_i^\top\tilde{\bm{\beta}}_n)\big\}^2-\big\{h^\prime(\bm{x}_i^\top\bm{\beta})\big\}^2\Big]*\Big\{(y_i-\mu_i)^2 - \mathbbm{E}(y_i-\mu_i)^2 \Big\}\Big{\|}$
Now $A_{51n}=o(1)\;\;w.p\;1$, due to assumptions (C.3), (C.4) and (C.5) and using Lemma \ref{lem:F-N} with $t=3$ and then Borel-Cantelli Lemma. $A_{52n}$ can be dealt with similarly to $A_{3n}$ and $A_{51n}$ and hence  
\begin{align}\label{eqn:var2.3}
A_{5n}=o(1)\;\;w.p\;1.
\end{align}
Combining (\ref{eqn:var2.1}), (\ref{eqn:var2.2}) and (\ref{eqn:var2.3}) the proof of Lemma \ref{lem:var2} is now complete. \hfill $\square$

\begin{lemma}\label{lem:var2prb}
  Under the assumptions (C.1)-(C.5) we have,
  \[
\Bigg\|\text{var}_*\Big(\frac{\tilde G^\top\bm{e}^*}{n^{1/2}}\Big)-\bm{L}\Bigg\|=o(1)\quad\text{with probability}\; 1.
\]
\end{lemma}

Proof of Lemma \ref{lem:var2prb} We recall that \[\bm{S}_n=\text{var}(\bm{W}_n)=\frac{1}{n}\sum_{i=1}^n\bm{x}_i\bm{x}_i^\top[h^\prime(\bm{x}_i^\top\bm{\beta})]^2\underbrace{b^{\prime\prime}[h(\bm{x}_i^\top\bm{\beta})]}_{:=\text{var}(y_i)}=\mathbbm{E}(L_n)=\frac{G^\top G}{n},
\]
where $G=V^{1/2}\Delta X$. On the other hand we write,
\begin{align*}
\text{var}_*\Big(\frac{\tilde G^\top\bm{e}^*}{n^{1/2}}\Big)&=\frac{1}{n}\tilde{G}^\top\text{var}_*(\bm{e}^*)\tilde{G}=\frac{1}{n}\tilde{G}^\top\Bigg[\frac{1}{n}\sum_{k=1}^{n}\Big(e_k^\dagger-\bar{e}^\dagger\Big)^2\Bigg]I_n\tilde{G}.
\end{align*}

Since, $\frac{1}{n}\sum_{k=1}^{n}\Big(e_k^\dagger-\bar{e}^\dagger\Big)^2$ is a scalar term and $\Bigg\|\frac{\tilde G^\top\tilde G}{n}-\mathbbm{E}(L_n)\Bigg\|=o(1)$ almost surely due to Lemma \ref{lem:prbvarest}, it's enough to prove that,
\[
\frac{1}{n}\sum_{k=1}^{n}\Big(e_k^\dagger-\bar{e}^\dagger\Big)^2\xrightarrow{a.s} 1\quad\text{as}\;\; n\to\infty.
\]
Now, $\frac{1}{n}\sum_{k=1}^{n}\Big(e_k^\dagger-\bar{e}^\dagger\Big)^2=\underbrace{\frac{1}{n}\sum_{k=1}^{n}\frac{[y_k - g^{-1}(\bm{x}_k^\top\tilde{\bm{\beta}}_n)]^2}{b^{\prime\prime}[h(\bm{x}_k^\top\tilde{\bm{\beta}}_n)]}}_{\text{Term I}}-\underbrace{\Bigg[\frac{1}{n}\sum_{k=1}^{n}\frac{y_k - g^{-1}(\bm{x}_k^\top\tilde{\bm{\beta}}_n)}{\sqrt{b^{\prime\prime}[h(\bm{x}_k^\top\tilde{\bm{\beta}}_n)]}}\Bigg]^2}_{\text{Term II}}$.\\

Now, we will show that $\text{Term I}\xrightarrow{a.s}1$ as $n\to\infty$. For that we denote,
\[
\eta_k:=x_k^\top\beta,\qquad
\tilde\eta_{k,n}:=x_k^\top\tilde\beta_n,\qquad
\mu_k:=g^{-1}(\eta_k).
\]
Also due to Lemma \ref{lem:asconcentration} we have,
\[
\|\tilde{\bm{\beta}}_n-\bm\beta\|
=o\!\left(\frac{\log n}{\sqrt n}\right)\quad a.s.
\]
Then due to assumption (C.4) uniformly in $k$,
\[
|\tilde\eta_{k,n}-\eta_k|
\le
\|x_k\|\,\|\tilde{\bm{\beta}}_n-\bm\beta\|
=
o\!\left(\frac{\log n}{\sqrt n}\right)=o(1)
\quad a.s..
\]
By the mean–value form of Taylor expansion, for every $k\in\{1,...,n\}$;
\[
g^{-1}(\tilde\eta_{k,n})
=
g^{-1}(\eta_k)
+
(\tilde\eta_{k,n}-\eta_k)(g^{-1})'(z_{k,n}),
\quad\text{for some}\;\;
|z_{k,n}-\eta_k|\le|\tilde\eta_{k,n}-\eta_k|.
\]
Hence
\[
y_k-g^{-1}(\tilde\eta_{k,n})
=
(y_k-\mu_k)
-
(\tilde\eta_{k,n}-\eta_k)(g^{-1})'(z_{k,n}),
\]
and therefore
\begin{align}\label{eqn:tag1}
\bigl[y_k-g^{-1}(\tilde\eta_{k,n})\bigr]^2
&=
(y_k-\mu_k)^2
-2(y_k-\mu_k)(\tilde\eta_{k,n}-\eta_k)(g^{-1})'(z_{k,n})
+(\tilde\eta_{k,n}-\eta_k)^2 (g^{-1})'(z_{k,n})^2 .
\end{align}

We handle the denominator by applying the mean--value theorem directly to the
inverse variance map.  
Using
\[
b''(h(u))=\frac{(g^{-1})'(u)}{h'(u)},\qquad h'(u)\neq 0,
\]
define
\[
\phi(u):=\frac{1}{b''(h(u))}=\frac{h'(u)}{(g^{-1})'(u)} .
\]

By the mean--value theorem, for each $k$,
\[
\phi(\tilde\eta_{k,n})
=
\phi(\eta_k)
+
(\tilde\eta_{k,n}-\eta_k)\,\phi'(\zeta_{k,n}),
\qquad
|\zeta_{k,n}-\eta_k|\le|\tilde\eta_{k,n}-\eta_k|.
\]

The derivative $\phi'(u)$ can be computed explicitly:
\[
\phi'(u)
=
\frac{h''(u)(g^{-1})'(u)-h'(u)(g^{-1})''(u)}
{\bigl[(g^{-1})'(u)\bigr]^2}.
\]
Therefore,
\[
\frac{1}{b''(h(\tilde\eta_{k,n}))}
=
\frac{1}{b''(h(\eta_k))}
+
(\tilde\eta_{k,n}-\eta_k)
\frac{h''(\zeta_{k,n})(g^{-1})'(\zeta_{k,n})
-
h'(\zeta_{k,n})(g^{-1})''(\zeta_{k,n})}
{\bigl[(g^{-1})'(\zeta_{k,n})\bigr]^2}.
\]

Under the standing regularity assumptions that $(g^{-1})'$ is bounded away from
zero and that $(g^{-1})''$, $h'$, and $h''$ are bounded on compact subsets of the
parameter space, we have
\[
\sup_k
\left|
\phi'(\zeta_{k,n})
\right|
<\infty
\quad\text{almost surely}.
\]
Consequently, we obtain the concise representation
\[
\frac{1}{b''(h(\tilde\eta_{k,n}))}
=
\frac{1}{b''(h(\eta_k))}
+
(\tilde\eta_{k,n}-\eta_k)\,\phi'(\zeta_{k,n}),
\qquad
\sup_k|\phi'(\zeta_{k,n})|<\infty\ \text{a.s.}
\]

Combining the numerator expression with the reciprocal expansion,
\[
\begin{aligned}
&\frac{\bigl[y_k-g^{-1}(\tilde\eta_{k,n})\bigr]^2}
{b''(h(\tilde\eta_{k,n}))}
=
\frac{(y_k-\mu_k)^2}{b''(h(\eta_k))}-2(y_k-\mu_k)
\frac{(g^{-1})'(z_{k,n})}{b''(h(\eta_k))}
(\tilde\eta_{k,n}-\eta_k)
\\
&\quad
+(y_k-\mu_k)^2
(\tilde\eta_{k,n}-\eta_k)\,\phi'(\zeta_{k,n})
+(\tilde\eta_{k,n}-\eta_k)^2
\frac{(g^{-1})'(z_{k,n})^2}{b''(h(\eta_k))}
\\
&\quad
-2(y_k-\mu_k)
(\tilde\eta_{k,n}-\eta_k)^2
(g^{-1})'(z_{k,n})\,\phi'(\zeta_{k,n})+(\tilde\eta_{k,n}-\eta_k)^3
(g^{-1})'(z_{k,n})^2\,\phi'(\zeta_{k,n}).
\end{aligned}
\]

Equivalently, grouping by powers of $(\tilde\eta_{k,n}-\eta_k)$,
\[
\begin{aligned}
\frac{\bigl[y_k-g^{-1}(\tilde\eta_{k,n})\bigr]^2}
{b''(h(\tilde\eta_{k,n}))}
&=
\frac{(y_k-\mu_k)^2}{b''(h(\eta_k))}
+(\tilde\eta_{k,n}-\eta_k)\,A_{k,n}
+(\tilde\eta_{k,n}-\eta_k)^2\,B_{k,n}
+(\tilde\eta_{k,n}-\eta_k)^3\,C_{k,n},
\end{aligned}
\]
where
\[
\begin{aligned}
A_{k,n}
&=
-2(y_k-\mu_k)
\frac{(g^{-1})'(z_{k,n})}{b''(h(\eta_k))}
+
(y_k-\mu_k)^2\,\phi'(\zeta_{k,n}),
\\[0.5em]
B_{k,n}
&=
\frac{(g^{-1})'(z_{k,n})^2}{b''(h(\eta_k))}
-2(y_k-\mu_k)(g^{-1})'(z_{k,n})\,\phi'(\zeta_{k,n}),
\\[0.5em]
C_{k,n}
&=
(g^{-1})'(z_{k,n})^2\,\phi'(\zeta_{k,n}).
\end{aligned}
\]

Since, $\mathbbm{E}(y-\mu)^2=\mathbbm{v}\text{ar}(y)=b''(h(\eta))$, by SLLN we have;
\[
\frac{1}{n}\sum_{k=1}^{n}\frac{(y_k-\mu_k)^2}{b''(h(\eta_k))}\xrightarrow{a.s} 1\quad\text{as}\;\; n\to\infty.
\]
All the remaining terms are $o(1)$ almost surely due to the assumptions (C.3), (C.4), (C.5) and application of Lemma \ref{lem:F-N}, Lemma  \ref{lem:asconcentration} and Borel-Cantelli Lemma.\\

Now similar handling and proof steps will lead us to,
\[
\frac{1}{n}\sum_{k=1}^{n}\frac{y_k - g^{-1}(\bm{x}_k^\top\tilde{\bm{\beta}}_n)}{\sqrt{b^{\prime\prime}[h(\bm{x}_k^\top\tilde{\bm{\beta}}_n)]}}\xrightarrow{a.s}0\quad\text{as}\;\; n\to\infty.
\]
Combining everything we have,
\[
\frac{1}{n}\sum_{k=1}^{n}\Big(e_k^\dagger-\bar{e}^\dagger\Big)^2\xrightarrow{a.s} 1\quad\text{as}\;\; n\to\infty.
\]
Therefore the proof is complete due to triangle inequality and Lemma \ref{lem:prbvarest}. \hfill $\square$.

\begin{lemma}\label{lem:bnormal}
Under the assumptions (C.2)-(C.5), we have
$$
\mathcal{L}\Big{(}\bm{W}_n\Big{)}\xrightarrow{d} N\big{(}0,\bm{L}\big{)}\;\; \text{and}\;\;\mathcal{L}\Big{(}\tilde{\bm{W}}_n^{*}\mid \mathcal{E}\Big{)}\xrightarrow{d_*} N\big{(}0,\bm{L}\big{)}, \;\;\text{w.p}\; 1.
$$
\end{lemma}
Proof of Lemma \ref{lem:bnormal}. First we are going to show $\mathcal{L}\Big{(}\bm{W}_n\Big{)}\xrightarrow{d} N\big{(}0,\bm{L}\big{)}$. Since $\mathbbm{V}\text{ar}(\bm{W}_n)=\bm{S}_n$ and $\bm{S}_n\rightarrow \bm{L}$, hence using Cramer-Wold device, it is enough to show that
\begin{align}\label{eqn:N.1}
\sup_{x\in \mathbb{R}}\Big|\mathbbm{P}\Big{(}\bm{t}^\top\bm{W}_n\leq x\Big)-\Phi\big(x s_n^{-1}(\bm{t})\big)\Big|=o(1),
\end{align}
where $s_n^2(\bm{t})=\bm{t}^\top\bm{S}_n\bm{t}$. Now due to Berry-Esseen Theorem, given as Theorem 12.4 in \citet{bhattacharya1986normal}, we have 
\begin{align*}
&\sup_{x\in \mathbb{R}}\Big|\mathbbm{P}\Big{(}\bm{t}^\top\bm{W}_n\leq x\Big)-\Phi(xs_n^{-1}(\bm{t}))\Big| \leq (2.75)\dfrac{\sum_{i=1}^{n}E\Big|n^{-1/2}\bm{t}^\top\bm{x}_i(y_i-\mu_i)h^\prime(\bm{x}_i^\top\bm{\beta})\Big|^3}{\Big(\bm{t}^\top\bm{S}_n \bm{t}\Big)^{3/2}}\\
& \leq (2.75)\eta_{1n}^{-3/2}n^{-1/2}\max\Big\{\|\bm{x}_i\|^3\mathbbm{E}|y_i-\mu_i|^3 |h^\prime(\bm{x}_i^\top\bm{\beta})|^3:i\in \{1\dots,n\}\Big\}=o(1),
\end{align*}
where $\eta_{1n}$ is the smallest eigen value of $\bm{S}_n$. The last equality follows since $\bm{S}_n$ converges to a p.d matrix $\bm{S}$. Therefore, we are done. \hfill $\square$\\
Now let us consider the Bootstrap version. Consider $\bm{A}\in \mathcal{E}$ such that $P(\bm{A})=1$ and on the the set $\bm{A}$, we have $\|\tilde{\bm{S}}_n-\bm{L}\|=o(1)$ and $\|\bm{T}_n\|=o(\log n)$. Hence due to Lemma \ref{lem:var2} and using Cramer-Wold device, it is enough to show that, on $A$,
\begin{align*}
\sup_{x\in \mathbb{R}}\Big|\mathbbm{P}_*\Big{(}\bm{t}^\top\tilde{\bm{W}}_n^{*}\leq x\Big)-\Phi\big(x \tilde{s}_n^{-1}(\bm{t})\big)\Big|=o(1)
\end{align*}
where $\tilde{s}_n^{2}(\bm{t})=\bm{t}^\top\tilde{\bm{S}}_n\bm{t}$. Now due to Berry-Esseen Theorem, given as Theorem 12.4 in \citet{bhattacharya1986normal}, we have on the set $A$,
\begin{align*}
\sup_{x\in \mathbb{R}}&\Big|\mathbbm{P}_*\Big{(}\bm{t}^\top\tilde{\bm{W}}_n^{*}\leq x\Big)-\Phi(x\tilde{s}_n^{-1}(\bm{t}))\Big|\\
&\leq (2.75)\dfrac{\sum_{i=1}^{n}E_*\Big|n^{-1/2}\big(y_i-\tilde{\mu}_i\big)h^\prime(\bm{x}_i^\top\tilde{\bm{\beta}}_n)\bm{t}^\top\bm{x}_i(G_i^*-\mu_{G^*})\mu_{G^*}^{-1}\Big|^3}{\Big(\bm{t}^\top\tilde{\bm{S}}_n\bm{t}\Big)^{3/2}}\\
& \leq 11*\tilde{\eta}_{1n}^{-3/2}E_*|G_1^*-\mu_{G^*}|^3\mu_{G^*}^{-3}\Big(A_{51n}+A_{52n}\Big),
\end{align*}
where $\tilde{\eta}_{1n}$ is the smallest eigen value of $\tilde{\bm{S}}_n$. Again, $A_{51n}=n^{-1/2}*\Bigg[\max_{i=1,\dots,n}\Big\{|h^\prime(\bm{x}_i^\top\tilde{\bm{\beta}}_n)|^3*\|\bm{x}_i\|^3\Big\}\Bigg]*\Big(n^{-1}\sum_{i=1}^{n}|y_i|^3\Big)$ and  $A_{52n}=n^{-1/2}*\Bigg[\max_{i=1,\dots,n}\Big\{|h^\prime(\bm{x}_i^\top\tilde{\bm{\beta}}_n)|^3*\|\bm{x}_i\|^3*|g^{-1}(\bm{x}_i^\top\tilde{\bm{\beta}}_{n})|^3\Big\}\Bigg].$
Now due to Lemma \ref{lem:F-N} with $t=2$ combined with Borel-Cantelli Lemma, assumptions (C.2), (C.4) \& (C.5), on the set $\bm{A}$ we have $(A_{51n}+A_{52n})=o(1)$ and $\tilde{\eta}_{1n}^{-3/2}=O(1)$. Again $E_*|G_1^*-\mu_{G^*}|^3\mu_{G^*}^{-3}<\infty$. Therefore we are done. \hfill $\square$

\begin{lemma}\label{lem:bnormalprb}
 Under the assumptions (C.2)-(C.5) we have,
 \[
 \mathcal{L}\Big(\frac{\tilde G^\top\bm{e}^*}{n^{1/2}}\Big|\mathcal{E}\Big)\xrightarrow{d_*} N\big{(}0,\bm{L}\big{)}, \;\;\text{w.p}\; 1.
 \]
\end{lemma}

Proof of Lemma \ref{lem:bnormalprb} This proof follows similarly from Lemma \ref{lem:bnormal} and \ref{lem:var2prb}. \hfill $\square$

\begin{lemma}\label{lem:stcontwrtu}
For $\bm{u}\in\mathbb{R}^p$, define $V_n(\bm{u},\lambda_{n})=(1/2)\bm{u}^\top \bm{L}_n\bm{u}-\bm{W}_n^\top \bm{u}+\lambda_{n}\sum_{j=1}^{p}\big(|\beta_j+n^{-1/2}u_j|-|\beta_j|\big)$ with $\bm{L}_n$, $\bm{W}_n$ defined earlier and $\bm{\beta}$ being a fixed quantity. Then under the assumption (C.2)-(C.6), $\{V_n(\cdot,\lambda_{n})\}_{n\geq 1}$ is stochastically equicontinuous on compacta. More precisely, for every $\varepsilon,\eta,\psi>0$, there exists a $\delta:=\delta(\varepsilon,\eta,\psi)>0$ such that
$$\mathbbm{P}\Big(\sup_{\{\|\bm{u}^\prime-\bm{u}\|<\delta\},\{\max\{\|\bm{u}^\prime\|,\|\bm{u}\|\}\leq \psi\}}|V_n(\bm{u}^\prime,\lambda_{n})-V_n(\bm{u},\lambda_{n})|>\eta\Big)<\varepsilon,$$
for large enough $n$.
\end{lemma}
Proof of Lemma \ref{lem:stcontwrtu}: Fix some $\varepsilon,\eta,\psi>0$, and choose a $\delta:=\delta(\varepsilon,\eta,\psi)>0$ such that we consider all such $\bm{u},\bm{u}^\prime$ with $\|\bm{u}\|,\|\bm{u}^\prime\|\le \psi$ for which $\|\bm{u}^\prime-\bm{u}\|<\delta$. Now due to Lemma \ref{lem:bnormal} $\bm{W}_n\xrightarrow{d} \bm{W}_\infty$, with $\bm{W}_\infty\sim N_p(\bm{0},\bm{L})$. Therefore for every $\varepsilon>0$, there exists $0<M^\prime:=M^\prime(\varepsilon)<\infty$ such that, 
\begin{align}\label{eqn:stcontwrtu}
 \mathbbm{P}\big(\|\bm{W}_n\|>M^\prime(\varepsilon)\big)<\varepsilon.   
\end{align}

Now we see that,
\begin{align}\label{eqn:difference}
&V_n(\bm{u}^\prime,\lambda_{n})-V_n(\bm{u},\lambda_{n})\nonumber\\
&=(1/2)\Big[(\bm{u}^\prime-\bm{u})^\top \bm{L}_n(\bm{u}^\prime-\bm{u})+2(\bm{u}^\prime-\bm{u})^\top \bm{L}_n\bm{u}\Big]-\bm{W}_n^\top [\bm{u}^\prime-\bm{u}]\nonumber\\
&\;\;\;\;\;\;\;\;\;\;\;\;\;\;\;\;\;\;\;\;\;\;\;\;\;\;\;\;\;\;\;\;\;\;\;\;\;\;\;\;\;\;\;\;\;\;\;\;\;\;\;\;\;\;+\lambda_{n}\sum_{j=1}^{p}\Big[|\beta_j+n^{-1/2}u^\prime_j|-|\beta_j+n^{-1/2}u_j|\Big].
\end{align}
 Therefore from (\ref{eqn:difference}) and assumptions (C.3) and (C.6) we have,
\begin{align}\label{eqn:towardstcont}
&\sup_{\{\|\bm{u}^\prime-\bm{u}\|<\delta\},\{\max\{\|\bm{u}^\prime\|,\|\bm{u}\|\}\leq \psi\}}|V_n(\bm{u}^\prime,\lambda_{n})-V_n(\bm{u},\lambda_{n})|\leq \tilde{\gamma}_1\delta(\delta+2\psi)+\delta\|\bm{W}_n\|+2\lambda_0p^{1/2}\delta
\end{align}
Now considering, $\delta=\frac{1}{2}\min\Big\{\Big(\frac{\eta}{4\tilde{\gamma}_1}\Big)^{1/2},\frac{\eta}{8[\tilde{\gamma}_1\psi+\lambda_0p^{1/2}]},\frac{\eta}{2M^\prime(\varepsilon)}\Big\}$ and from (\ref{eqn:stcontwrtu}), for large enough $n$ we have,
\begin{align}\label{eqn:final1232}
\mathbbm{P}\Big(\sup_{\{\|\bm{u}^\prime-\bm{u}\|<\delta\},\{\max\{\|\bm{u}^\prime\|,\|\bm{u}\|\}\leq \psi\}}|V_n(\bm{u}^\prime,\lambda_{n})-V_n(\bm{u},\lambda_{n})|>\eta\Big)<\varepsilon. 
\end{align}
Therefore our proof is complete. \hfill $\square$

\begin{lemma}\label{lem:argmin}
Suppose that $\{\bm{U}_n(\cdot)\}_{n\geq 1}$ and $\bm{U}_{\infty}(\cdot)$ are convex stochastic processes on $\mathbb{R}^p$ such that $\bm{U}_{\infty}(\cdot)$ has almost surely unique minimum $\xi_{\infty}$. Also assume that\\
(a) every finite dimensional distribution of $\bm{U}_n(\cdot)$ converges to that of $\bm{U}_{\infty}(\cdot)$, that is, for any natural number $k$ and for any $\{\bm{t}_1,..,\bm{t}_k\}\subset \mathbb{R}^p$, we have  $(\bm{U}_n(\bm{t}_1),...\bm{U}_n(\bm{t}_k))\xrightarrow{d}(\bm{U}_{\infty}(\bm{t}_1),...\bm{U}_{\infty}(\bm{t}_k)).$\\ 
(b) $\{\bm{U}_n(\cdot)\}_{n\geq 1}$ is equicontinuous on compact sets, in probability, i.e., for every $\epsilon, \eta, M>0$, there exists a $\delta>0$ such that $$\limsup_{n\to\infty}\mathbbm{P}\Bigg(\sup_{\{\|\bm{r}-\bm{s}\|<\delta, \max \{\|\bm{r}\|,\|\bm{s}\|\}<M\}}\big|\bm{U}_n(r)-\bm{U}_n(s)\big|>\eta\Bigg)<\epsilon.$$
Then we have $\operatorname*{Argmin}_{\bm{t}}\bm{U}_n(\bm{t}) \xrightarrow{d} \xi_{\infty}$.
\end{lemma}
Proof of lemma \ref{lem:argmin}. We have ignored the measurability issues in defining stochastic equicontinuity above, to keep the representation simple, and will utilize the results of \citet{kim1990cube} and \citet{davis1992m} to reach the conclusion. As in \citet{kim1990cube}, let $\mathbb{B}_{loc}(\mathbb{R}^p)$ be the space of all locally bounded functions on $\mathbb{R}^p$. We equip this space with the metric $\tau(\cdot,\cdot)$, defined by 
$$\tau(u,v)=\sum_{k=1}^{\infty}2^{-k}\min \{1,\tau_k(u,v)\},$$ where, $\tau_k(u,v)=\sup_{||\bm{t}||\leq k}|u(\bm{t})-v(\bm{t})|$. This metric generates the topology of uniform convergence on compacta, on $\mathbb{B}_{loc}(\mathbb{R}^p)$.\\ 
First, note that $\bm{U}_{n}(\cdot)$ and $\bm{U}_{\infty}(\cdot)$ belong to $\mathbb{B}_{loc}(\mathbb{R}^p)$. Therefore, Theorem 2.3 of \citet{kim1990cube} implies that $\bm{U}_n(\cdot)\xrightarrow{d}\bm{U}_{\infty}(\cdot)$ under the topology of uniform convergence on compacta. Now apply Theorem 2.2 of \citet{kim1990cube}, Dudley's almost sure representation theorem, to get a new probability space $(\tilde{\Omega},\tilde{\mathcal{F}},\tilde{\mathbb{P}})$ such that we can define $\tilde{\bm{U}}_n(\cdot,\tilde{\omega})=\bm{U}_n(\cdot,\phi_n(\tilde{\omega}))$ and $\tilde{\bm{U}}_{\infty}(\cdot,\tilde{\omega})=\bm{U}_{\infty}(\cdot,\phi_{\infty}(\tilde{\omega}))$ for each $\tilde{\omega}\in \tilde{\Omega}$ based on perfect maps $\phi_n:\tilde{\Omega}\to \Omega$ and $\phi_{\infty}:\tilde{\Omega}\to \Omega$. Moreover, 
\begin{enumerate}[label=(\Alph*)]
    \item $\tilde{\bm{U}}_n(\cdot)$ and $\bm{U}_n(\cdot)$ have same finite dimensional distributions and $\tilde{\bm{U}}(\cdot)$ and $\bm{U}_{\infty}(\cdot)$ have same finite dimensional distributions.
    \item there exists a sequence of random variables $\{\tilde{\varepsilon}_n\}_{n\geq 1}$ on $(\tilde{\Omega},\tilde{\mathcal{F}},\tilde{\mathbb{P}})$ such that $$\tau\Big(\tilde{\bm{U}}_n(\cdot,\tilde{\omega}),\tilde{\bm{U}}_{\infty}(\cdot,\tilde{\omega})\Big)\leq \tilde{\varepsilon}_n(\tilde{\omega}),\; \text{for every}\; \tilde{\omega}\in \tilde{\Omega},\;\text{and}\; \tilde{\varepsilon}_n\rightarrow 0,\;\text{a.s}\;\mathbb{\tilde{P}}.$$
\end{enumerate}
Define, $\tilde{\bm{\xi}}_n = Argmin_{\bm{s}\in \mathbb{R}^p}\tilde{\bm{U}}_n(\bm{s})$ for all $n$ and $\tilde{\bm{\xi}}_{\infty}=Argmin_{\bm{s}\in \mathbb{R}^p}\tilde{\bm{U}}_{\infty}(\bm{s})$. Clearly, $\tilde{\bm{U}}_n(\cdot)$ and $\tilde{\bm{U}}_{\infty}(\cdot)$ are convex stochastic processes and $\tilde{\bm{\xi}}_{\infty}$ is the almost sure unique minimum of $\tilde{\bm{U}}_{\infty}(\cdot)$, due to the properties of perfect maps. Now to establish that $\bm{\xi}_n\xrightarrow{d}\bm{\xi}_{\infty}$, it suffices to prove that, for every uniformly continuous and bounded function $g$ on $\mathbb{R}^p$, $\mathbbm{E}(g(\bm{\xi}_{n}))\rightarrow \mathbbm{E}(g(\bm{\xi}_{\infty}))$ as $n\to\infty$. However for such a function $g$, due to the properties of perfect maps,
\begin{align}\label{eqn:dudley}
\big|\mathbbm{E}(g(\bm{\xi}_{n}))- \mathbbm{E}(g(\bm{\xi}_{\infty}))\big|=\big|\tilde{\mathbbm{E}}(g(\tilde{\bm{\xi}}_{n}))- \tilde{\mathbbm{E}}(g(\tilde{\bm{\xi}}_{\infty}))\big|\leq \tilde{\mathbbm{E}} \big|g(\tilde{\bm{\xi}}_{n})-g(\tilde{\bm{\xi}}_{\infty})\big|,
 \end{align}
and hence it is enough to establish that $\tilde{\bm{\xi}}_{n}\rightarrow\tilde{\bm{\xi}}_{\infty},\;\text{a.s}\;\tilde{\mathbb{P}}$. We will establish this by following the arguments of Lemma 2.2 of \citet{davis1992m}.\\
Suppose that $\tilde{A}\in \tilde{\mathcal{F}}$ with $\tilde{\mathbb{P}}(\tilde{A})=1$, and on $\tilde{A}$, $\tilde{\xi}_{\infty}$ is the unique minimum of $\tilde{\bm{U}}_{\infty}$ and $\tilde{\varepsilon}_n\rightarrow 0$. Fix an $\tilde{\omega} \in \tilde{A}$ and assume, if possible, that $\big{\|}\tilde{\bm{\xi}}_{n}(\tilde{\omega})-\tilde{\bm{\xi}}_{\infty}(\tilde{\omega})\big{\|}>\gamma$ for infinitely many $n$, for some $\gamma > 0$. Now consider the compact subset $K_\gamma(\tilde{\omega})=\big\{\bm{s}:\|\bm{s}-\tilde{\bm{\xi}}_{\infty}(\tilde{\omega})\|=\gamma\big\}$ of $\mathcal{R}^p$. Hence due to the properties of the set $\tilde{A}$, we have as $n\to\infty$,
\begin{align}\label{eqn:davis1}
 & \sup_{\bm{s}\in K_\gamma(\tilde{\omega})} \Big|\tilde{\bm{U}}_n(\bm{s},\tilde{\omega})-\tilde{\bm{U}}(\bm{s},\tilde{\omega})\Big|\to 0\; \text{and}\; \tilde{\bm{U}}_n(\tilde{\bm{\xi}}_{\infty}(\tilde{\omega}),\tilde{\omega})\rightarrow\tilde{\bm{U}}(\tilde{\bm{\xi}}_{\infty}(\tilde{\omega}),\tilde{\omega}).  
\end{align}
Again, $\tilde{\xi}_{\infty}(\tilde{\omega})$ is the unique minimizer of $\tilde{\bm{U}}_{\infty}(\cdot, \tilde{\omega})$ implying that for any $\bm{s} \in K_\gamma(\tilde{\omega})$,
\begin{align}\label{eqn:uniformconveqn}
 \tilde{\bm{U}}_{n}(\bm{s},\tilde{\omega})>\tilde{\bm{U}}_n(\tilde{\bm{\xi}}_{\infty}(\tilde{\omega}),\tilde{\omega})\geq \tilde{\bm{U}}_{n}(\tilde{\bm{\xi}}_{n}(\tilde{\omega}),\tilde{\omega}),
 \end{align}
 for infinitely many $n$. This contradicts the convexity of $\tilde{\bm{U}}_n(\cdot, \tilde{\omega})$ by by choosing $\bm{s} \in K_\gamma(\tilde{\omega})$, such that the points
$\bm{s}, \tilde{\bm{\xi}}_{\infty}(\tilde{\omega}), \tilde{\bm{\xi}}_n(\tilde{\omega})$ are collinear.  Therefore, for any $\tilde{\omega} \in \tilde{A}$ and $\gamma > 0$, $\big{\|}\tilde{\bm{\xi}}_{n}(\tilde{\omega})-\tilde{\bm{\xi}}_{\infty}(\tilde{\omega})\big{\|}\leq\gamma$ for all but finitely many $n$, implying that $\tilde{\bm{\xi}}_{n}\rightarrow\tilde{\bm{\xi}}_{\infty},\;\text{a.s}\;\tilde{\mathbb{P}}$. \hfill $\square$

\section{Proofs of Main Results}\label{sec:proofmain}
In this section, we provide the proofs of our main results, i.e. proofs of Theorem \ref{thm:lassolimit},\ref{prop:ntheo}, \ref{thm:ptheo} and \ref{thm:prbpos}.

\subsection{Proof of Theorem \ref{thm:lassolimit}}\label{sec:thm4.1}
Note that
$$n^{1/2}(\hat{\bm{\beta}}_{n}- \bm{\beta}) = \mbox{Argmin}_{\bm{u}}V_n(\bm{u}) = \mbox{Argmin}_{\bm{u}} \Big\{ \ell_{1n}(\bm{u}) + \ell_{2n}(\bm{u}) \Big \},$$
where,
$\ell_{1n}(\bm{u}) = \sum_{i=1}^{n}\Big[-y_{i}\big[ h\big\{\bm{x}_{i}^\top(\bm{\beta}+\frac{\bm{u}}{n^{1/2}})\big\}-h\big(\bm{x}_{i}^\top\bm{\beta}\big)\big] +\big[h_{1}\big\{\bm{x}_{i}^\top(\bm{\beta}+\frac{\bm{u}}{n^{1/2}})\big\}-h_{1}\big(\bm{x}_{i}^\top\bm{\beta}\big)\big] \Big]$,
with $h_{1} = b \circ h$ and $\ell_{2n}(u) = \lambda_{n}\sum_{j=1}^{p}\Big(|\beta_{j}+\frac{u_{j}}{n^{1/2}}|-|\beta_{j}|\Big)$.
Now, by Taylor's theorem,
\begin{align*}
&h\big\{x_{i}^\top(\beta+\frac{u}{n^{1/2}})\big\}-h\big(x_{i}^\top\beta\big)= n^{-1/2}(\bm{u}^\top\bm{x}_i)h^\prime(\bm{x}_i^\top\bm{\beta}) + (2n)^{-1}(\bm{u}^\top\bm{x}_i)^2h^{\prime\prime}(\bm{x}_i^\top\bm{\beta})\\
&\qquad\qquad\qquad\qquad\qquad\qquad\qquad\qquad\qquad\qquad\qquad\qquad+ (6n^{3/2})^{-1}(\bm{u}^\top\bm{x}_i)^3h^{\prime\prime\prime}(z_i),\\
&h_1\big\{x_{i}^\top(\beta+\frac{u}{n^{1/2}})\big\}-h_1\big(x_{i}^\top\beta\big)= n^{-1/2}(\bm{u}^\top\bm{x}_i)h_1^\prime(\bm{x}_i^\top\bm{\beta}) + (2n)^{-1}(\bm{u}^\top\bm{x}_i)^2h_1^{\prime\prime}(\bm{x}_i^\top\bm{\beta})\\
&\qquad\qquad\qquad\qquad\qquad\qquad\qquad\qquad\qquad\qquad\qquad\qquad+ (6n^{3/2})^{-1}(\bm{u}^\top\bm{x}_i)^3h_1^{\prime\prime\prime}(z_i),
\end{align*}
for some $z_i$'s such that $|z_i-\bm{x}_i^\top\bm{\beta}|\leq n^{-1/2}(\bm{u}^\top\bm{x}_i)$, $i\in \{1,\dots,n\}$. Now note that $h=(g \circ b^\prime)^{-1}$ and hence $h_1^\prime = (g^{-1})h^\prime$ and $h_1^{\prime\prime}=(g^{-1})^\prime h^\prime + (g^{-1})h^{\prime\prime}$. Therefore,
\begin{align*}
\ell_{1n}(\bm{u})= (1/2)\bm{u}^\top\bm{L}_n\bm{u} - \bm{W}_n^\top \bm{u} + R_{1n}(\bm{u}),
\end{align*}
where 
$R_{1n}(\bm{u})=(6n^{3/2})^{-1}\sum_{i=1}^{n}\Bigg\{-y_ih^{\prime\prime\prime}(z_i)+h_1^{\prime\prime\prime}(z_i)\Bigg\}(\bm{u}^\top\bm{x}_i)^3.$
Now note that $h_1^{\prime\prime\prime}=(g^{-1})^{\prime\prime}h^\prime + 2(g^{-1})^{\prime}h^{\prime\prime} + (g^{-1})h^{\prime\prime\prime}$. Hence using assumptions (C.3) and (C.4), we can claim that $\big\{|h^{\prime\prime\prime}(z_i)|+|h_1^{\prime\prime\prime}(z_i)|\big\}$ is bounded uniformly for all $i\in \{1,\dots,n\}$, for sufficiently large $n$. Again by using Markov's inequality we have $n^{-1}\sum_{i=1}^{n}|y_i|=O_p(1)$.  Therefore, $\|R_{1n}\|=o_P(1)$. Hence due to Lemma \ref{lem:varest} and Lemma \ref{lem:bnormal}, $$\ell_{1n}(\bm{u})\xrightarrow{d} \Big[(1/2)\bm{u}^\top\bm{L}\bm{u}-\bm{Z}_1^\top \bm{u}\Big],$$ where $\bm{Z}_1\sim N_p(\bm{0},\bm{L})$. Again as $\mathcal{A}=\{1,\dots,p_0\}$ and $n^{-1/2}\lambda_n\rightarrow \lambda_0$, for $n\rightarrow \infty$ we have
\begin{align*}
\ell_{2n}(\bm{u})=\lambda_n\sum_{j=1}^{p}\Big(|\beta_{j}+\dfrac{u_{j}}{n^{1/2}}|-|\beta_{j}|\Big) \rightarrow \lambda_0\Big[\sum_{j=1}^{p_0}sgn({{\beta}_{j}})u_j+\sum_{j=p_0+1}^{p}|u_j|\Big].
\end{align*}
Therefore, $V_n(\bm{u})\xrightarrow{d} V(\bm{u})= \bigg[\Big\{(1/2)\bm{u}^\top\bm{L}\bm{u}-\bm{Z}_1^\top \bm{u}\Big\} + \lambda_0\Big\{\sum_{j=1}^{p_0}sgn({{\beta}_{j}})u_j+\sum_{j=p_0+1}^{p}|u_j|\Big\}\bigg]. $\\
Hence we have finite dimensional distributional convergence of $V_n(\cdot)$ to $V(\cdot)$. Now due to Lemma \ref{lem:stcontwrtu}, $\{V_n(\cdot)\}_{n\geq 1}$ is equi-continuous on compact sets in probability. Since $\bm{L}$ is a p.d matrix, $V(\cdot)$ has almost sure unique minimum. Therefore, we can apply Lemma \ref{lem:argmin}, to claim that,
\begin{align*}
n^{1/2}\big(\hat{\bm{\beta}}_n-\bm{\beta}\big) \xrightarrow{d} \mbox{Argmin}_{\bm{u}} V(\bm{u}),
\end{align*}
Therefore the proof is complete. \hfill $\square$

\subsection{Proof of Theorem \ref{prop:ntheo}}\label{appAthm4.1}

We first define the set :
\begin{align*}
\bm{B}=&\Big\{n^{1/2}\|\hat{\bm{\beta}}_n-\bm{\beta}\|=o(\log n)\Big\}\cap \Big\{\|\hat{\bm{L}}_n^*-\bm{L}\|=o_{P_*}(1)\Big\}\\
&\cap \Big\{\mathcal{L}\big(\hat{\bm{W}}_n^*|\mathcal{E}\big)\xrightarrow{d}N(\bm{0},\bm{S})\Big\}\cap \Big\{(n^{-3/2})\sum_{i=1}^{n}\big(|y_i|-\mathbbm{E}|y_i|\big)=o(1)\Big\}
\end{align*}
We are going to show that 
\begin{align}\label{eqn:T.3}
    \mathbbm{P}\Big[\lim_{n\rightarrow \infty}\rho\big\{\hat{F}_n^{(PB)}(\cdot),G_{\infty}(\hat{\bm{T}}_{\infty},\cdot)\big\}= 0\Big]=1,
\end{align}
where $\hat{F}_n^{(PB)}(\cdot)$ is the conditional distribution of $n^{1/2}\big(\hat{\bm{\beta}}_n^{*(PB)}-\hat{\bm{\beta}}_n\big)$. Note that by Lemma \ref{lem:F-N}, $P\Big[(n^{-3/2})\sum_{i=1}^{n}\big(|y_i|-\mathbbm{E}|y_i|\big)=o(1)\Big]=1$. This fact together with Lemma \ref{lem:asconcentration}, \ref{lem:varest} and \ref{lem:bnormal}, imply $\mathbbm{P}(\bm{B})=1$. Then to prove (\ref{eqn:T.3}), it's enough to show that 
\begin{align}\label{eqn:T.4}
     \lim_{n\rightarrow \infty}\rho\big\{\hat{F}_n^{(PB)}(\omega, \cdot),G_{\infty}(\hat{\bm{T}}_{\infty}(\omega),\cdot)\big\}= 0,\; 
     \text{for all}\; \omega \in \bm{B}.
\end{align} 
Now note that for each $\omega \in \bm{B}$,
\begin{align}\label{eqn:T.5}
n^{1/2}(\hat{\bm{\beta}}^{*(PB)}_{n}- \hat{\bm{\beta}}_n) \equiv n^{1/2}\big\{\hat{\bm{\beta}}^{*(PB)}_{n}(\omega,\cdot)- \hat{\bm{\beta}}_n(\omega)\big\} = \mbox{Argmin}_{\bm{u}} \Big\{ \hat{\ell}^*_{1n}(\bm{u},\omega,\cdot) + \hat{\ell}^*_{2n}(\bm{u},\omega,\cdot) \Big \},
\end{align}
where, $\hat{\ell}_{2n}^*(\bm{u},\omega,\cdot) = \lambda_{n}\sum_{j=1}^{p}\big\{ \big|\hat{\beta}_{j,n}(\omega)+\frac{u_{j}}{n^{1/2}}\big|-\big|\hat{\beta}_{j,n}(\omega)\big|\big\}$
and
\begin{align*}
\hat{\ell}^*_{1n}(\bm{u},\omega,\cdot) &= \; \sum_{i=1}^{n}\Big[-y_{i}\Big\{h\big\{\bm{x}_{i}^\top(\hat{\bm{\beta}}_n(\omega)+\frac{\bm{u}}{n^{1/2}})\big\}-h\big\{\bm{x}_{i}^\top\hat{\bm{\beta}}_n(\omega)\big\}\Big\}\\ 
&+\Big\{h_{1}\big\{\bm{x}_{i}^\top(\hat{\bm{\beta}}_n(\omega)+\frac{\bm{u}}{n^{1/2}})\big\}-h_{1}\big\{\bm{x}_{i}^\top\hat{\bm{\beta}}_n(\omega)\big\}\Big\} \Big]G_i^*\mu_{G^*}^{-1}\\
&+ n^{-1/2}\sum_{i=1}^{n}\big\{y_i-\hat{\mu}_i(\omega)\big\}[h^\prime\{\bm{x}_i^\top\hat{\bm{\beta}}_n(\omega)\}](\bm{x}_i^\top\bm{u})
\end{align*}

Similar to original case, using Taylor's theorem we have
\begin{align*}
\hat{\ell}^*_{1n}(\bm{u},\omega,\cdot)= (1/2)\bm{u}^\top \big[\hat{\bm{L}}_n^*(\omega,\cdot)\big]\bm{u} - \bm{u}^\top\big[\hat{\bm{W}}_n^*(\omega,\cdot)\big]+\hat{R}_{1n}^*(\bm{u},\omega,\cdot),
\end{align*}
where $\hat{R}_{1n}^*(\bm{u},\omega,\cdot)=(6n^{3/2})^{-1}\sum_{i=1}^{n}\Bigg[-y_ih^{\prime\prime\prime}(\hat{z}_i^*)+h_1^{\prime\prime\prime}(\hat{z}_i^*)\Bigg](\bm{u}^\top\bm{x}_i)^3G_i^*\mu_{G^*}^{-1},$
for some $\hat{z}_i^*\equiv z_i^*(\bm{u},\omega,\cdot)$ such that $|\hat{z}_i^*-\bm{x}_i^\top\hat{\bm{\beta}}_n|\leq n^{-1/2}(\bm{u}^\top\bm{x}_i)$, $i\in \{1,\dots,n\}$.  Again use assumption (C.3) and $\mathbbm{E}(G_1^{*3})< \infty$ alongwith Lemma \ref{lem:asconcentration}, to claim that $\max\Big\{\big[|h^{\prime\prime\prime}(\hat{z}_i^*)|+|h_1^{\prime\prime\prime}(\hat{z}_i^*)|\big]: i\in \{1,\dots,n\}\Big\}=O(1)$ for all $\omega \in \bm{B}$. Again by Markov's inequality, we have
$n^{-3/2}\sum_{i=1}^{n}|y_i(\omega)|G_i^* =o_{P_*}(1)$ for all $\omega \in \bm{B}$. Therefore for all $\omega \in \bm{B}$, $\|\hat{R}_{1n}^*(\bm{u},\omega,\cdot)\|=o_{P_*}(1)$ and hence $$\hat{\ell}_{1n}^*(\bm{u},\omega,\cdot)\xrightarrow{d}\big\{(1/2)\bm{u}^\top \bm{L}\bm{u}-\bm{u}^\top \bm{Z}_2\big\}.$$
Using this fact along with Lemma \ref{lem:argmin}, it is remaining to show that
\begin{align}\label{eqn:T.6}
\hat{\ell}_{2n}^*(\bm{u},\omega,\cdot)  \rightarrow\; & \lambda_0\sum_{j=1}^{p_0}u_j sgn(\beta_j)+\lambda_0\sum_{j=p_0+1}^{p} \Big[sgn(\hat{T}_{\infty,j}(\omega)) \Big\{\hat{T}_{\infty,j}(\omega)-2\{u_j+\hat{T}_{\infty,j}(\omega)\}\nonumber \\
&\times \mathbbm{1}\big\{sgn(\hat{T}_{\infty,j}(\omega))(u_j+\hat{T}_{\infty,j}(\omega))<0\big\}\Big\}+|u_j|\mathbbm{1}\{\hat{T}_{\infty,j}(\omega)=0\}\Big],
\end{align}
for any $\omega \in \bm{B}$. Actually (\ref{eqn:T.6}) follows exactly through the same line as in case of Residual Bootstrap in the proof of Theorem 3.1 of \citet{chatterjee2010asymptotic} given at pages 4506-4507. Therefore we are done. \hfill $\square$ 

\subsection{Proof of Theorem \ref{thm:ptheo}}\label{appAthm4.2}
In Theorem \ref{thm:lassolimit}, we have already shown that, $\rho\big\{F_n(\cdot),F_{\infty}(\cdot)\big\}\rightarrow 0\;\text{as}\; n\rightarrow \infty.$
Hence it's enough to show that for any $\omega \in \bm{B}$, 
\begin{align}\label{T.7}
    \rho\big\{\tilde{F}_n^{(PB)}(\omega,\cdot),F_{\infty}(\omega)\big\}\rightarrow 0\;\text{as}\; n\rightarrow \infty.
\end{align}
 The definition of the set $\bm{B}$ is given in the proof of Theorem \ref{prop:ntheo}. To that end, note that for each $\omega \in \bm{B}$,
\begin{align}\label{eqn:T.8}
n^{1/2}(\hat{\bm{\beta}}^{*(PB)}_{n}- \tilde{\bm{\beta}}_n) \equiv n^{1/2}\big\{\hat{\bm{\beta}}^{*(PB)}_{n}(\omega,\cdot)- \tilde{\bm{\beta}}_n(\omega)\big\} = \mbox{Argmin}_{\bm{u}} \Big\{ \tilde{\ell}^*_{1n}(\bm{u},\omega,\cdot) + \tilde{\ell}^*_{2n}(\bm{u},\omega,\cdot) \Big \},
\end{align}
where
\begin{align*}
\tilde{\ell}^*_{1n}(\bm{u},\omega,\cdot) &= \; \sum_{i=1}^{n}\Bigg[-y_{i}\Big\{ h\big\{\bm{x}_{i}^\top(\tilde{\bm{\beta}}_n(\omega)+\frac{\bm{u}}{n^{1/2}})\big\}-h\big\{\bm{x}_{i}^\top\tilde{\bm{\beta}}_n(\omega)\big\}\Big\}\\
&+\Big\{h_{1}\big\{\bm{x}_{i}^\top(\tilde{\bm{\beta}}_n(\omega)+\frac{u}{n^{1/2}})\big\}-h_{1}\big\{\bm{x}_{i}^\top\tilde{\bm{\beta}}_n(\omega)\big\}\Big\} \Bigg]G_i^*\mu_{G^*}^{-1}\\
&+ n^{-1/2}\sum_{i=1}^{n}\big\{y_i-\tilde{\mu}_i(\omega)\big\}[h^\prime\{\bm{x}_i^\top\tilde{\bm{\beta}}_n(\omega)\}](\bm{x}_i^\top\bm{u})
\end{align*}
and $\tilde{\ell}_{2n}^*(\bm{u},\omega,\cdot) = \lambda_{n}\sum_{j=1}^{p}\Big\{ \big|\tilde{\beta}_{j,n}(\omega)+\frac{u_{j}}{n^{1/2}}\big|-\big|\tilde{\beta}_{j,n}(\omega)\big|\Big\}.$
Similar to original case, using Taylor's theorem we have
\begin{align*}
\tilde{\ell}^*_{1n}(\bm{u},\omega,\cdot)= (1/2)\bm{u}^\top \big\{\tilde{\bm{L}}_n^*(\omega,\cdot)\big\}\bm{u} - \bm{u}^\top\big\{\tilde{\bm{W}}_n^*(\omega,\cdot)\big\}+\tilde{R}_{1n}^*(\bm{u},\omega,\cdot),
\end{align*}
where $\tilde{R}_{1n}^*(\bm{u},\omega,\cdot)=(6n^{3/2})^{-1}\sum_{i=1}^{n}\Bigg[\Big\{-y_ih^{\prime\prime\prime}(\tilde{z}_i^*)(\bm{u}^\top\bm{x}_i)^3G_i^*\mu_{G^*}^{-1}\Big\}+ \Big\{h_1^{\prime\prime\prime}(\tilde{z}_i^*)(\bm{u}^\top\bm{x}_i)^3G_i^*\mu_{G^*}^{-1}\Big\}\Bigg],$
for some $\tilde{z}_i^*\equiv z_i^*(\bm{u},\omega,\cdot)$ such that $|\tilde{z}_i^*-\bm{x}_i^\top\tilde{\bm{\beta}}_n|\leq n^{-1/2}(\bm{u}^\top\bm{x}_i)$, $i\in \{1,\dots,n\}$. Again use definition of $\tilde{\bm{\beta}}_n$, the assumption (C.3), (C.6) and Lemma \ref{lem:asconcentration}, to claim that $\max\Big\{\big[|h^{\prime\prime\prime}(\tilde{z}_i^*)|+|h_1^{\prime\prime\prime}(\tilde{z}_i^*)|\big]: i\in \{1,\dots,n\}\Big\}=O(1)$ for all $\omega \in \bm{B}$. Again by Markov's inequality, we have\\
$n^{-3/2}\sum_{i=1}^{n}|y_i(\omega)|G_i^* =o_{P_*}(1)$ for all $\omega \in \bm{B}$. Therefore for all $\omega \in \bm{B}$ we have $\|\tilde{R}_{1n}^*(\bm{u},\omega,\cdot)\|=o_{P_*}(1)$ and hence $$\tilde{\ell}_{1n}^*(\bm{u},\omega,\cdot)\xrightarrow{d}\big\{(1/2)\bm{u}^\top \bm{L}\bm{u}-\bm{u}^\top \bm{Z}_2\big\}.$$
Now like the first part of Theorem \ref{prop:ntheo}, the equi-continuity on compact sets also holds here in probability. Using this fact along with Lemma \ref{lem:argmin}, it is remaining to show that
\begin{align}\label{eqn:T.9}
\tilde{\ell}_{2n}^*(\bm{u},\omega,\cdot)  \rightarrow\;  \lambda_0\Big\{\sum_{j=1}^{p_0}sgn({{\beta}_{j}})u_j+\sum_{j=p_0+1}^{p}|u_j|\Big\},
\end{align}
for any $\omega \in \bm{B}$. Again for $\omega \in\bm{B}$ there exists $N(\omega)$ such that for $n>N(\omega)$,
\begin{align*}
\bigg{\{} \begin{array}{ll} \tilde{\beta}_{j,n}(\omega)=\hat{\beta}_{j,n}(\omega)\;\;\; \text{and}\;\;\; sgn({\tilde{\beta}_{j,n}}(\omega))=sgn({\beta_{j}})\;  \text{for}\;j\in \mathcal{A}\;\\ 
\tilde{\beta}_{j,n}(\omega)=0\; \text{for}\;j\in \{1,\dots,p\}\setminus \mathcal{A},  \end{array} 
\end{align*}
due to the definition of $\tilde{\bm{\beta}}_n$. Therefore (\ref{eqn:T.9}) is true and we are done. \hfill $\square$

\subsection{Proof of Theorem \ref{thm:prbpos}}\label{appAthm6.1}
To establish the almost sure consistency of PRB method, we define the set;
\begin{align*}
\bm{D}=&\Big\{n^{1/2}\|\hat{\bm{\beta}}_n-\bm{\beta}\|=o(\log n)\Big\}\cap \Bigg\{\Bigg\|\frac{\tilde{G}_n^\top\tilde{G}_n}{n}-\bm L\Bigg\|=o(1)\Bigg\}\cap \Bigg\{\mathcal{L}\Big(\frac{\tilde{G}_n^\top\bm{e}^*}{n^{1/2}}\Big|\mathcal{E}\Big)\xrightarrow{d_*} N\big{(}0,\bm{L}\big{)}\Bigg\}    
\end{align*}
Now it is straight forward to verify that,
\begin{align}\label{eqn:prbcen}
&n^{1/2}(\hat{\bm{\beta}}_n^{*(PRB)}-\tilde{\bm{\beta}}_n)\nonumber\\
&=\operatorname*{arg\,min}_{\bm{u}}\Bigg[\frac{1}{2}\bm{u}^\top\Big(\frac{\tilde{G}_n^\top\tilde{G}_n}{n}\Big)\bm{u}-\bm{u}^\top\Big(\frac{\tilde{G}_n^\top\bm{e}^*}{n^{1/2}}\Big)+\lambda_n\sum_{j=1}^{p}\Big\{\Big|\tilde\beta_{n,j}+\frac{u_j}{n^{1/2}}\Big|-|\tilde\beta_{n,j}|\Big\}\Bigg]
\end{align}
We observe the following:
\begin{itemize}
    \item Lemma \ref{lem:prbvarest} will give us that,
    \[
    \Bigg\|\frac{\tilde{G}_n^\top\tilde{G}_n}{n}-\bm L\Bigg\|=o(1)\;\quad\text{with probability}\;\; 1
    \]
    \item Lemma \ref{lem:var2prb} implies that,
     \[
\Bigg\|\text{var}_*\Big(\frac{\tilde{G}_n^\top\bm{e}^*}{n^{1/2}}\Big)-\bm{L}\Bigg\|=o(1)\quad\text{with probability}\; 1.
\]
\item In Lemma \ref{lem:bnormalprb} we have derived that,
\[
 \mathcal{L}\Big(\frac{\tilde{G}_n^\top\bm{e}^*}{n^{1/2}}\Big|\mathcal{E}\Big)\xrightarrow{d_*} N\big{(}0,\bm{L}\big{)}\quad\text{with probability}\; 1.
 \]
\end{itemize} 
Combining these facts and Lemma \ref{lem:asconcentration}, we can conclude that, $\mathbbm{P}(D)=1$. Therefore for any $\omega\in D$, it's enough to establish that;
\begin{align}\label{T.47}
    \rho\big\{\tilde{F}_n^{(PRB)}(\omega,\cdot),F_{\infty}(\omega)\big\}\rightarrow 0\;\text{as}\; n\rightarrow \infty.
\end{align}
The rest of the proof is similar to that of Theorem \ref{thm:ptheo}, and hence we skip the details.\hfill $\square$

\section{Additional Simulation Studies}\label{sec:appA}
This section is mainly devoted to additional simulation results pertaining to PB methods mentioned in section \ref{sec:sim}. To be precise, in section \ref{subsec:lincvfixpfixan}, we have presented empirical coverage probabilities for nominal $90\%$ both-sided and one-sided confidence intervals for gamma and linear regression under comparative analysis scheme (i) when penalty parameter $\lambda_n$ is chosen through $K-$ fold CV in R and $(p,p_0)=(7,4)$, $a_n=n^{-1/3}$ and $n\in \{50,100,150,300,500\}$ are employed. In section \ref{subsec:fixcvsim}, we present the same thing but $\lambda_n$ is no longer chosen in a CV-based way. Rather we predefine $\lambda_n=n^{1/2}\lambda_0$ with $\lambda_0=0.025$ and analyze the results over $n\in \{50,100,150,300,500\}$ for logistic, gamma and linear regressions. In section \ref{subsec:varyingan}, we keep $(p,p_0)=(7,4)$ as fixed, but vary our thresholding parameter $a_n$ over $n\in \{50,100,150,300,500\}$, and perform simulation results for logistic regression. All reproducible codes are available at \url{https://github.com/mayukhc13/Bootstrapping-Lasso-in-GLM.git}.

\subsection{Simulation Study for Gamma and Linear Regression for Cross-validated choice of Penalty Parameter}\label{subsec:lincvfixpfixan}
Table~\ref{tab:1s} summarizes the empirical coverage performance of 90\% confidence intervals under Gamma and Linear regression models for both two-sided and right-sided inference. 
\begin{table}[H]
\centering
\caption{Empirical coverage probabilities of 90\% confidence intervals under Gamma and Linear regression models.}
\label{tab:1s}

\vspace{0.3cm}

\begin{minipage}[t]{0.48\textwidth}
\centering
\textbf{(a) Gamma regression: Two-sided}

\vspace{0.1cm}

\resizebox{\textwidth}{!}{%
\begin{tabular}{lccccc}
\toprule
$\beta_j$ & 50 & 100 & 150 & 300 & 500 \\
\midrule
$-0.5$ & 0.866 & 0.868 & 0.874 & 0.880 & 0.892 \\
 & (0.489) & (0.359) & (0.287) & (0.195) & (0.151) \\
$1.0$ & 0.856 & 0.866 & 0.876 & 0.884 & 0.898 \\
 & (0.594) & (0.425) & (0.281) & (0.202) & (0.159) \\
$-1.5$ & 0.854 & 0.862 & 0.882 & 0.896 & 0.904 \\
 & (0.705) & (0.398) & (0.288) & (0.201) & (0.161) \\
$2.0$ & 0.840 & 0.856 & 0.872 & 0.886 & 0.898 \\
 & (0.552) & (0.381) & (0.301) & (0.188) & (0.164) \\
$0$ & 0.812 & 0.838 & 0.870 & 0.880 & 0.906 \\
 & (0.485) & (0.354) & (0.263) & (0.204) & (0.169) \\
$0$ & 0.818 & 0.838 & 0.856 & 0.870 & 0.894 \\
 & (0.568) & (0.369) & (0.274) & (0.220) & (0.150) \\
$0$ & 0.826 & 0.854 & 0.884 & 0.896 & 0.912 \\
 & (0.574) & (0.326) & (0.279) & (0.201) & (0.153) \\
\bottomrule
\end{tabular}}
\end{minipage}
\hfill
\begin{minipage}[t]{0.48\textwidth}
\centering
\textbf{(b) Linear regression: Two-sided}

\vspace{0.1cm}

\resizebox{\textwidth}{!}{%
\begin{tabular}{lccccc}
\toprule
$\beta_j$ & 50 & 100 & 150 & 300 & 500 \\
\midrule
$-0.5$ & 0.852 & 0.868 & 0.876 & 0.888 & 0.894 \\
 & (0.545) & (0.389) & (0.294) & (0.181) & (0.151) \\
$1.0$ & 0.864 & 0.878 & 0.884 & 0.898 & 0.908 \\
 & (0.568) & (0.374) & (0.307) & (0.213) & (0.144) \\
$-1.5$ & 0.858 & 0.862 & 0.878 & 0.890 & 0.902 \\
 & (0.548) & (0.381) & (0.297) & (0.192) & (0.164) \\
$2.0$ & 0.844 & 0.866 & 0.886 & 0.896 & 0.902 \\
 & (0.553) & (0.346) & (0.274) & (0.210) & (0.162) \\
$0$ & 0.838 & 0.852 & 0.872 & 0.888 & 0.904 \\
 & (0.593) & (0.417) & (0.257) & (0.204) & (0.154) \\
$0$ & 0.824 & 0.852 & 0.878 & 0.890 & 0.906 \\
 & (0.563) & (0.387) & (0.276) & (0.222) & (0.164) \\
$0$ & 0.844 & 0.866 & 0.872 & 0.890 & 0.896 \\
 & (0.480) & (0.319) & (0.280) & (0.198) & (0.137) \\
\bottomrule
\end{tabular}}
\end{minipage}

\vspace{0.4cm}

\begin{minipage}[t]{0.48\textwidth}
\centering
\textbf{(c) Gamma regression: Right-sided}

\vspace{0.1cm}

\resizebox{\textwidth}{!}{%
\begin{tabular}{lccccc}
\toprule
$\beta_j$ & 50 & 100 & 150 & 300 & 500 \\
\midrule
$-0.5$ & 0.850 & 0.864 & 0.872 & 0.882 & 0.898 \\
$1.0$ & 0.848 & 0.866 & 0.872 & 0.884 & 0.918 \\
$-1.5$ & 0.856 & 0.862 & 0.880 & 0.902 & 0.906 \\
$2.0$ & 0.828 & 0.864 & 0.886 & 0.894 & 0.896 \\
$0$ & 0.848 & 0.856 & 0.870 & 0.878 & 0.892 \\
$0$ & 0.828 & 0.836 & 0.860 & 0.882 & 0.902 \\
$0$ & 0.868 & 0.872 & 0.882 & 0.890 & 0.898 \\
\bottomrule
\end{tabular}}
\end{minipage}
\hfill
\begin{minipage}[t]{0.48\textwidth}
\centering
\textbf{(d) Linear regression: Right-sided}

\vspace{0.1cm}

\resizebox{\textwidth}{!}{%
\begin{tabular}{lccccc}
\toprule
$\beta_j$ & 50 & 100 & 150 & 300 & 500 \\
\midrule
$-0.5$ & 0.852 & 0.868 & 0.884 & 0.898 & 0.904 \\
$1.0$ & 0.854 & 0.874 & 0.882 & 0.889 & 0.900 \\
$-1.5$ & 0.862 & 0.874 & 0.886 & 0.904 & 0.912 \\
$2.0$ & 0.858 & 0.878 & 0.880 & 0.886 & 0.896 \\
$0$ & 0.854 & 0.868 & 0.872 & 0.886 & 0.908 \\
$0$ & 0.834 & 0.862 & 0.886 & 0.888 & 0.898 \\
$0$ & 0.840 & 0.844 & 0.878 & 0.888 & 0.904 \\
\bottomrule
\end{tabular}}
\end{minipage}

\end{table}

 Across all settings, coverage probabilities systematically improve with increasing sample size $n$, approaching the nominal level as $n$ grows, while the corresponding interval widths decrease monotonically, reflecting increasing estimation precision. For two-sided intervals (Panels (a) and (b)), both models exhibit mild undercoverage at small sample sizes, particularly for zero and moderate-effect coefficients, with more pronounced deviations in the Gamma regression. However, by $n \ge 300$, coverage stabilizes close to 0.90 for all coefficients in both models. Linear regression consistently attains slightly higher coverage than Gamma regression for comparable $n$, accompanied by marginally narrower confidence intervals. For right-sided intervals (Panels (c) and (d)), coverage probabilities are uniformly closer to the nominal level across all coefficients and sample sizes, indicating improved finite-sample behavior relative to two-sided intervals. This improvement is particularly notable for nonzero coefficients under the Gamma model. As in the two-sided case, Linear regression demonstrates marginally superior coverage accuracy, though differences between the two models diminish as $n$ increases.

\begin{table}[H]
\centering
\caption{Empirical Coverage Probabilities of 90\% Confidence region of $\beta$}
\label{tab:2s}
\begin{tabular}{@{}lcrcrrr@{}}
\hline
 &\multicolumn{5}{c}{Coverage Probability} \\
\cline{2-6}
Regression Type &
\multicolumn{1}{c}{$n=50$} &
\multicolumn{1}{c}{$n=100$}&
\multicolumn{1}{c}{$n=150$}&
\multicolumn{1}{c}{$n=300$}&
\multicolumn{1}{c@{}}{$n=500$}\\
\hline

Gamma & 0.832 & 0.860 & 0.878 & 0.886 & 0.898 \\
Linear & 0.872 & 0.876 & 0.889 & 0.898 & 0.906\\
\hline
\end{tabular}
\end{table}

\begin{figure*}[ht]
\centering

\textbf{Gamma Regression}\par\vspace{0.15cm}

\subfloat[$\beta_1=-0.5$]{%
\includegraphics[width=.3\linewidth]{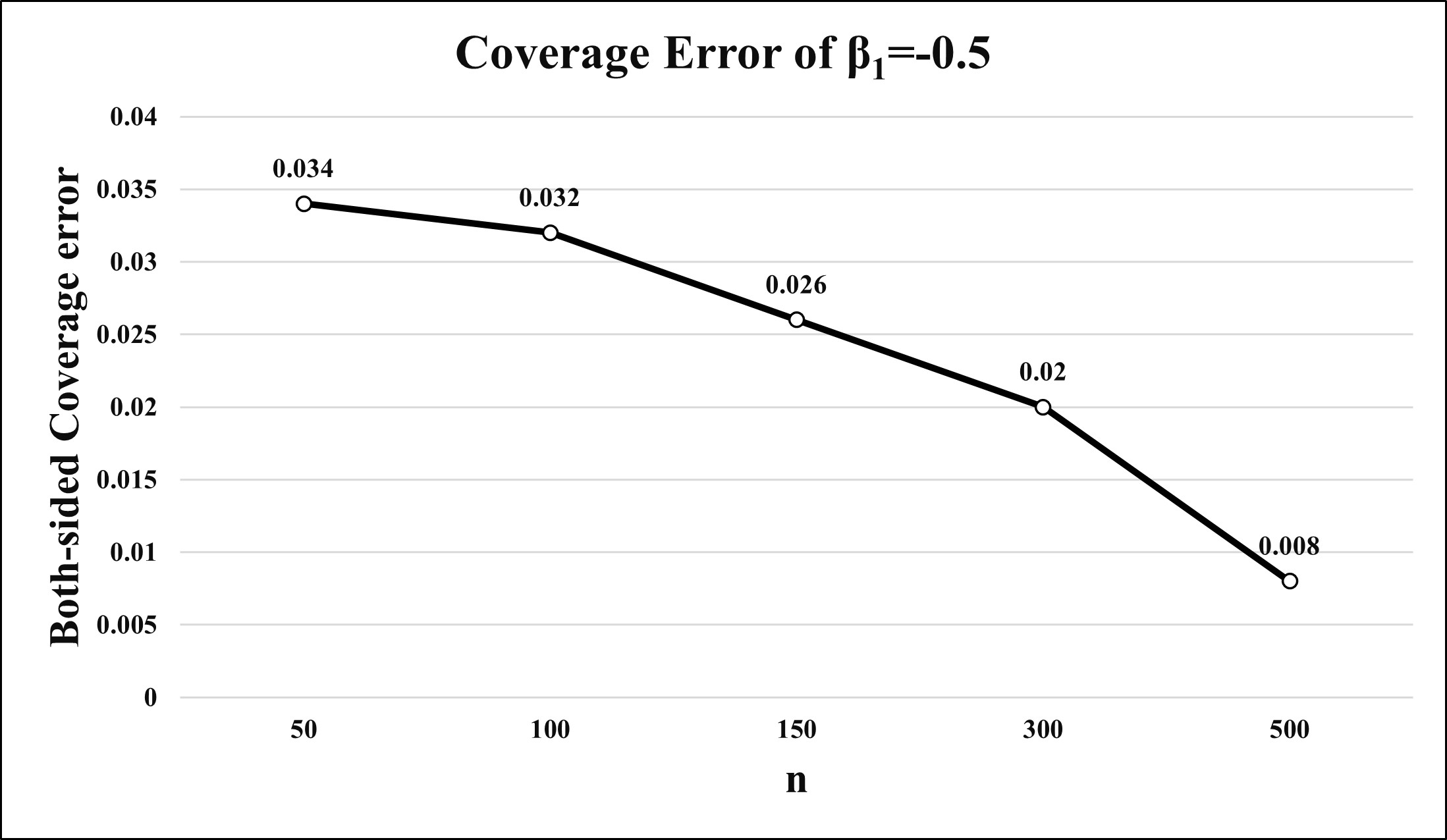}}
\hfill
\subfloat[$\beta_5=0$]{%
\includegraphics[width=.3\linewidth]{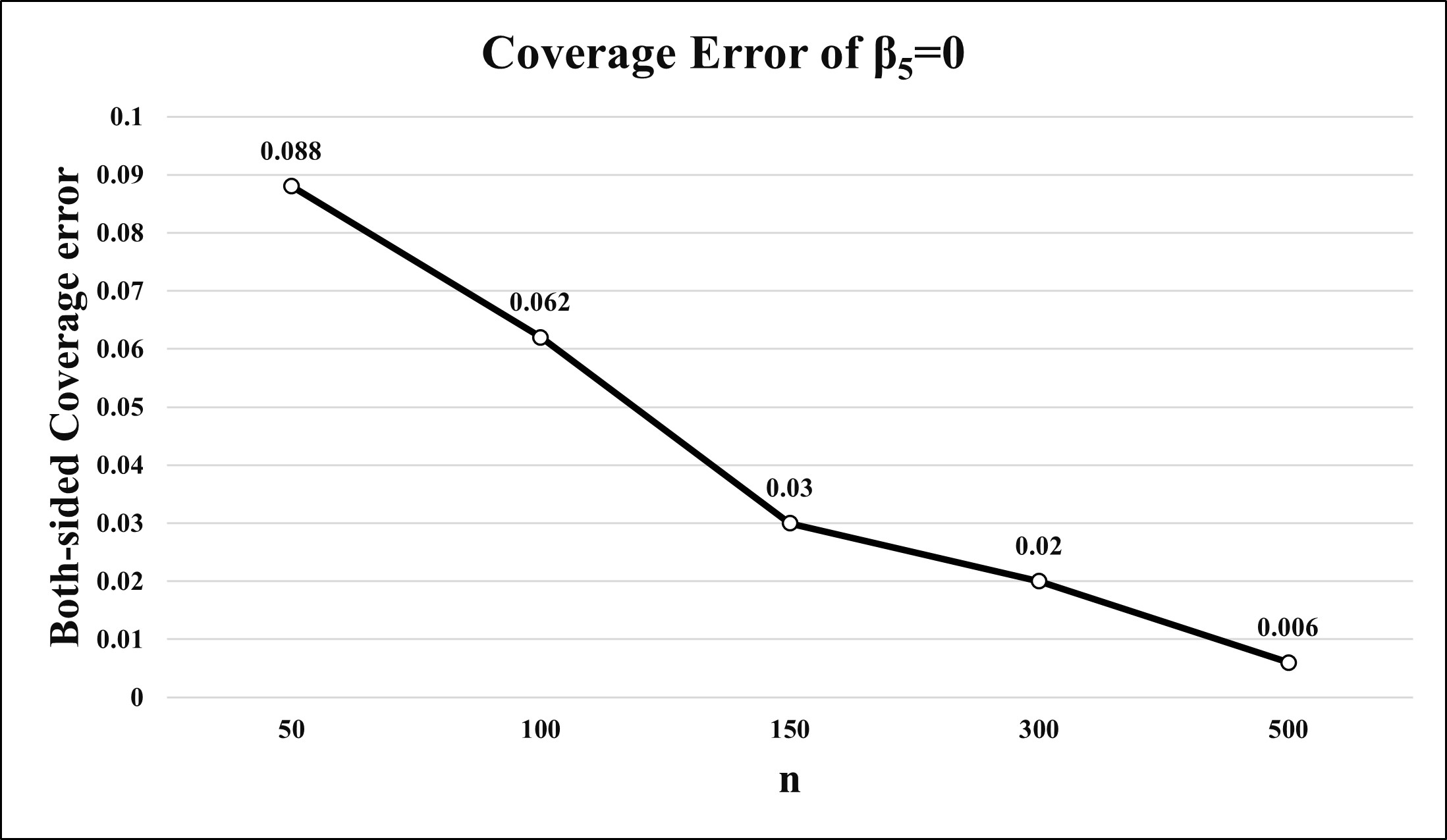}}
\hfill
\subfloat[$\beta_4=2$]{%
\includegraphics[width=.3\linewidth]{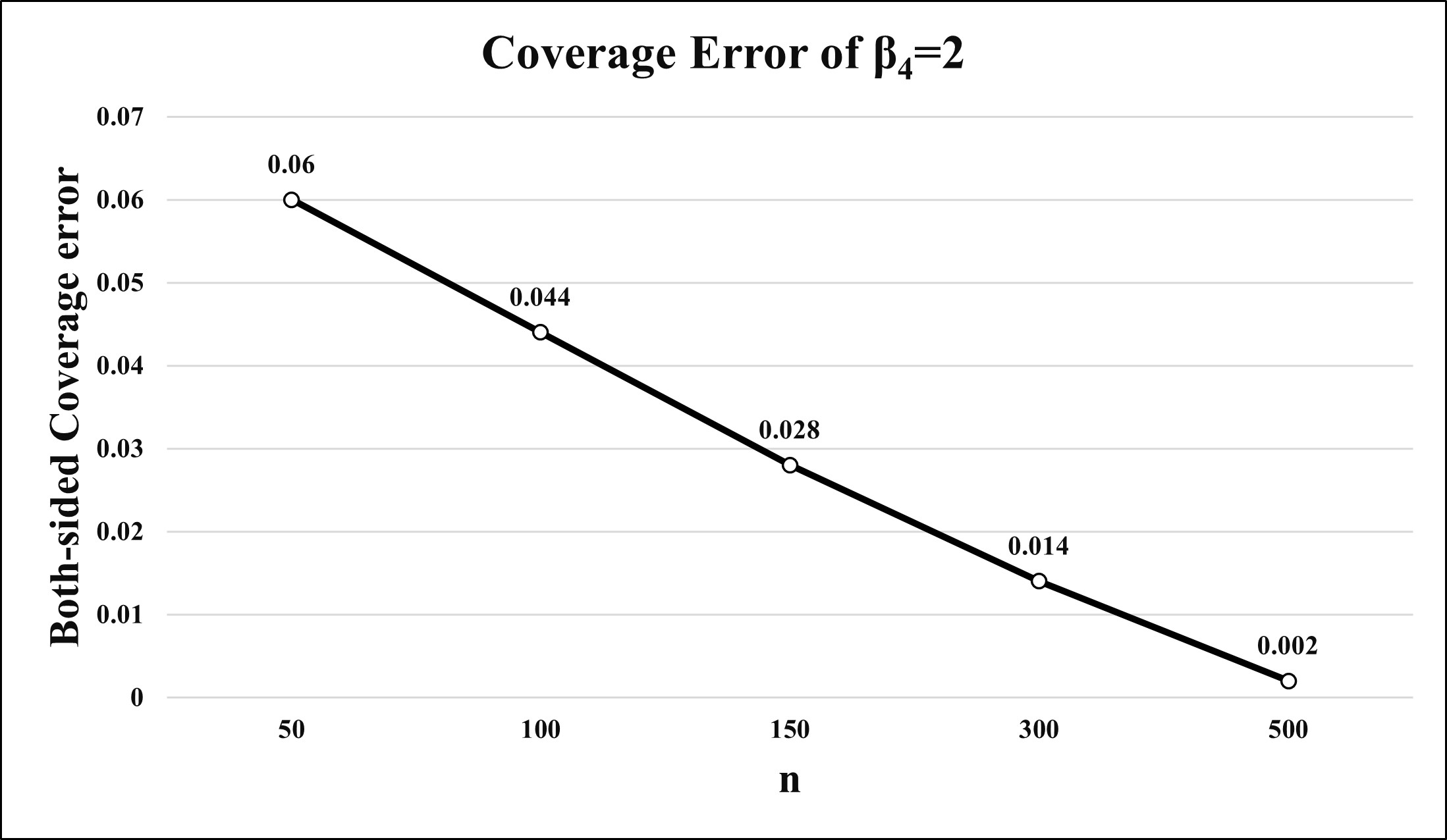}}

\vspace{0.4cm}

\textbf{Linear Regression}\par\vspace{0.15cm}

\subfloat[$\beta_1=-0.5$]{%
\includegraphics[width=.3\linewidth]{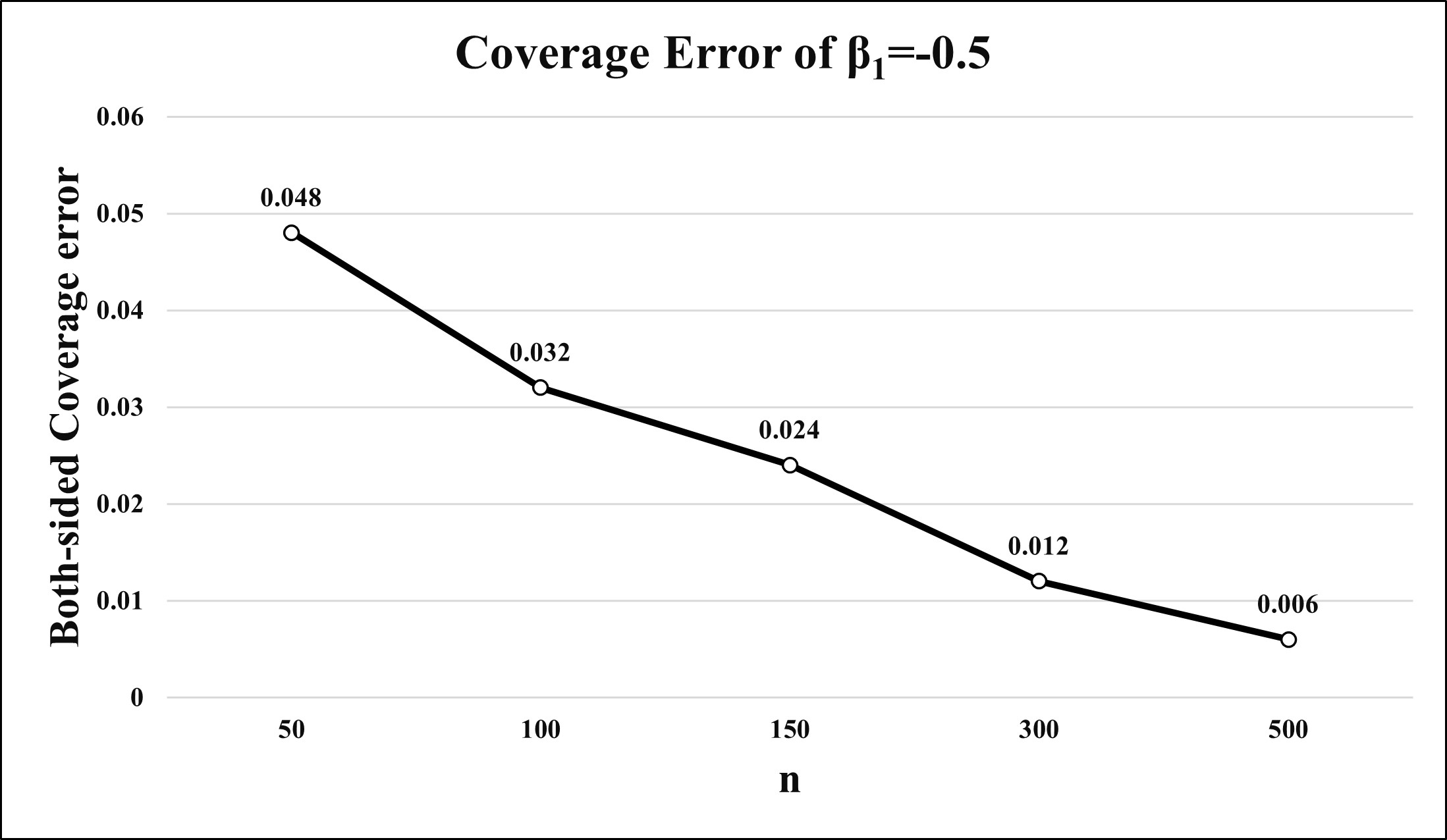}}
\hfill
\subfloat[$\beta_5=0$]{%
\includegraphics[width=.3\linewidth]{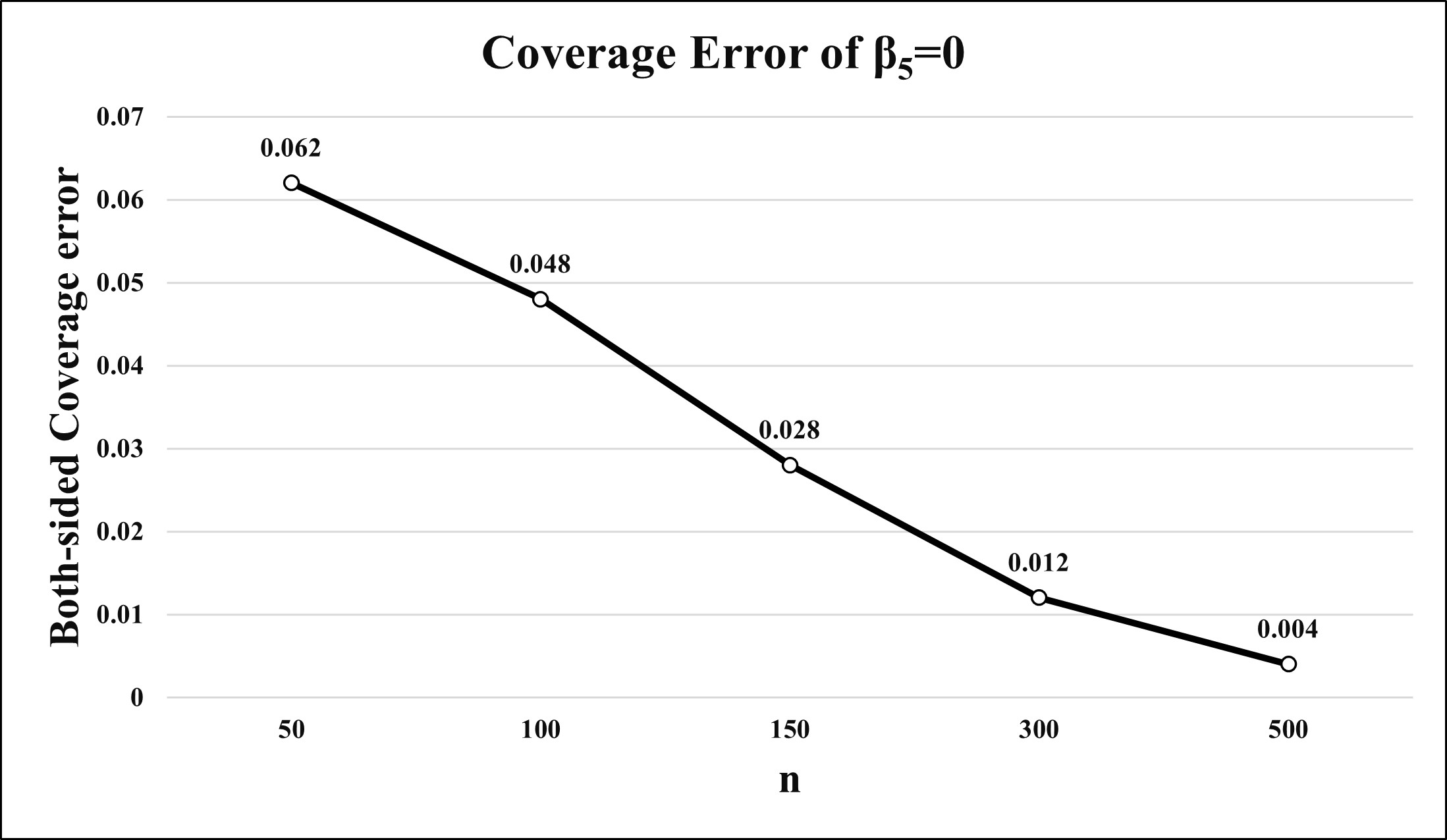}}
\hfill
\subfloat[$\beta_4=2$]{%
\includegraphics[width=.3\linewidth]{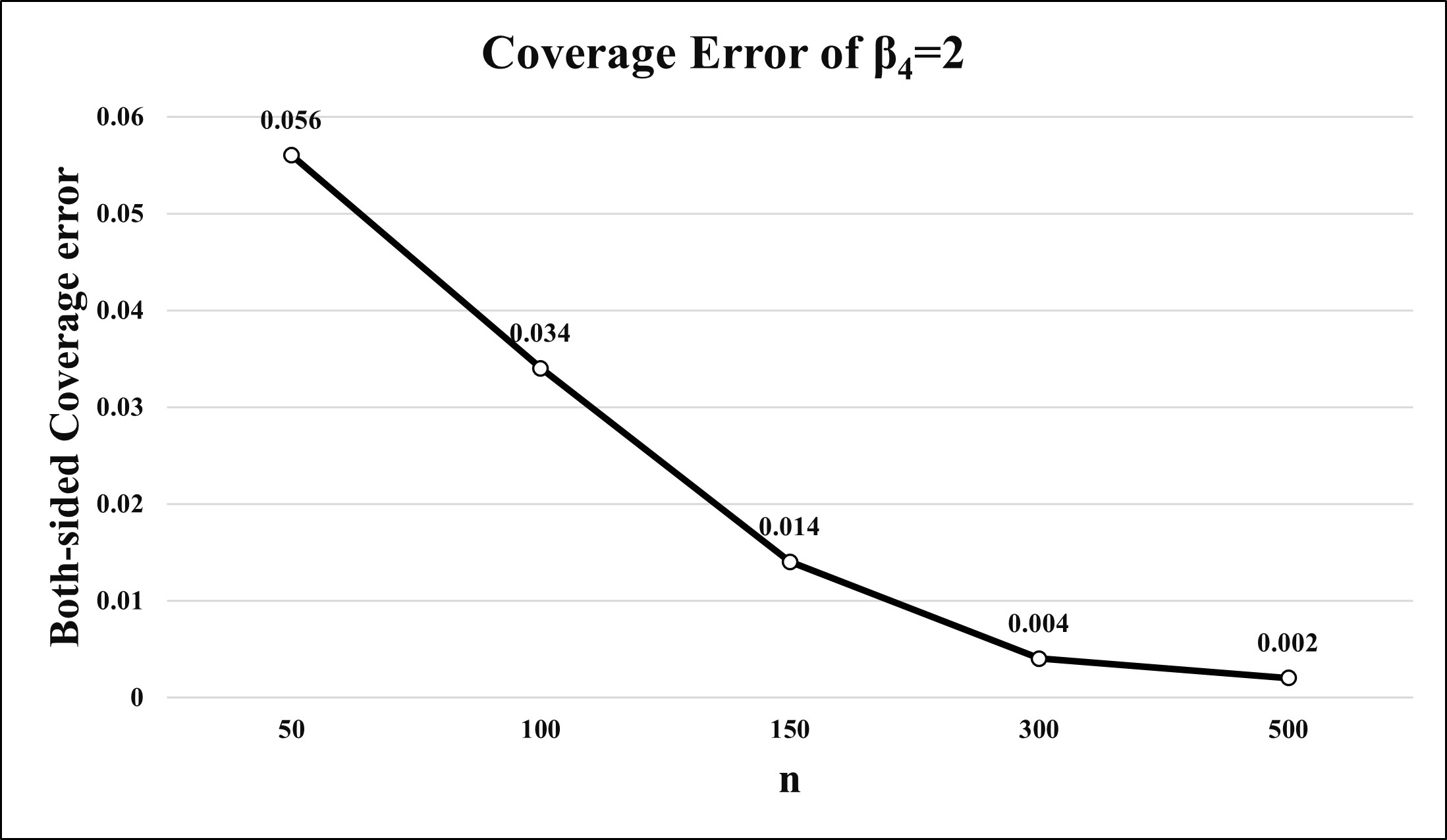}}

\caption{Coverage error of two-sided $90\%$ confidence intervals as a function of sample size $n$ in Gamma and Linear regression with $K$-fold CV.}
\label{fig:coverage_combined}
\end{figure*}

Overall, the results indicate that while both regression models achieve satisfactory asymptotic performance, Linear regression exhibits more stable finite-sample coverage, especially for two-sided inference. Right-sided intervals provide a practical improvement in coverage accuracy under both models, particularly in small to moderate samples. Table~\ref{tab:2s} shows that the linear regression model consistently attains slightly higher empirical coverage than the Gamma regression across all sample sizes, with the gap narrowing as $n$ increases and both methods approaching the nominal level. Figure \ref{fig:coverage_combined} shows the plots between both-sided coverage error versus n for $\beta_1=-0.5$, $\beta_5=0$ and $\beta_4=2$ in gamma and linear regression, where
$$\text{coverage error}\ =\ |\text{empirical coverage probability} -\text{ nominal confidence level}|.$$

\subsection{Simulation Study for Predefined Choice of Penalty Parameter }\label{subsec:fixcvsim}
Now we report additional simulation setup under the cases when we no longer consider the regularisation parameter $\lambda_n$ through cross validation method which is technically data dependent reported earlier as in our main paper. We choose our penalty parameter as $\lambda_n=n^{1/2}\lambda_0$, where the choices of fixed $\lambda_0$ under respective choices of $n$ are as follows :
$$(\lambda_0,n) \in \big\{(0.025,50),(0.025,100),(0.025,150),(0.025,300),(0.025,500)\big\}.$$

Now for each of $(n,p,p_0)$, the design matrix is once and initially generated from some structure outside the loop and kept fixed throughout the entire simulation. By that we mean, generation of $n$ i.i.d design vectors say, $\bm{x}_i = (x_{i1},\ldots, x_{ip})^\top \ \text{for  all} \ i \in \{1,\ldots, n\}$ from zero mean $p$-variate normal distribution such that it has same covariance structure as earlier. The entire data generation process is similar to original case.
\begin{table}[H]
\centering
\caption{Empirical Coverage Probabilities \& Average Widths of 90\% Confidence Intervals in Logistic Regression}
\label{tabs3}
\begin{tabular}{@{}lcrcrrr@{}}
\hline
 &\multicolumn{5}{c}{Both Sided} \\
\cline{2-6}
$\beta_j$ &
\multicolumn{1}{c}{$n=50$} &
\multicolumn{1}{c}{$n=100$}&
\multicolumn{1}{c}{$n=150$}&
\multicolumn{1}{c}{$n=300$}&
\multicolumn{1}{c@{}}{$n=500$}\\
\hline
-0.5 & 0.972 & 0.956 & 0.924 & 0.914 & 0.896\\
 & (1.925) & (1.201) & (0.840) & (0.587) & (0.439) \\  
1.0 & 0.966   & 0.942 & 0.926 & 0.908 & 0.896\\
& (2.179) & (1.268) & (0.919) & (0.685) & (0.513) \\ 
-1.5 & 0.970 & 0.964 & 0.938 & 0.916 & 0.906\\
& (2.217) & (1.429) & (1.152) & (0.724) & (0.548) \\
2.0    & 0.946 & 0.932 & 0.912 & 0.906 & 0.900\\
& (2.753) & (1.739) & (1.311) & (0.843) & (0.637) \\
0 & 0.946 & 0.924 & 0.910 & 0.904 & 0.896\\
& (1.645) & (1.129) & (0.783) & (0.555) & (0.402) \\
0 & 0.966 & 0.946 & 0.924 & 0.908 & 0.898 \\
& (1.747) & (1.012) & (0.826) & (0.574) & (0.427) \\
0 & 0.926 & 0.920 & 0.910 & 0.902 & 0.898\\
& (1.971) & (1.018) & (0.813) & (0.501) & (0.396) \\

\hline
\end{tabular}
\end{table}

\begin{table}[H]
\centering
\caption{Empirical Coverage Probabilities of 90\% Right-sided Confidence Intervals in Logistic Regression}
\label{tabs5}
\begin{tabular}{@{}lcrcrrr@{}}
\hline
 &\multicolumn{5}{c}{Right Sided} \\
\cline{2-6}
$\beta_j$ &
\multicolumn{1}{c}{$n=50$} &
\multicolumn{1}{c}{$n=100$}&
\multicolumn{1}{c}{$n=150$}&
\multicolumn{1}{c}{$n=300$}&
\multicolumn{1}{c@{}}{$n=500$}\\
\hline
-0.5 & 0.978 & 0.958 & 0.930 & 0.916 & 0.906  
\\
1.0 & 0.946 & 0.944 & 0.920 & 0.904 & 0.896 
\\
-1.5 & 0.948 & 0.922 & 0.914 & 0.906 & 0.900 
\\
2.0 & 0.932 & 0.924 & 0.918 & 0.905 & 0.902
\\
0 & 0.968 & 0.944 & 0.924 & 0.914 & 0.904
\\
0 & 0.952 & 0.934 & 0.912 & 0.904 & 0.898 
\\
0 & 0.944 & 0.912 & 0.904 & 0.901 & 0.892
\\
\hline
\end{tabular}
\end{table}

\begin{figure}[H]
\centering
\subfloat[Coverage Error of $\beta_1=-0.5$]{
  \label{figS2.1}
  \includegraphics[width=.3\linewidth]{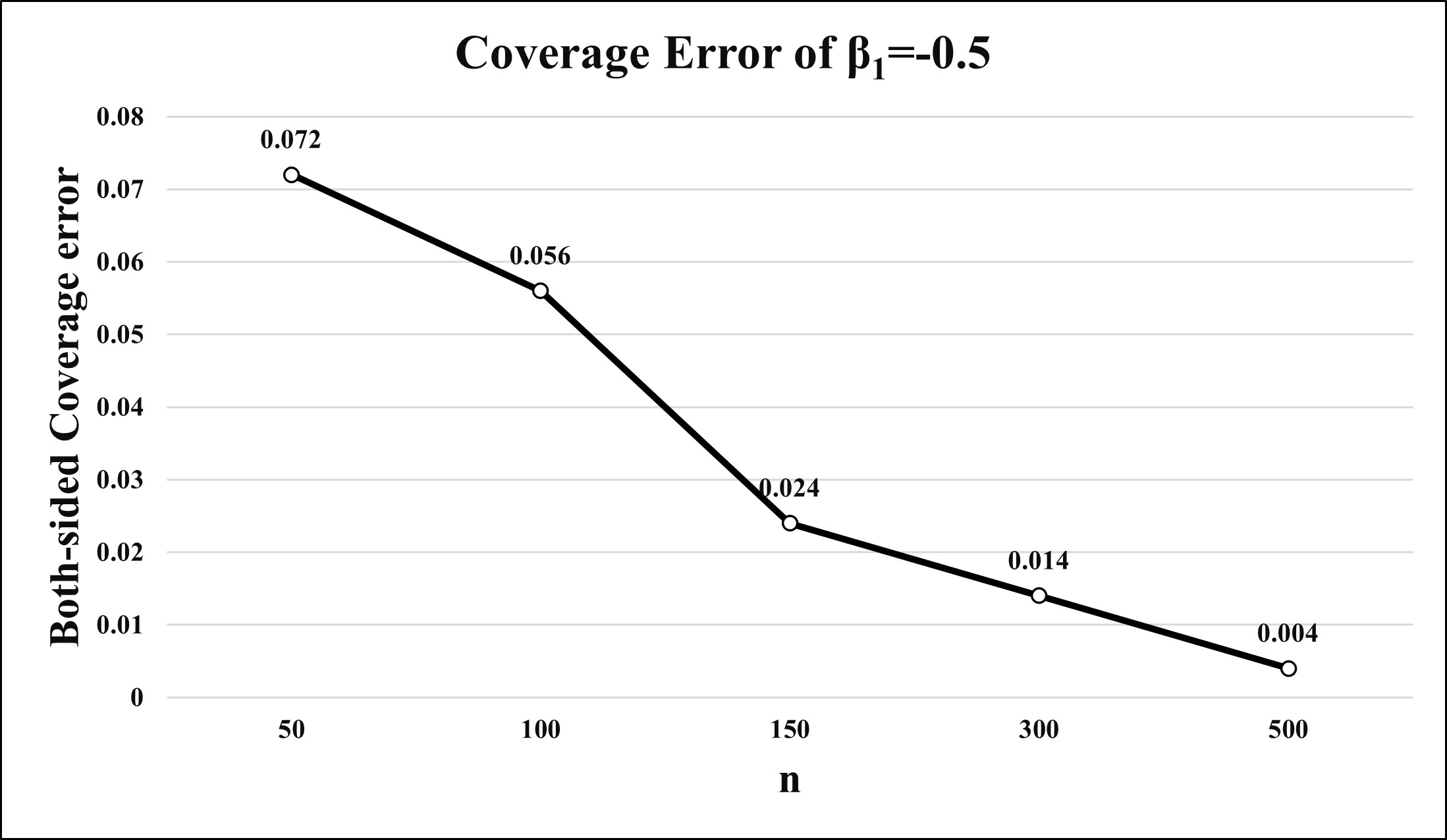}
}\hspace{0.5em}
\subfloat[Coverage Error of $\beta_5=0$]{
  \label{figS2.2}
  \includegraphics[width=.3\linewidth]{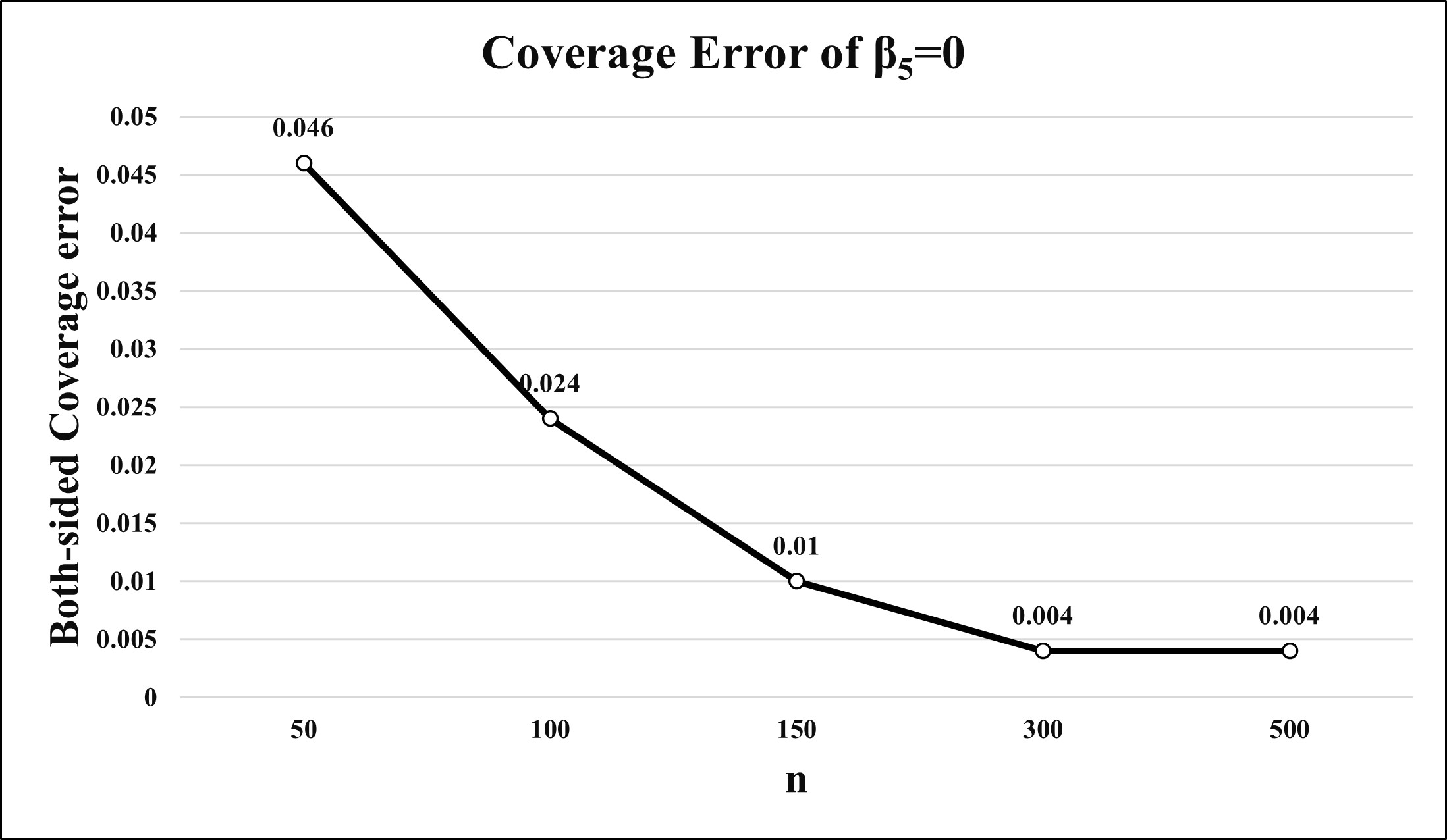}
}\hspace{0.5em}
\subfloat[Coverage Error of $\beta_4=2$]{
  \label{figS2.3}
  \includegraphics[width=.3\linewidth]{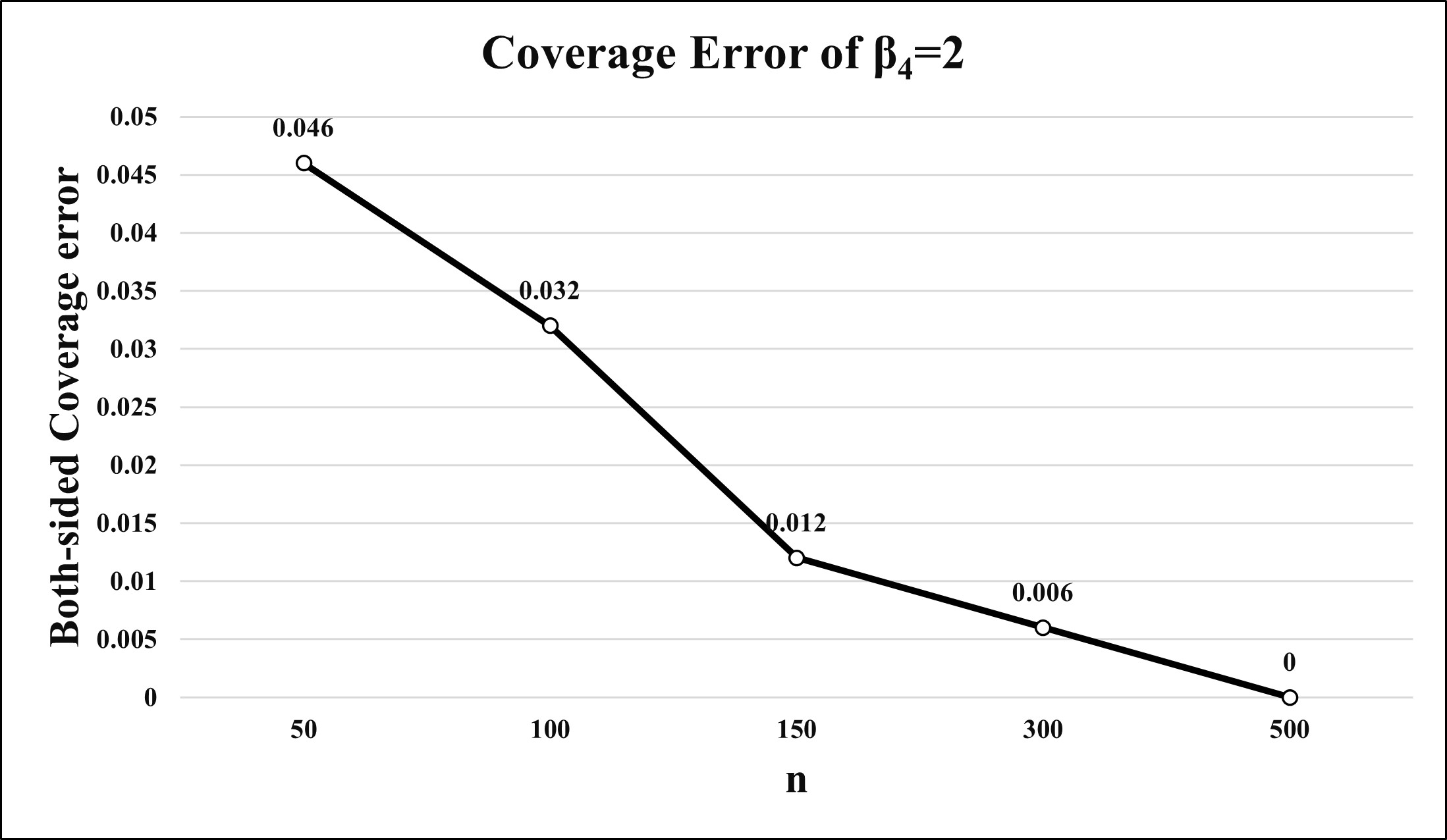}
}
\caption{Coverage Error of both-sided $90\%$ confidence intervals over $n$ in logistic regression.}
\label{fig S2}
\end{figure}

For logistic regression, we observe that as $n$ increases over the course, the empirical coverage probabilities get closer and closer to nominal confidence level of 0.90 (see Table \ref{tabs3}, Table \ref{tabs5} and Figure \ref{fig S2}) than earlier choices for all regression coefficients. In Table \ref{tabs3}, note that the average width of the intervals become smaller and smaller as $n$ increases for all the individual parameter.

\begin{table}[htbp]
\centering
\caption{Empirical coverage probabilities of 90\% confidence intervals under Gamma and Linear regression models.}
\label{tab:SA6}

\begin{minipage}[t]{0.48\textwidth}
\centering
\textbf{(a) Gamma regression: Two-sided}

\vspace{0.1cm}

\resizebox{\textwidth}{!}{%
\begin{tabular}{lccccc}
\toprule
$\beta_j$ & $n=50$ & $n=100$ & $n=150$ & $n=300$ & $n=500$ \\
\midrule
-0.5 & 0.812 & 0.844 & 0.872 & 0.898 & 0.906\\
 & (0.532) & (0.376) & (0.241) & (0.207) & (0.182) \\  
1.0 & 0.850   & 0.868 & 0.870 & 0.888 & 0.898\\
& (0.633) & (0.350) & (0.312) & (0.234) & (0.206) \\ 
-1.5 & 0.864 & 0.868 & 0.886 & 0.898 & 0.912\\ 
& (0.546) & (0.338) & (0.279) & (0.224) & (0.124) \\  
2.0    & 0.844 & 0.872 & 0.880 & 0.896 & 0.900\\ 
& (0.558) & (0.406) & (0.257) & (0.210) & (0.196) \\ 
0 & 0.828 & 0.842 & 0.868 & 0.884 & 0.892\\ 
& (0.620) & (0.339) & (0.259) & (0.232) & (0.202) \\
0 & 0.822 & 0.834 & 0.858 & 0.888 & 0.896 \\
& (0.601) & (0.361) & (0.278) & (0.254) & (0.178) \\
0 & 0.818 & 0.824 & 0.846 & 0.878 & 0.890\\
& (0.489) & (0.328) & (0.276) & (0.204) & (0.167) \\
\bottomrule
\end{tabular}}
\end{minipage}\hfill
\begin{minipage}[t]{0.48\textwidth}
\centering
\textbf{(b) Linear regression: Two-sided}

\vspace{0.1cm}

\resizebox{\textwidth}{!}{%
\begin{tabular}{lccccc}
\toprule
$\beta_j$ & $n=50$ & $n=100$ & $n=150$ & $n=300$ & $n=500$ \\
\midrule
-0.5 & 0.856 & 0.872 & 0.885 & 0.892 & 0.904\\
 & (0.492) & (0.358) & (0.295) & (0.199) & (0.164) \\
1.0 & 0.850   & 0.868 & 0.872 & 0.884 & 0.892\\
& (0.509) & (0.331) & (0.291) & (0.209) & (0.159) \\ 
-1.5 & 0.852 & 0.864 & 0.886 & 0.889 & 0.906\\
& (0.454) & (0.349) & (0.334) & (0.213) & (0.158) \\ 
2.0    & 0.832 & 0.872 & 0.884 & 0.890 & 0.898\\
& (0.508) & (0.377) & (0.340) & (0.211) & (0.152) \\
0 & 0.816 & 0.842 & 0.874 & 0.882 & 0.900\\
& (0.542) & (0.376) & (0.261) & (0.197) & (0.152) \\
0 & 0.818 & 0.836 & 0.866 & 0.870 & 0.896 \\
& (0.631) & (0.363) & (0.290) & (0.207) & (0.164) \\
0 & 0.818 & 0.822 & 0.844 & 0.886 & 0.896\\
& (0.581) & (0.367) & (0.279) & (0.208) & (0.144) \\
\bottomrule
\end{tabular}}
\end{minipage}

\vspace{0.4cm}

\begin{minipage}[t]{0.48\textwidth}
\centering
\textbf{(c) Gamma regression: Right-sided}

\vspace{0.1cm}

\resizebox{\textwidth}{!}{%
\begin{tabular}{lccccc}
\toprule
$\beta_j$ & $n=50$ & $n=100$ & $n=150$ & $n=300$ & $n=500$ \\
\midrule
-0.5 & 0.854 & 0.876 & 0.880 & 0.886 & 0.898
\\
1.0 & 0.860   & 0.866 & 0.870 & 0.882 & 0.896
\\
-1.5 & 0.860 & 0.874 & 0.882 & 0.906 & 0.900
\\
2.0 & 0.816 & 0.832 & 0.874 & 0.888 & 0.896
\\
0 & 0.812 & 0.850 & 0.858 & 0.878 & 0.890
\\
0 &  0.824 & 0.870 & 0.882 & 0.896 & 0.908  
\\
0 &  0.844 & 0.854 & 0.880 & 0.898 & 0.910
\\
\bottomrule
\end{tabular}}
\end{minipage}\hfill
\begin{minipage}[t]{0.48\textwidth}
\centering
\textbf{(d) Linear regression: Right-sided}

\vspace{0.1cm}

\resizebox{\textwidth}{!}{%
\begin{tabular}{lccccc}
\toprule
$\beta_j$ & $n=50$ & $n=100$ & $n=150$ & $n=300$ & $n=500$ \\
\midrule
-0.5 & 0.864 & 0.868 & 0.880 & 0.892 & 0.900 
\\
1.0 & 0.864   & 0.878 & 0.890 & 0.896 & 0.902
\\
-1.5 & 0.860 & 0.876 & 0.878 & 0.882 & 0.906 
\\
2.0 & 0.864 & 0.880 & 0.884 & 0.890 & 0.912
\\
0 & 0.826 & 0.846 & 0.878 & 0.884 & 0.890
\\
0 &   0.848 & 0.862 & 0.888 & 0.892 & 0.908 
\\
0 &  0.840 & 0.856 & 0.860 & 0.894 & 0.900
\\
\bottomrule
\end{tabular}}
\end{minipage}

\end{table}
The empirical results in Table \ref{tab:SA6} indicate that the coverage probabilities of the proposed confidence intervals converge toward the nominal 90\% level as the sample size increases under both Gamma and Linear regression models. Two-sided intervals exhibit slight under-coverage for small samples, which diminishes steadily with increasing $n$. Right-sided intervals demonstrate more stable finite-sample performance, particularly in the linear regression setting. The average widths of the intervals decrease monotonically with $n$, reflecting improved estimation precision. Overall, the observed finite-sample behavior is consistent with the theoretical asymptotic properties.

\begin{figure*}[ht]
\centering

\textbf{Gamma Regression}\par\vspace{0.15cm}

\subfloat[$\beta_1=-0.5$]{%
\includegraphics[width=.3\linewidth]{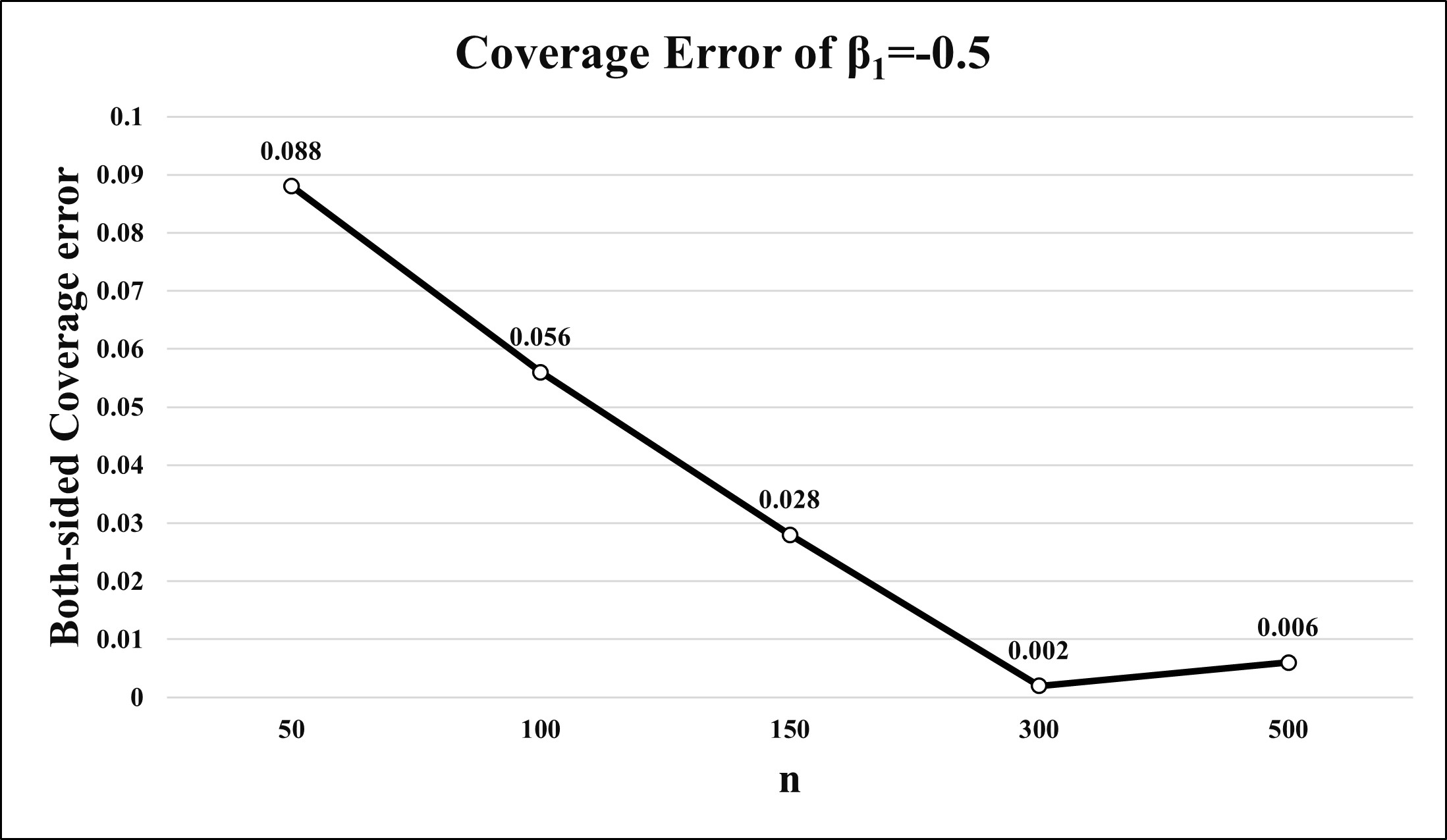}}
\hfill
\subfloat[$\beta_5=0$]{%
\includegraphics[width=.3\linewidth]{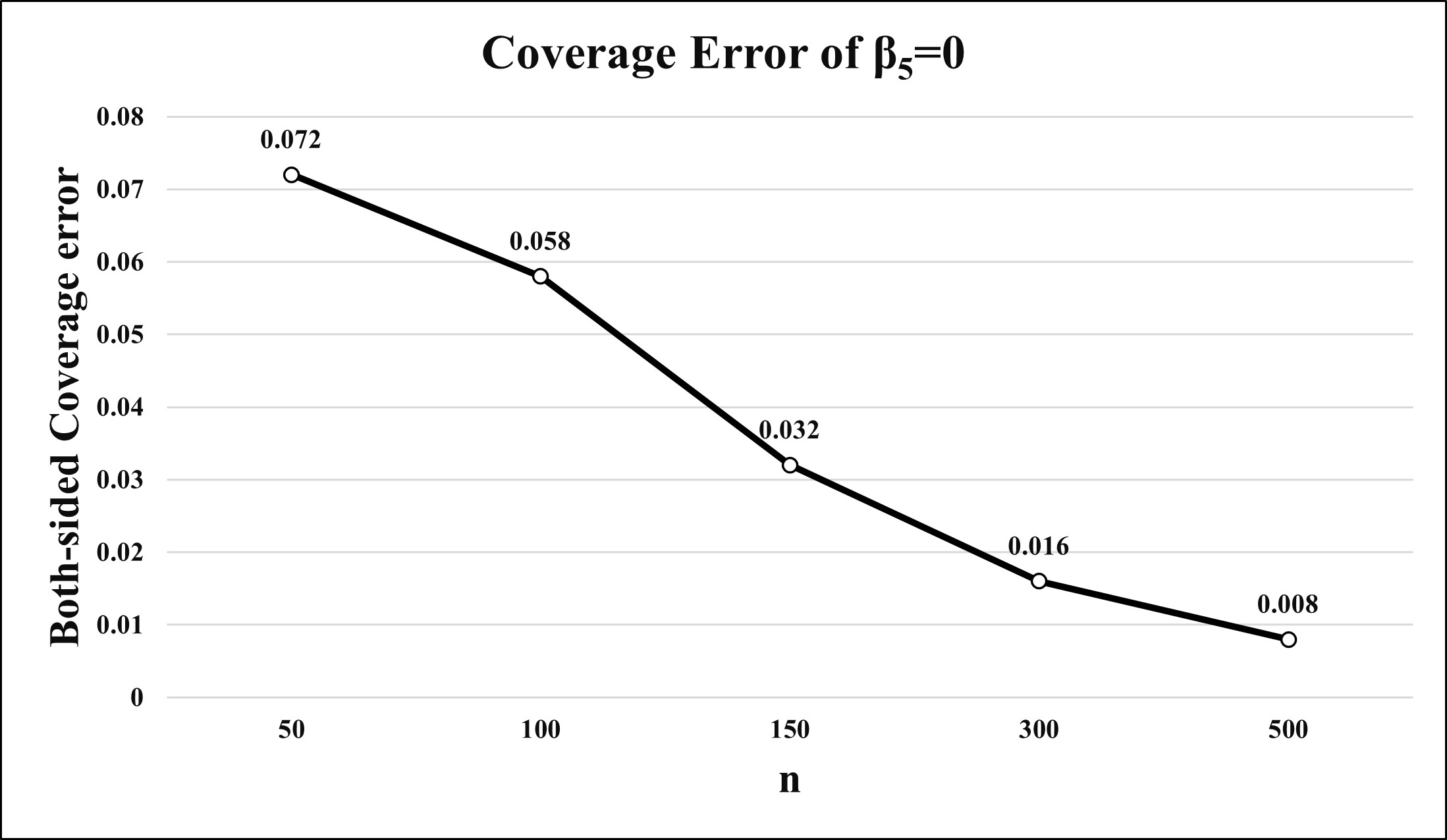}}
\hfill
\subfloat[$\beta_4=2$]{%
\includegraphics[width=.3\linewidth]{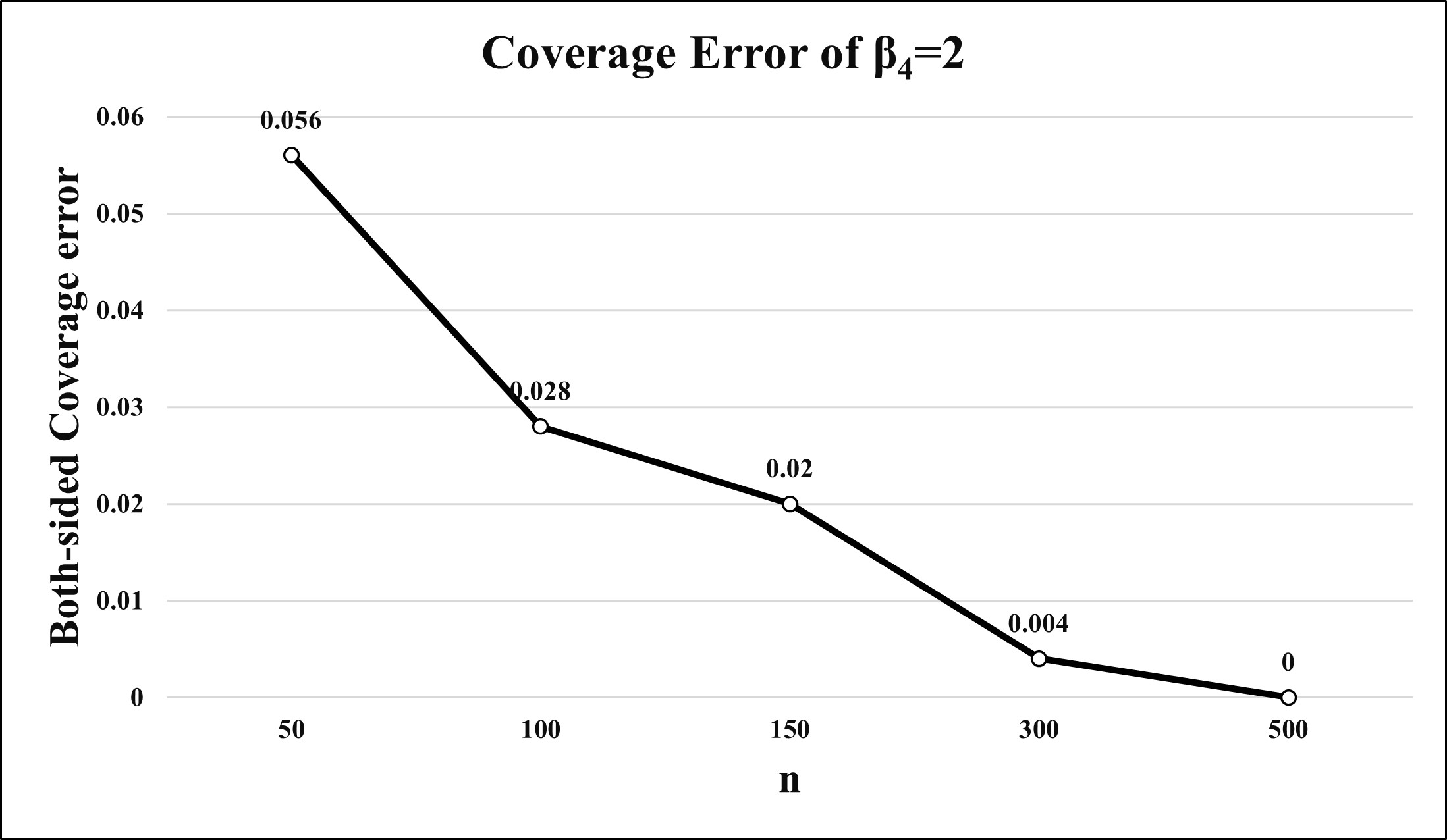}}

\vspace{0.4cm}

\textbf{Linear Regression}\par\vspace{0.15cm}

\subfloat[$\beta_1=-0.5$]{%
\includegraphics[width=.3\linewidth]{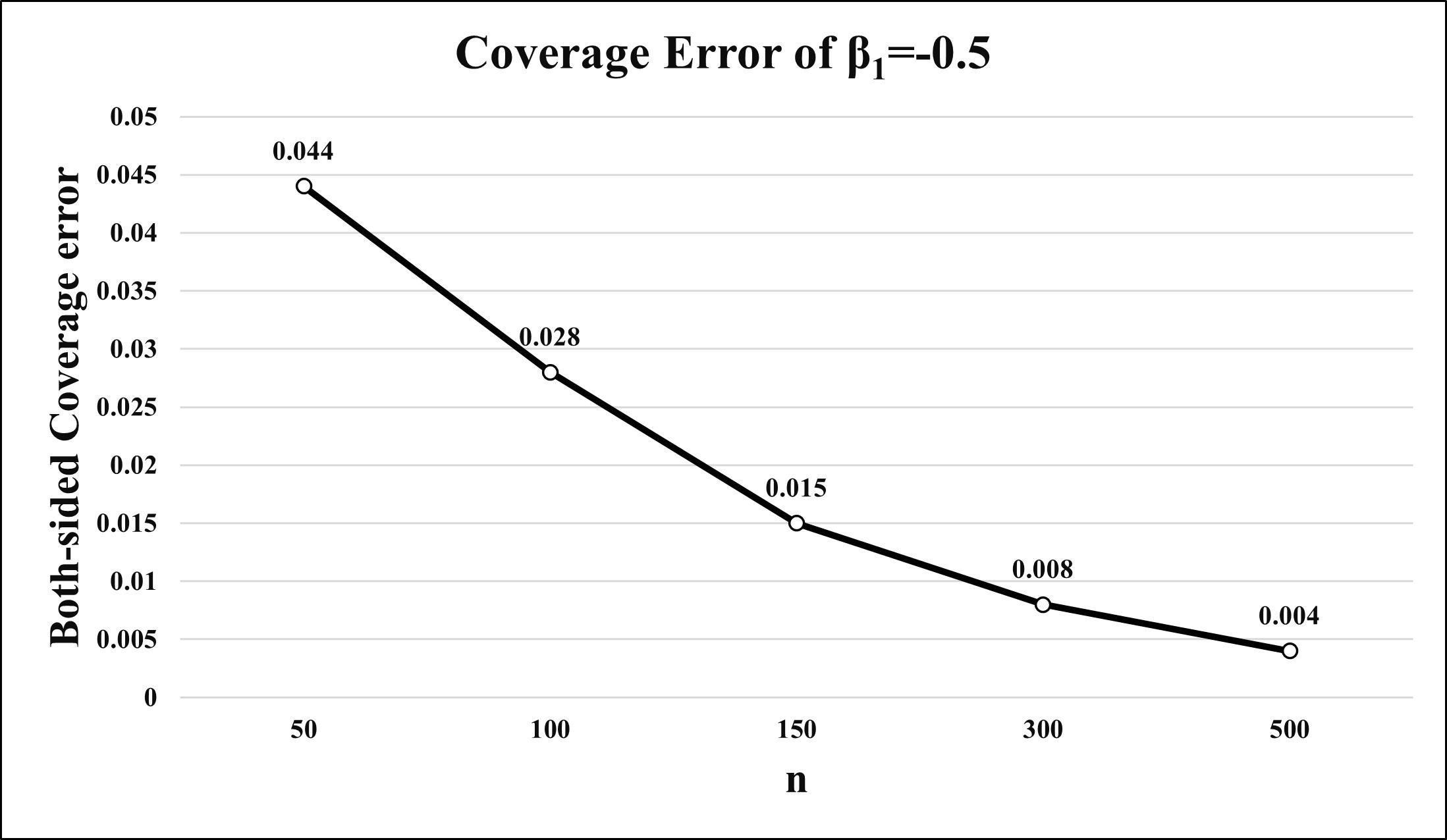}}
\hfill
\subfloat[$\beta_5=0$]{%
\includegraphics[width=.3\linewidth]{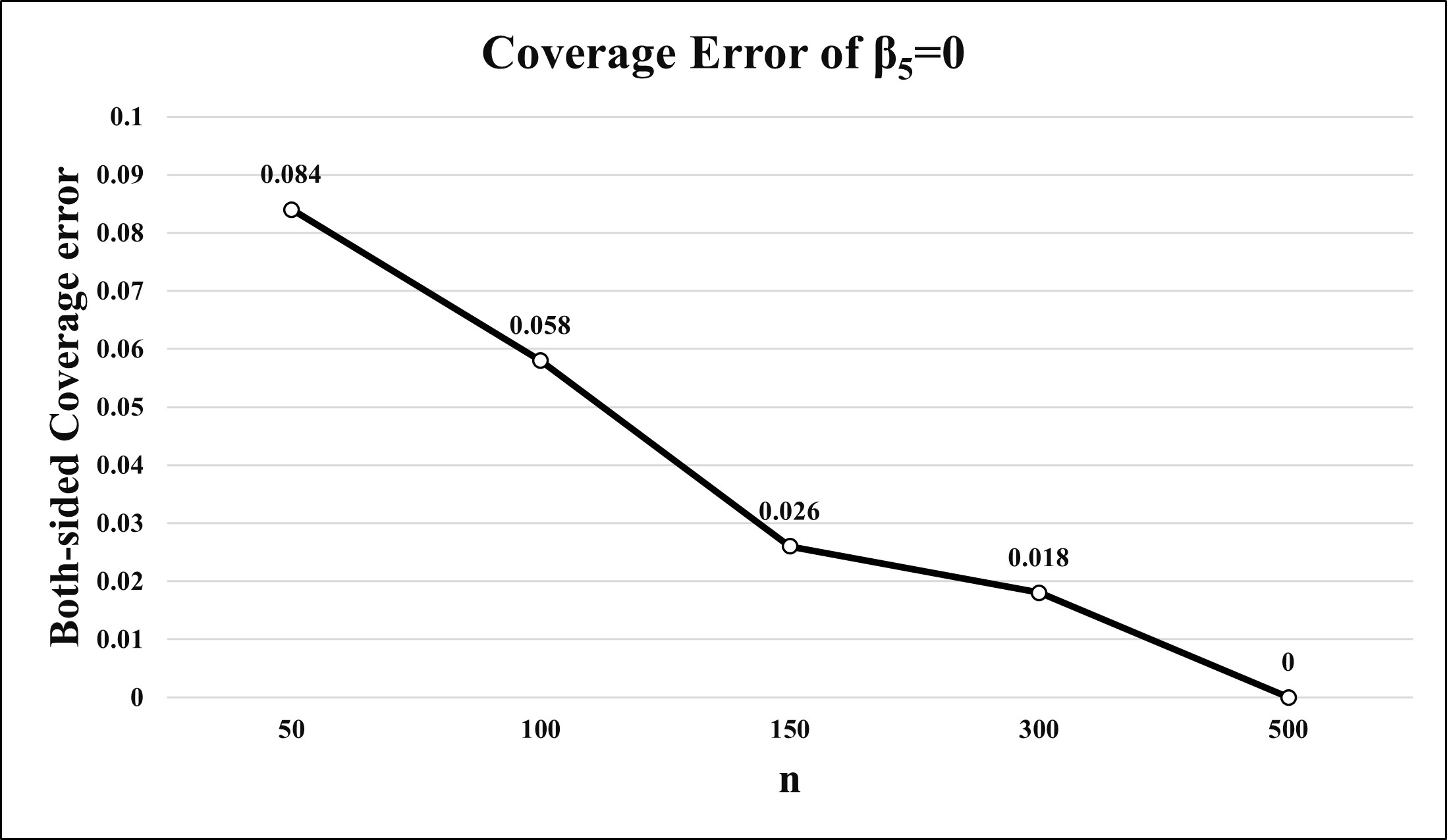}}
\hfill
\subfloat[$\beta_4=2$]{%
\includegraphics[width=.3\linewidth]{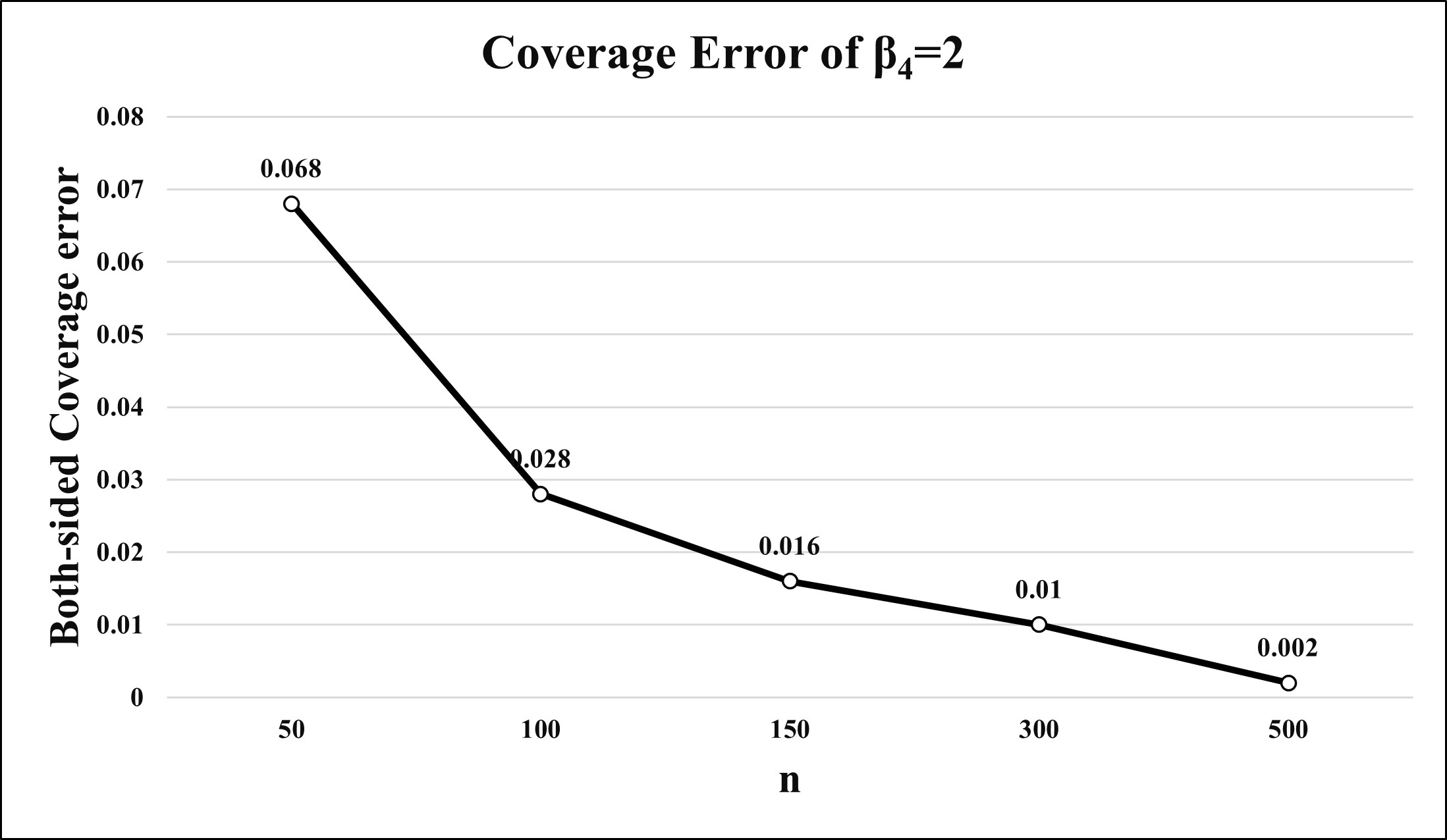}}

\caption{Coverage error of two-sided $90\%$ confidence intervals as a function of sample size $n$ in Gamma and Linear regression with fixed CV choice.}
\label{fig:A.3}
\end{figure*}

\begin{table}[H]
\centering
\caption{Empirical Coverage Probabilities of 90\% Confidence region of $\beta$}
\label{tabS9}
\begin{tabular}{@{}lcrcrrr@{}}
\hline
 &\multicolumn{5}{c}{Coverage Probability} \\
\cline{2-6}
Regression Type &
\multicolumn{1}{c}{$n=50$} &
\multicolumn{1}{c}{$n=100$}&
\multicolumn{1}{c}{$n=150$}&
\multicolumn{1}{c}{$n=300$}&
\multicolumn{1}{c@{}}{$n=500$}\\
\hline
Logistic & 0.986 & 0.960 & 0.932 & 0.912 & 0.908 \\
Gamma & 0.802 & 0.856 & 0.874 & 0.889 & 0.904 \\
Linear & 0.846 & 0.868 & 0.870 & 0.882 & 0.896 \\
\hline
\end{tabular}
\end{table}

We also observe the empirical coverage probabilities of 90\% confidence intervals of $\bm{\beta}$ using the Euclidean norm of the vectors $\bm{T}_n = n^{1/2}(\hat{\bm{\beta}}_n-\bm{\beta})$ and $\tilde{\bm{T}}_n^{*(PB)} = n^{1/2}(\hat{\bm{\beta}}_n^{*(PB)}-\tilde{\bm{\beta}}_n)$ (see Table \ref{tabS9}). We observe that as $n$ increases over the course, the simulation results get better in the sense that the empirical coverage probabilities get closer and closer to nominal confidence level of 0.90 for all regression coefficients in case of all three regression methods. Average width of both sided intervals for each component of $\bm{\beta}$ is mentioned in parentheses under empirical coverage probability.

\subsection{Simulation Study for varying choices of $a_n$ over $n$ in Logistic Regression}\label{subsec:varyingan}
Recall that, to establish Theorem \ref{thm:ptheo} regarding PB approximation, we need to threshold the original Lasso estimator to incorporate it as centering term in PB pivotal quantity. Now this thresholding heavily relies on the sequence $\{a_n\}_{n\geq 1}$ such that we require $a_n+(n^{-1/2}\log n)a_n^{-1}\to 0$ as $n\to\infty.$ Now any sequence $\{a_n\}_{n\geq 1}$ that satisfies the above condition, should be eligible to result in better approximation as $n$ increases. In this section, our aim is to produce, finite sample results in terms of empirical coverage probabilities of nominal $90\%$ confidence intervals for $$(n,p,p_0)\in \{(50,7,4), (100,7,4), (150,7,4), (300,7,4), (500,7,4)\}$$ under the varying choices of $a_n=n^{-c}$ with $0<c<1/2$ as: 
$c\in \{0.0015, 1/6, 1/5, 1/4, 0.485\}.$

\begin{table}[H]
\centering
\caption{Empirical Coverage Probabilities of 90\% Confidence region of $\beta$ over $a_n$ and $n$}
\label{tabS10}
\begin{tabular}{@{}lcrcrrr@{}}
\hline
 &\multicolumn{6}{c}{Coverage Probability for $a_n=n^{-c}$} \\
\cline{2-7}
$n$ &
\multicolumn{1}{c}{$c=0.0015$} &
\multicolumn{1}{c}{$c=1/6$}&
\multicolumn{1}{c}{$c=1/5$}&
\multicolumn{1}{c}{$c=1/4$}&
\multicolumn{1}{c}{$c=1/3$}&
\multicolumn{1}{c@{}}{$c=0.485$}\\
\hline
50& 0.720 & 0.986 & 0.970 & 0.982 & 0.988 & 0.810 \\
100 & 0.820 & 0.974 & 0.956 & 0.964 & 0.978 & 0.842 \\
150& 0.842 & 0.942 & 0.928 & 0.934 & 0.942 & 0.869 \\
300 & 0.884 & 0.920 & 0.906 & 0.914 & 0.910 & 0.886 \\
500 & 0.926 & 0.910 & 0.902 & 0.898 & 0.901 & 0.897 \\
\hline
\end{tabular}
\end{table}

Table \ref{tabS10} gives us a synopsis about how the finite sample coverages of $\bm{\beta}$ perform, once we vary the choices of $a_n$. As evident from the table, there's no surprise that empirical coverage probabilities get better and better within the proximity of nominal confidence level $0.90$ as $n$ increases when we employ different theoretical choices of $a_n$ just like the earlier results. Next we will present the same results for individual coefficients of $\bm{\beta}$ along with their coverage errors. To reduce space consumption, for each choice of $c$ mentioned above, we have plotted coverage error versus $n$ for $\beta_1=-0.5$, $\beta_5=0$ and $\beta_4=2$ within a single graph differentiated by three different indicators.

\begin{figure}[H]\centering
\includegraphics[width=.8\linewidth]{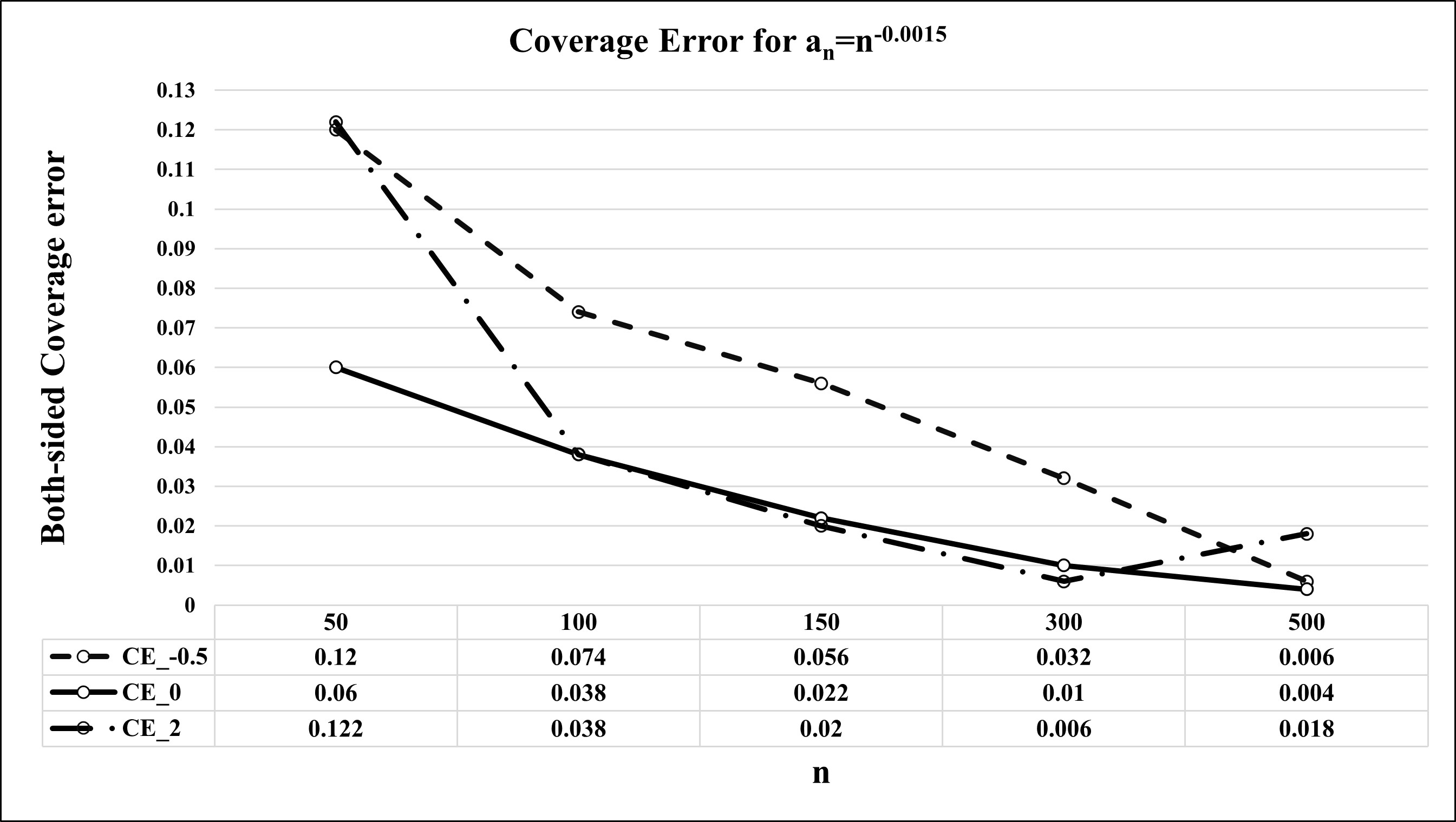}
\caption{Coverage Error of Both sided $90\%$ Confidence Interval for $a_n=n^{-0.0015}$.}
\label{fig S5}
\end{figure}
For $a_n=n^{-0.0015}$, as it can be seen from Table \ref{tab:SA7} that, the empirical coverage probabilities for each of the components of $\bm{\beta}$ are getting closer and closer to nominal confidence level $0.90$ as we increase the sample size $n$. Now also the average widths of the confidence intervals are denoted within the parentheses. These widths are getting smaller as we move towards larger $n$. Now Figure \ref{fig S5}, depicts that the coverage error gets closer to $0$ as $n$ increases. The indicators $CE\_-0.5$, $CE\_0$ and $CE\_2$ respectively denote the coverage errors corresponding to $\beta_1=-0.5$, $\beta_5=0$ and $\beta_4=2$.

\begin{table}[htbp]
\centering
\caption{Empirical coverage probabilities of 90\% confidence intervals when $a_n = n^{-0.0015}$.}
\label{tab:SA7}

\begin{minipage}[t]{0.48\textwidth}
\centering
\textbf{(a) Both-sided CI}

\vspace{0.1cm}

\resizebox{\textwidth}{!}{%
\begin{tabular}{lccccc}
\hline
$\beta_j$ & $n=50$ & $n=100$ & $n=150$ & $n=300$ & $n=500$\\
\hline
-0.5 & 0.780 & 0.826 & 0.844 & 0.868 & 0.894\\
 & (1.615) & (1.101) & (0.640) & (0.582) & (0.477) \\  
1.0 & 0.792 & 0.802 & 0.856 & 0.908 & 0.922\\
 & (2.079) & (1.368) & (0.926) & (0.785) & (0.539) \\ 
-1.5 & 0.800 & 0.846 & 0.858 & 0.884 & 0.916\\
 & (2.212) & (1.229) & (1.102) & (0.824) & (0.614) \\ 
2.0 & 0.778 & 0.862 & 0.880 & 0.894 & 0.918\\
 & (2.253) & (1.239) & (1.011) & (0.873) & (0.719) \\
0 & 0.840 & 0.862 & 0.878 & 0.890 & 0.904\\
 & (1.545) & (1.029) & (0.683) & (0.545) & (0.460) \\
0 & 0.836 & 0.848 & 0.870 & 0.878 & 0.884\\
 & (1.647) & (0.912) & (0.828) & (0.674) & (0.448) \\
0 & 0.826 & 0.862 & 0.880 & 0.889 & 0.904\\
 & (1.981) & (1.028) & (0.803) & (0.506) & (0.421) \\
\hline
\end{tabular}}
\end{minipage}\hfill
\begin{minipage}[t]{0.48\textwidth}
\centering
\textbf{(b) Right-sided CI}

\vspace{0.1cm}

\resizebox{\textwidth}{!}{%
\begin{tabular}{lccccc}
\hline
$\beta_j$ & $n=50$ & $n=100$ & $n=150$ & $n=300$ & $n=500$\\
\hline
-0.5 & 0.778 & 0.858 & 0.872 & 0.876 & 0.898\\
& & & & & \\
1.0  & 0.816 & 0.844 & 0.856 & 0.884 & 0.912\\
& & & & & \\
-1.5 & 0.748 & 0.812 & 0.864 & 0.896 & 0.920\\
& & & & & \\
2.0  & 0.782 & 0.824 & 0.868 & 0.895 & 0.906\\
& & & & & \\
0 & 0.808 & 0.814 & 0.840 & 0.864 & 0.886\\
& & & & & \\
0 & 0.792 & 0.838 & 0.858 & 0.888 & 0.904\\
& & & & & \\
0 & 0.802 & 0.842 & 0.882 & 0.890 & 0.902\\
& & & & & \\
\hline
\end{tabular}}
\end{minipage}

\end{table}

\begin{figure}[H]\centering
\vspace{0.6em}\includegraphics[width=.8\linewidth]{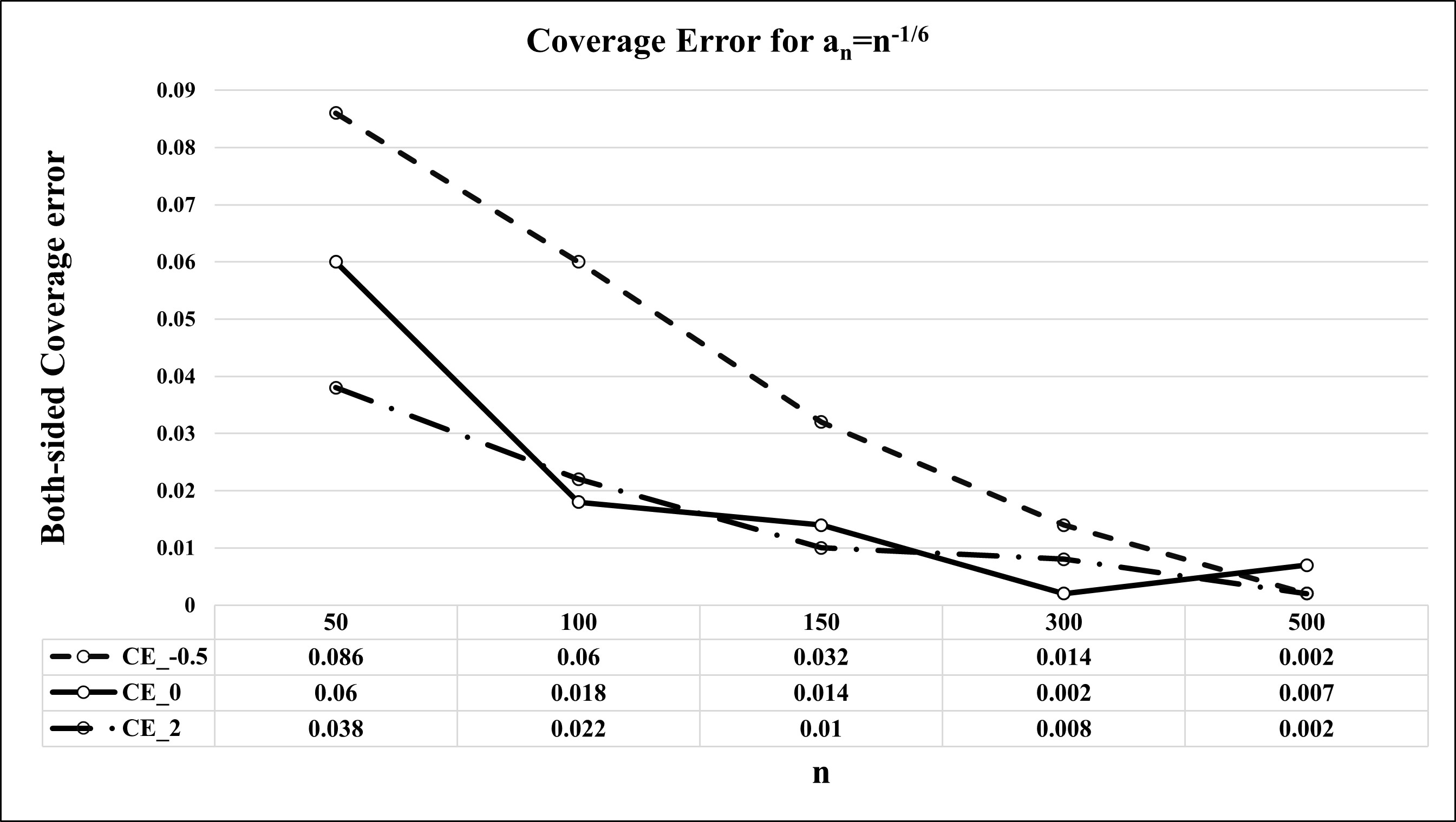}
\caption{Coverage Error of Both sided $90\%$ Confidence Interval for $a_n=n^{-1/6}$.}
\label{fig S6}
\end{figure}
For $a_n=n^{-1/6}$, as it can be seen from Table \ref{tab:SA8} that, the empirical coverage probabilities for each of the components of $\bm{\beta}$ are getting closer and closer to nominal confidence level $0.90$ as we increase the sample size $n$. Now also the average widths of the confidence intervals are denoted within the parentheses. These widths are getting smaller as we move towards larger $n$. Now Figure \ref{fig S6}, depicts that the coverage error gets closer to $0$ as $n$ increases. The indicators $CE\_-0.5$, $CE\_0$ and $CE\_2$ respectively denote the coverage errors corresponding to $\beta_1=-0.5$, $\beta_5=0$ and $\beta_4=2$.

\begin{table}[htbp]
\centering
\caption{Empirical coverage probabilities of 90\% confidence intervals when $a_n=n^{-1/6}$.}
\label{tab:SA8}

\begin{minipage}[t]{0.48\textwidth}
\centering
\textbf{(a) Both-sided CI}

\vspace{0.1cm}

\resizebox{\textwidth}{!}{%
\begin{tabular}{lccccc}
\hline
$\beta_j$ & $n=50$ & $n=100$ & $n=150$ & $n=300$ & $n=500$\\
\hline
-0.5 & 0.986 & 0.960 & 0.932 & 0.914 & 0.898\\
 & (1.514) & (1.203) & (0.840) & (0.682) & (0.447) \\
1.0 & 0.976 & 0.962 & 0.926 & 0.918 & 0.902\\
 & (2.009) & (1.568) & (0.726) & (0.585) & (0.439) \\
-1.5 & 0.980 & 0.946 & 0.928 & 0.906 & 0.896\\
 & (2.202) & (1.529) & (1.024) & (0.624) & (0.514) \\
2.0 & 0.938 & 0.922 & 0.910 & 0.908 & 0.902\\
 & (2.153) & (1.839) & (1.001) & (0.573) & (0.419) \\
0 & 0.840 & 0.882 & 0.886 & 0.898 & 0.907\\
 & (1.550) & (1.022) & (0.643) & (0.525) & (0.464) \\
0 & 0.936 & 0.928 & 0.910 & 0.898 & 0.900\\
 & (1.247) & (0.902) & (0.728) & (0.574) & (0.348) \\
0 & 0.876 & 0.882 & 0.888 & 0.896 & 0.914\\
 & (1.881) & (1.328) & (0.603) & (0.562) & (0.424) \\
\hline
\end{tabular}}
\end{minipage}\hfill
\begin{minipage}[t]{0.48\textwidth}
\centering
\textbf{(b) Right-sided CI}

\vspace{0.1cm}

\resizebox{\textwidth}{!}{%
\begin{tabular}{lccccc}
\hline
$\beta_j$ & $n=50$ & $n=100$ & $n=150$ & $n=300$ & $n=500$\\
\hline
-0.5 & 0.978 & 0.948 & 0.922 & 0.916 & 0.896\\
& & & & & \\
1.0  & 0.946 & 0.924 & 0.916 & 0.884 & 0.902\\
& & & & & \\
-1.5 & 0.970 & 0.934 & 0.924 & 0.916 & 0.890\\
& & & & & \\
2.0  & 0.852 & 0.864 & 0.888 & 0.896 & 0.916\\
& & & & & \\
0 & 0.968 & 0.944 & 0.940 & 0.904 & 0.898\\
& & & & & \\
0 & 0.928 & 0.916 & 0.910 & 0.898 & 0.902\\
& & & & & \\
0 & 0.934 & 0.942 & 0.912 & 0.906 & 0.898\\
& & & & & \\
\hline
\end{tabular}}
\end{minipage}

\end{table}

\begin{figure}[H]\centering
\vspace{0.9em}\includegraphics[width=.58\linewidth]{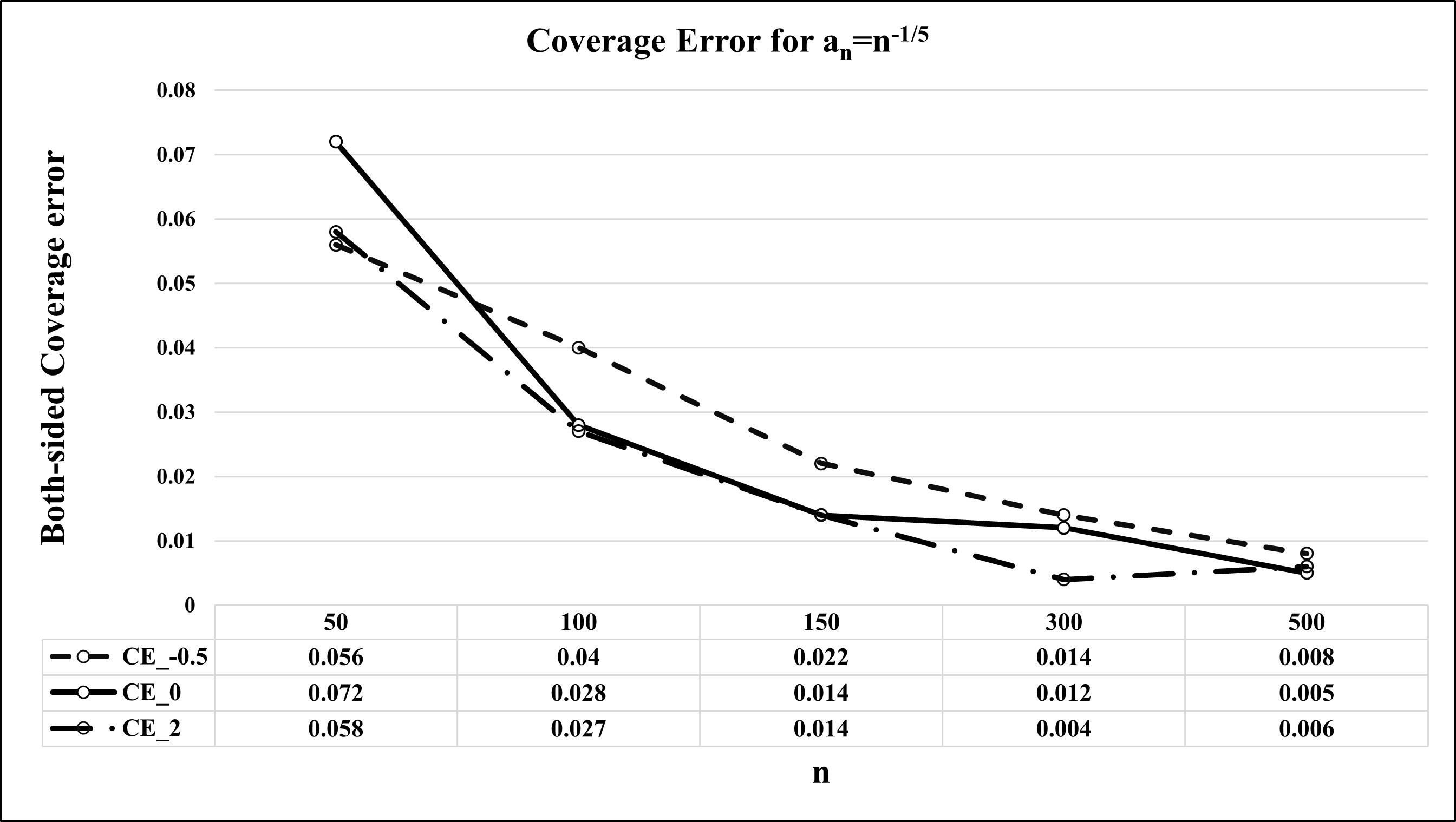}
\caption{Coverage Error of Both sided $90\%$ Confidence Interval for $a_n=n^{-1/5}$.}
\label{fig S7}
\end{figure}

\begin{table}[htbp]
\centering
\caption{Empirical coverage probabilities of 90\% confidence intervals when $a_n = n^{-1/5}$.}
\label{tab:SA9}

\begin{minipage}[t]{0.48\textwidth}
\centering
\textbf{(a) Both-sided CI}

\vspace{0.1cm}

\resizebox{\textwidth}{!}{%
\begin{tabular}{lccccc}
\hline
$\beta_j$ & $n=50$ & $n=100$ & $n=150$ & $n=300$ & $n=500$\\
\hline
-0.5 & 0.956 & 0.940 & 0.922 & 0.914 & 0.908\\
 & (1.314) & (1.103) & (0.740) & (0.642) & (0.547) \\
1.0 & 0.964 & 0.922 & 0.920 & 0.908 & 0.898\\
 & (2.012) & (1.768) & (0.926) & (0.885) & (0.539) \\
-1.5 & 0.980 & 0.936 & 0.928 & 0.916 & 0.906\\
 & (1.902) & (1.629) & (1.028) & (0.724) & (0.614) \\
2.0 & 0.958 & 0.927 & 0.914 & 0.904 & 0.894\\
 & (1.853) & (1.339) & (1.011) & (0.673) & (0.449) \\
0 & 0.828 & 0.872 & 0.886 & 0.888 & 0.905\\
 & (1.450) & (1.012) & (0.543) & (0.505) & (0.454) \\
0 & 0.936 & 0.920 & 0.910 & 0.898 & 0.896\\
 & (1.147) & (0.982) & (0.828) & (0.674) & (0.448) \\
0 & 0.878 & 0.888 & 0.898 & 0.902 & 0.916\\
 & (1.891) & (1.428) & (0.803) & (0.662) & (0.524) \\
\hline
\end{tabular}}
\end{minipage}\hfill
\begin{minipage}[t]{0.48\textwidth}
\centering
\textbf{(b) Right-sided CI}

\vspace{0.1cm}

\resizebox{\textwidth}{!}{%
\begin{tabular}{lccccc}
\hline
$\beta_j$ & $n=50$ & $n=100$ & $n=150$ & $n=300$ & $n=500$\\
\hline
-0.5 & 0.974 & 0.958 & 0.932 & 0.926 & 0.906\\
& & & & & \\
1.0  & 0.948 & 0.924 & 0.910 & 0.889 & 0.902\\
& & & & & \\
-1.5 & 0.960 & 0.948 & 0.926 & 0.902 & 0.898\\
& & & & & \\
2.0  & 0.812 & 0.864 & 0.898 & 0.896 & 0.910\\
& & & & & \\
0 & 0.962 & 0.942 & 0.930 & 0.914 & 0.898\\
& & & & & \\
0 & 0.928 & 0.914 & 0.910 & 0.898 & 0.902\\
& & & & & \\
0 & 0.934 & 0.942 & 0.912 & 0.908 & 0.900\\
& & & & & \\
\hline
\end{tabular}}
\end{minipage}

\end{table}

For $a_n=n^{-1/5}$, as it can be seen from Table \ref{tab:SA9} that, the empirical coverage probabilities for each of the components of $\bm{\beta}$ are getting closer and closer to nominal confidence level $0.90$ as we increase the sample size $n$. Now Figure \ref{fig S7}, depicts that the coverage error gets closer to $0$ as $n$ increases. As $n$ increases, these coverage results are not at all surprising.The pattern is more or less similar as far as better approximation is concerned.

\begin{figure}[H]\centering
\includegraphics[width=.98\linewidth]{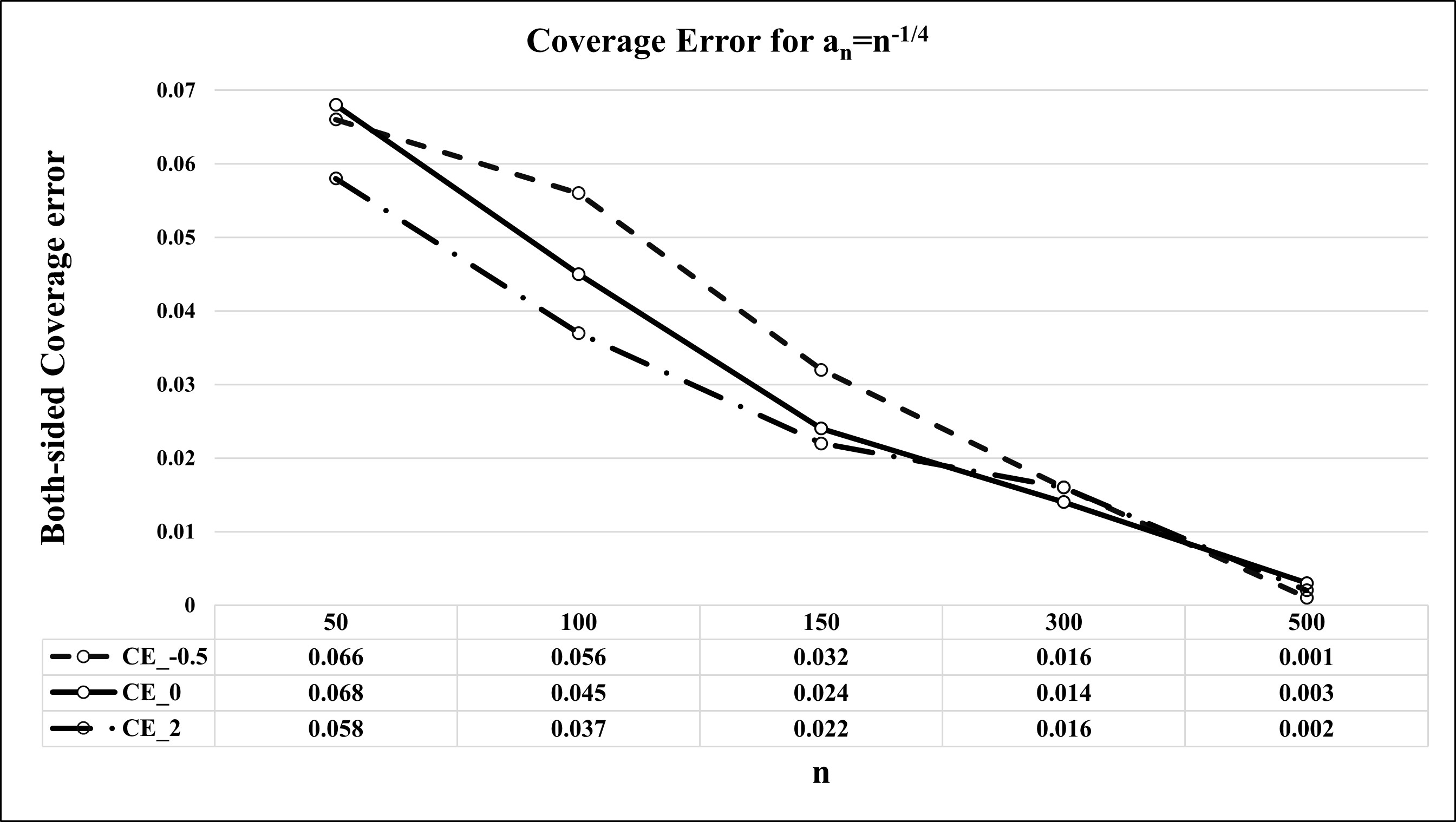}
\caption{Coverage Error of Both sided $90\%$ Confidence Interval for $a_n=n^{-1/4}$.}
\label{fig S8}
\end{figure}

\begin{table}[htpb]
\centering
\caption{Empirical coverage probabilities of 90\% confidence intervals when $a_n = n^{-1/4}$.}
\label{tabs17-18}

\begin{minipage}[t]{0.48\textwidth}
\centering
\textbf{(a) Both-sided 90\% CI}

\vspace{0.1cm}

\resizebox{\textwidth}{!}{%
\begin{tabular}{@{}lccccc@{}}
\hline
$\beta_j$ & $n=50$ & $n=100$ & $n=150$ & $n=300$ & $n=500$ \\
\hline
-0.5 & 0.966 & 0.956 & 0.932 & 0.916 & 0.901\\
 & (1.324) & (1.163) & (0.840) & (0.742) & (0.617) \\  
1.0 & 0.954   & 0.942 & 0.928 & 0.910 & 0.899\\
& (2.212) & (1.868) & (0.906) & (0.825) & (0.639) \\ 
-1.5 & 0.960 & 0.946 & 0.938 & 0.916 & 0.904\\
& (1.922) & (1.329) & (1.008) & (0.824) & (0.514) \\
2.0    & 0.958 & 0.937 & 0.922 & 0.916 & 0.898\\
& (1.753) & (1.369) & (1.031) & (0.683) & (0.459) \\
0 & 0.968 & 0.945 & 0.924 & 0.914 & 0.903\\
& (1.250) & (1.002) & (0.863) & (0.605) & (0.434) \\
0 & 0.934 & 0.940 & 0.924 & 0.918 & 0.896 \\
& (1.247) & (0.988) & (0.826) & (0.684) & (0.548) \\
0 & 0.968 & 0.948 & 0.926 & 0.902 & 0.900\\
& (1.715) & (1.128) & (0.843) & (0.562) & (0.424) \\

\hline
\end{tabular}}
\end{minipage}
\hfill
\begin{minipage}[t]{0.48\textwidth}
\centering
\textbf{(b) Right-sided 90\% CI}

\vspace{0.1cm}

\resizebox{\textwidth}{!}{%
\begin{tabular}{@{}lccccc@{}}
\hline
$\beta_j$ & $n=50$ & $n=100$ & $n=150$ & $n=300$ & $n=500$ \\
\hline
-0.5 & 0.944 & 0.938 & 0.922 & 0.910 & 0.902 
\\
& & & & & \\
1.0 & 0.948 & 0.926 & 0.912 & 0.889 & 0.904
\\
& & & & & \\
-1.5 & 0.950 & 0.928 & 0.924 & 0.912 & 0.899 
\\
& & & & & \\
2.0 & 0.932 & 0.924 & 0.898 & 0.896 & 0.905
\\
& & & & & \\
0 & 0.962 & 0.952 & 0.938 & 0.914 & 0.908
\\
& & & & & \\
0 & 0.948 & 0.924 & 0.914 & 0.898 & 0.906 
\\
& & & & & \\
0 & 0.954 & 0.947 & 0.916 & 0.908 & 0.900
\\
& & & & & \\
\hline
\end{tabular}}
\end{minipage}

\end{table}

For $a_n=n^{-1/4}$, as it can be seen from Table \ref{tabs17-18} that, the empirical coverage probabilities for each of the components of $\bm{\beta}$ are getting closer and closer to nominal confidence level $0.90$ as we increase the sample size $n$. Now Figure \ref{fig S8}, depicts that the coverage error gets closer to $0$ as $n$ increases.

\begin{figure}[H]\centering
\vspace{0.9em}\includegraphics[width=.8\linewidth]{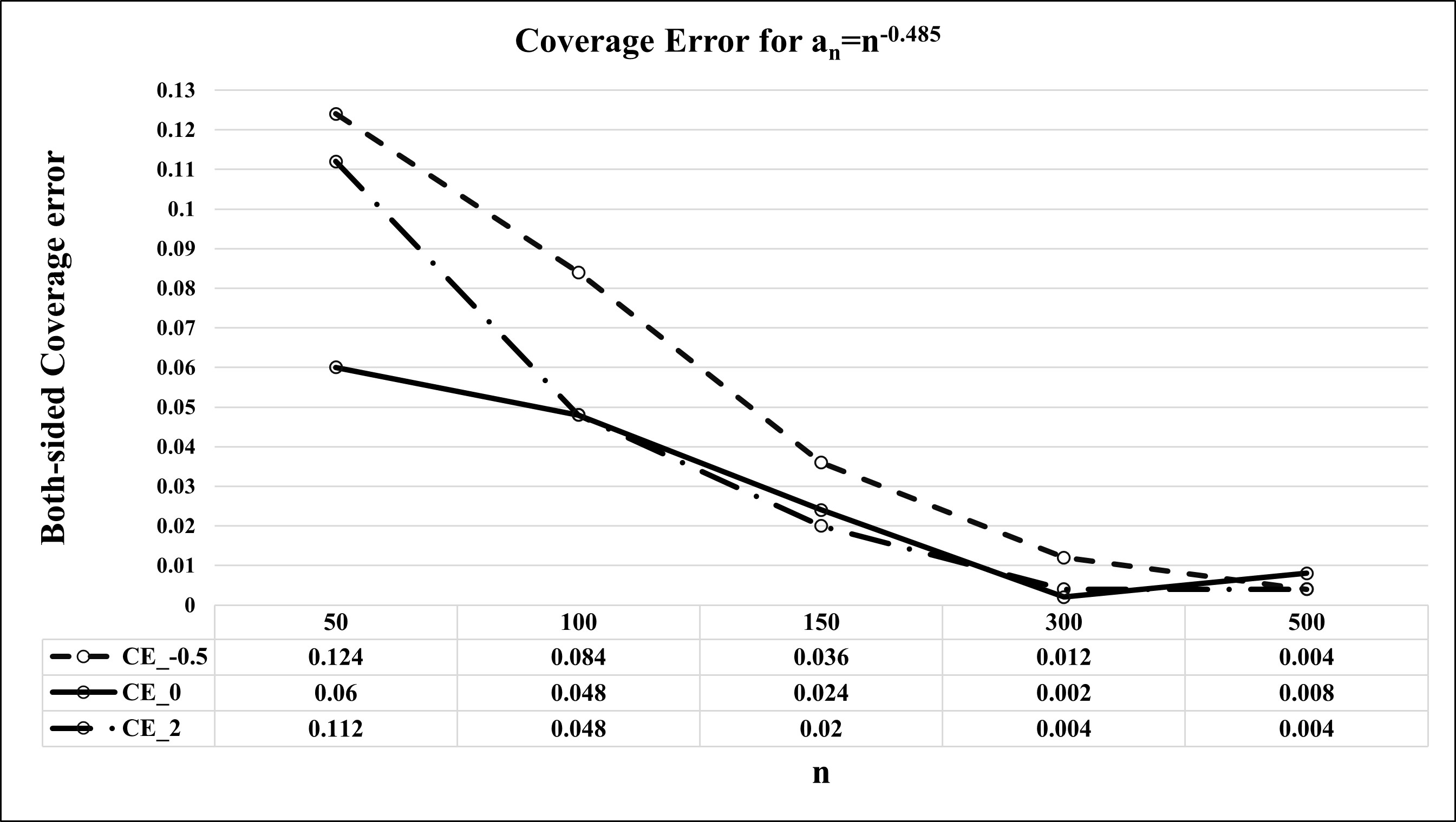}
\caption{Coverage Error of Both sided $90\%$ Confidence Interval for $a_n=n^{-0.485}$.}
\label{fig S10}
\end{figure}

\begin{table}[H]
\centering
\caption{Empirical coverage probabilities of 90\% confidence intervals when $a_n = n^{-0.485}$.}
\label{tabs19-120}

\begin{minipage}[t]{0.48\textwidth}
\centering
\textbf{(a) Both-sided 90\% CI}

\vspace{0.1cm}

\resizebox{\textwidth}{!}{%
\begin{tabular}{@{}lccccc@{}}
\hline
$\beta_j$ & $n=50$ & $n=100$ & $n=150$ & $n=300$ & $n=500$ \\
\hline
-0.5 & 0.776 & 0.816 & 0.864 & 0.888 & 0.904\\
 & (1.815) & (1.121) & (0.740) & (0.682) & (0.487) \\ 
1.0 & 0.794   & 0.822 & 0.854 & 0.880 & 0.902\\
& (2.092) & (1.668) & (0.916) & (0.705) & (0.639) \\ 
-1.5 & 0.820 & 0.848 & 0.868 & 0.894 & 0.914\\
& (1.812) & (1.029) & (0.802) & (0.624) & (0.414) \\
2.0    & 0.788 & 0.852 & 0.880 & 0.896 & 0.904\\
& (2.232) & (1.209) & (1.012) & (0.773) & (0.619) \\
0 & 0.840 & 0.852 & 0.876 & 0.898 & 0.908\\
& (1.245) & (0.829) & (0.682) & (0.525) & (0.468) \\
0 & 0.816 & 0.868 & 0.878 & 0.888 & 0.898 \\
& (1.672) & (0.942) & (0.858) & (0.684) & (0.458) \\
0 & 0.816 & 0.842 & 0.868 & 0.889 & 0.900\\
& (1.881) & (1.428) & (0.823) & (0.606) & (0.428) \\

\hline
\end{tabular}}
\end{minipage}
\hfill
\begin{minipage}[t]{0.48\textwidth}
\centering
\textbf{(b) Right-sided 90\% CI}

\vspace{0.1cm}

\resizebox{\textwidth}{!}{%
\begin{tabular}{@{}lccccc@{}}
\hline
$\beta_j$ & $n=50$ & $n=100$ & $n=150$ & $n=300$ & $n=500$ \\
\hline
-0.5 & 0.768 & 0.818 & 0.842 & 0.876 & 0.894 
\\
& & & & & \\
1.0 & 0.816 & 0.854 & 0.886 & 0.894 & 0.912
\\
& & & & & \\
-1.5 & 0.788 & 0.822 & 0.864 & 0.888 & 0.900 
\\
& & & & & \\
2.0 & 0.782 & 0.834 & 0.858 & 0.894 & 0.898
\\
& & & & & \\
0 & 0.808 & 0.824 & 0.846 & 0.874 & 0.896
\\
& & & & & \\
0 & 0.798 & 0.839 & 0.878 & 0.898 & 0.908 
\\
& & & & & \\
0 & 0.802 & 0.842 & 0.862 & 0.890 & 0.912
\\
& & & & & \\
\hline
\end{tabular}}
\end{minipage}

\end{table}
For $a_n=n^{-0.485}$, as it can be seen from Table \ref{tabs19-120} that, the empirical coverage probabilities for each of the components of $\bm{\beta}$ are getting closer and closer to nominal confidence level $0.90$ as we increase the sample size $n$. Now also the average widths of the confidence intervals are denoted within the parentheses. These widths are getting smaller as we move towards larger $n$. Now Figure \ref{fig S10}, depicts that the coverage error gets closer to $0$ as $n$ increases. The indicators $CE\_-0.5$, $CE\_0$ and $CE\_2$ respectively denote the coverage errors corresponding to $\beta_1=-0.5$, $\beta_5=0$ and $\beta_4=2$. We also mention the coverage errors over $n$, for these three components.

\bibliographystyle{amsplain}

\begin{thebibliography}{10}

\bibitem[{Berkson(1944)}]{berkson1944application}
\textsc{Berkson, J.} (1944).
\newblock \textit{Application of the logistic function to bio-assay}.
\newblock \textit{Journal of the American statistical association}.
\newblock \textbf{39(227)}, 357--365.

\bibitem[{Bhattacharya and Rao(1986)}]{bhattacharya1986normal}
\textsc{Bhattacharya, R. N.} \& 
\textsc{Rao, R. R.} (1986).
\newblock \textit{{Normal Approximation and Asymptotic Expansions}}.
\newblock vol. {\textbf{64}} SIAM.


\bibitem[{B{\"u}hlmann and van de Geer(2011)}]{buhlmann2011statistics}
\textsc{B{\"u}hlmann, P.} \&
\textsc{van de Geer, S.} (2011).
\newblock \textit{Statistics for High-Dimensional Data: Methods, Theory and Applications}.
\newblock Springer, Heidelberg.

\bibitem[{Bunea(2008)}]{bunea2008honest}
\textsc{Bunea, F.} (2008).
\newblock \textit{Honest variable selection in linear and logistic regression models via $l_1$ and $l_1$+$l_2$ penalization}.
\newblock \textit{Electronic Journal of Statistics}.
\newblock \textbf{2}, 1153--1194.

\bibitem[{Camponovo(2015)}]{camponovo2015validity}
\textsc{Camponovo, L.} (2015).
\newblock \textit{On the validity of the pairs bootstrap for lasso estimators}.
\newblock \textit{Biometrika}.
\newblock {\textbf{102(4)}}, 981--987.

\bibitem[{Chatterjee and Lahiri(2010)}]{chatterjee2010asymptotic}
\textsc{Chatterjee, A.} \&
\textsc{Lahiri, S. N.} (2010).
\newblock \textit{Asymptotic properties of the residual bootstrap for lasso estimators}.
\newblock \textit{Proceedings of the American Mathematical Society}.
\newblock \textbf{138(12)}, 4497--4509.

\bibitem[{Chatterjee and Lahiri(2011)}]{chatterjee2011bootstrapping}
\textsc{Chatterjee, A.} \&
\textsc{Lahiri, S. N.} (2011).
\newblock \textit{Bootstrapping lasso estimators}.
\newblock \textit{Journal of the American Statistical Association}.
\newblock \textbf{106(494)}, 608--625.

\bibitem[{Chatterjee and Lahiri(2011)}]{chatterjee2011strong}
\textsc{Chatterjee, A.} \&
\textsc{Lahiri, S. N.} (2011).
\newblock \textit{Strong consistency of Lasso estimators}.
\newblock \textit{Sankhya A}.
\newblock \textbf{73}, 55--78.

\bibitem[{Claeskens et al. (2003)}]{claeskens2003quadratic}
\textsc{Claeskens, G.}, \textsc{Aerts, M.} \& \textsc{Molenberghs, G.} (2003).
\newblock \textit{A quadratic bootstrap method and improved estimation in logistic regression}.
\newblock \textit{Statistics \& Probability Letters}.
\newblock \textbf{61(4)}, 383--394. 


\bibitem[{Cox(1958)}]{cox1958regression}
\textsc{Cox, D. R.} (1958).
\newblock \textit{The regression analysis of binary sequences}.
\newblock \textit{Journal of the Royal Statistical Society: Series B (Methodological)}.
\newblock \textbf{20(2)}, 215--232.

\bibitem[{Das et al.(2019)}]{das2019perturbation}
\textsc{Das, D.} \&
\textsc{Gregory, K.} \&
\textsc{Lahiri, S. N.} (2019).
\newblock \textit{Perturbation bootstrap in adaptive lasso}.
\newblock \textit{The Annals of Statistics}.
\newblock \textbf{47(4)}, 2080--2116.


\bibitem[{Das and Lahiri(2019)}]{das2019distributional}
\textsc{Das, D.} \&
\textsc{Lahiri, S. N.} (2019).
\newblock \textit{Distributional consistency of the lasso by perturbation bootstrap}.
\newblock \textit{Biometrika}.
\newblock \textbf{106(4)}, 957--964.

\bibitem[{Das et al. (2025)}]{das2025higher}
\textsc{Das, D.}, \textsc{Chatterjee, A.} \& \textsc{Lahiri, S. N.} (2025).
\newblock \textit{Higher order accurate symmetric bootstrap confidence intervals in high-dimensional penalized regression}.
\newblock \textit{Journal of the American Statistical Association}.
\newblock \textbf{120}(551), 1645--1656.


\bibitem[{Davis et al.(1992)}]{davis1992m}
\textsc{Davis, R. A.}, 
\textsc{Knight, K.} \&
\textsc{Liu, J.} (1992).
\newblock \textit{{M-estimation for autoregressions with infinite variance}}.
\newblock \textit{Stochastic Processes and Their Applications}.
\newblock {\textbf{40(1)}}, 145--180.


\bibitem[{Efron et al. (2004)}]{efron2004least}
\textsc{Efron, B.},
\textsc{Hastie, T.},
\textsc{Johnstone, I.}\&
\textsc{Tibshirani, R.} (2004).
\newblock \textit{Least angle regression}.
\newblock \textit{Annals of Statistics}.
\newblock \textbf{32(2)}, 407--499.


\bibitem[{Ferger (2021)}]{ferger2021continuous}
\textsc{Ferger, D.} (2021).
\newblock \textit{A continuous mapping theorem for the argmin-set functional with applications to convex stochastic processes}.
\newblock \textit{Kybernetika}.
\newblock {\textbf{57(3)}}, 426--445.



\bibitem[{Freedman(1981)}]{freedman1981bootstrapping}
\textsc{Freedman, D. A.} (1981).
\newblock \textit{Bootstrapping regression models}.
\newblock \textit{The annals of statistics}.
\newblock \textbf{9(6)}, 1218--1228.


\bibitem[{Friedl(2011)}]{friedl2011}
\textsc{Friedl, H.} (2011).
\newblock \textit{Variance estimation in generalized linear models using the bootstrap}.
\newblock \textit{Statistical Science}.
\newblock \textbf{26}, 217--232.


\bibitem[{Friedman et al. (2007)}]{friedman2007pathwise}
\textsc{Friedman, J.},
\textsc{Hastie, T.},
\textsc{H{\"o}fling, H.}\&
\textsc{Tibshirani, R.} (2007).
\newblock \textit{Pathwise coordinate optimization}.
\newblock \textit{Annals of Applied Statistics}.
\newblock \textbf{1(2)}, 302--332.



\bibitem[{Friedman et al.(2010)}]{friedman2010regularization}
\textsc{Friedman, J.} \&
\textsc{Hastie, T.} \&
\textsc{Tibshirani, R.} (2010).
\newblock \textit{Regularization paths for generalized linear models via coordinate descent}.
\newblock \textit{Journal of Statistical Software}.
\newblock \textbf{33(1)}, 1--22.


\bibitem[{Fuk and Nagaev(1971)}]{fuk1971probability}
\textsc{Fuk, D. K.} \& 
\textsc{Nagaev, S. V.} (1971).
\newblock \textit{Probability inequalities for sums of independent random variables}.
\newblock \textit{Theory of Probability \& Its Applications}.
\newblock {\textbf{16(4)}}, 643--660.


 


\bibitem[{Geyer (1996)}]{geyer1996asymptotics}
\textsc{Geyer, C, J.} (1996).
\newblock \textit{On the asymptotics of convex stochastic optimization}.
\newblock \textit{Unpublished manuscript}.
\newblock {\textbf{37}}.


\bibitem[{Hjort and Pollard(1993)}]{hjort1993asymptotics}
\textsc{Hjort, N. L.} \& 
\textsc{Pollard, D.} (1993).
\newblock \textit{Asymptotics for minimisers of convex processes Technical Report}.
\newblock \textit{Yale University}.



\bibitem[{Jin et al.(2001)}]{jin2001simple}
\textsc{Jin, Z.},
\textsc{Ying, Z.} \&
\textsc{ Wei, L. J.} (2001).
\newblock \textit{A simple resampling method by perturbing the minimand}.
\newblock \textit{Biometrika}.
\newblock \textbf{88(2)}, 381--390.

\bibitem[{Kato (2009)}]{kato2009asymptotics}
\textsc{Kato, K.} (2009).
\newblock \textit{Asymptotics for argmin processes: Convexity arguments}.
\newblock \textit{Journal of Multivariate Analysis}.
\newblock {\textbf{100(8)}}, 1816--1829.


\bibitem[{Kim and Pollard (1990)}]{kim1990cube}
\textsc{Kim, J.} \&
\textsc{Pollard, D.} (1990).
\newblock \textit{Cube root asymptotics}.
\newblock \textit{The Annals of Statistics}.
\newblock \textbf{18(1)}, 191--219.


\bibitem[{Knight and Fu(2000)}]{knight2000asymptotics}
\textsc{Knight, K.} \&
\textsc{Fu, W.} (2000).
\newblock \textit{Asymptotics for Lasso-Type estimators}.
\newblock \textit{The Annals of Statistics}.
\newblock \textbf{28(5)}, 1356--1378.

\bibitem[{Lahiri (2021)}]{lahiri2021necessary}
\textsc{Lahiri, S. N.} (2021).
\newblock \textit{NECESSARY AND SUFFICIENT CONDITIONS FOR VARIABLE SELECTION CONSISTENCY OF THE LASSO IN HIGH DIMENSIONS}.
\newblock \textit{The Annals of Statistics}.
\newblock {\textbf{49(2)}}, 820--844.

\bibitem[{Lee(1990)}]{lee1990bootstrapping}
\textsc{Lee, K.-W.} (1990).
\newblock \textit{Bootstrapping logistic regression models with random regressors}.
\newblock \textit{Communications in Statistics---Theory and Methods}.
\newblock \textbf{19}(7), 2527--2539.


\bibitem[{Liu (1988)}]{liu1988bootstrap}
\textsc{Liu, R. Y.} (1988).
\newblock \textit{Bootstrap procedures under some non-i.i.d.\ models}.
\newblock \textit{Annals of Statistics}.
\newblock \textbf{16(4)}, 1696--1708.




\bibitem[{Ma and Kosorok(2005)}]{ma2005robust}
\textsc{Ma, S.} \& 
\textsc{Kosorok, M. R.} (2005).
\newblock \textit{{Robust semiparametric M-estimation and the weighted bootstrap}}.
\newblock \textit{Journal of Multivariate Analysis}.
\newblock {\textbf{96(1)}}, 190--217.


\bibitem[{Minnier et al.(2011)}]{minnier2011perturbation}
\textsc{Minnier, J. A.}, 
\textsc{Tian, L.} \&
\textsc{Cai, T.} (2011).
\newblock \textit{{A perturbation method for inference on regularized regression estimates}}.
\newblock \textit{Journal of the American Statistical Association}.
\newblock {\textbf{106(496)}}, 1371--1382.

\bibitem[{Moulton(1986)}]{moulton1986bootstrapping}
\textsc{Moulton, L. H.} (1986).
\newblock \textit{Bootstrapping generalized linear models with application to longitudinal data (logit)}.
\newblock Ph.D. thesis, The Johns Hopkins University.


\bibitem[{Moulton and Zeger(1989)}]{moulton1989}
\textsc{Moulton, L. H.} \&
\textsc{Zeger, S. L.} (1989).
\newblock \textit{Analyzing repeated measures on generalized linear models via the bootstrap}.
\newblock \textit{Biometrics}.
\newblock \textbf{45(2)}, 381--394.

\bibitem[{Moulton and Zeger(1991)}]{moulton1991}
\textsc{Moulton, L. H.} \&
\textsc{Zeger, S. L.} (1991).
\newblock \textit{Bootstrapping generalized linear models}.
\newblock \textit{Computational Statistics \& Data Analysis}.
\newblock \textbf{11(1)}, 53--63.




\bibitem[{Nelder and Wedderburn(1972)}]{nelder1972generalized}
\textsc{Nelder, J. A.} \&
\textsc{Wedderburn, R. W. M.} (1972).
\newblock \textit{Generalized linear models}.
\newblock \textit{Journal of the Royal Statistical Society: Series A (General)}.
\newblock \textbf{135(3)}, 370--384.

\bibitem[{Ng and Newton(2022)}]{ng2022random}
\textsc{Ng, T. L.} \& 
\textsc{Newton, M. A.} (2022).
\newblock \textit{{Random weighting in Lasso regression}}.
\newblock \textit{Electronic Journal of Statistics}.
\newblock {\textbf{16(1)}}, 3430--3481.

\bibitem[{Patr{\'\i}cio et al. (2018)}]{patricio2018using}
\textsc{Patr{\'\i}cio, M.},
\textsc{Pereira, J.},
\textsc{Cris{\'o}stomo, J.},
\textsc{Matafome, P.},
\textsc{Gomes, M.},
\textsc{Sei{\c{c}}a, R.}\&
\textsc{Caramelo, F.} (2018).
\newblock \textit{Using Resistin, glucose, age and BMI to predict the presence of breast cancer}.
\newblock \textit{BMC cancer}.
\newblock \textbf{18(1)}, 1--8.

\bibitem[{Pollard (1990)}]{pollard1990empirical}
\textsc{Pollard, D.} (1990).
\newblock \textit{Empirical processes: theory and applications}.
\newblock vol. {\textbf{2}}  CBMS Conference Series in Probability 
and Statistics,  Vol. 2. Hayward, CA: Institute of Mathematical Statistics. 

\bibitem[{Rockafellar(1997)}]{rockafellar1997convex}
\textsc{Rockafellar, R. T.} (1997).
\newblock \textit{Convex analysis}.
\newblock vol. {\textbf{11}} Princeton university press.




\bibitem[{Simonoff and Tsai(1988)}]{simonoff1988jackknifing}
\textsc{Simonoff, J. S.} \&
\textsc{Tsai, C.-L.} (1988).
\newblock \textit{Jackknifing and bootstrapping quasi--likelihood estimators}.
\newblock \textit{Journal of Statistical Computation and Simulation}.
\newblock \textbf{30}(3), 213--232.



\bibitem[{Tibshirani(1996)}]{tibshirani1996regression}
\textsc{Tibshirani, R.} (1996).
\newblock \textit{Regression shrinkage and selection via the lasso}.
\newblock \textit{Journal of the Royal Statistical Society: Series B (Methodological)}.
\newblock \textbf{58(1)}, 267--288.


\bibitem[{Winkler(1996)}]{winkler1996}
\textsc{Winkler, G.} (1996).
\newblock \textit{Bootstrapping goodness of fit statistics in loglinear Poisson models}.
\newblock \textit{Sonderforschungsbereich 386, Paper 53}, University of Munich.




\end{thebibliography}

\end{document}